\documentclass[aps,reprint,onecolumn,floatfix,nofootinbib,12pt]{revtex4-2}
\usepackage{hyperref}
\hypersetup{
  breaklinks=true, % splits links across lines
  colorlinks=true, % displays links as colored text instead of blocks
  pdfusetitle=true, % \title and \author values into pdf metadata
}
\usepackage{float}
\usepackage{epsfig}
\usepackage{graphicx}
\usepackage{amsmath}
\usepackage{amsfonts}
\usepackage{amssymb}
\usepackage{bm}
\usepackage{bbold}
\usepackage{epsfig}
\usepackage{graphicx}
\usepackage{color}
\usepackage{pictex}
\usepackage{mathtools}
\usepackage{extarrows}
\usepackage{footnote}
\usepackage{footmisc}
\usepackage{dsfont}
\usepackage{subcaption}
\usepackage{centernot}
\usepackage{setspace}
\newcommand{\s}{\scriptscriptstyle}
\newcommand{\footlabel}[2]{%
    \addtocounter{footnote}{1}%
    \footnotetext[\thefootnote]{%
        \addtocounter{footnote}{-1}%
        \refstepcounter{footnote}\label{#1}%
        #2%
    }%
    $^{\ref{#1}}$%
}
\begin{document}
\title{Topology of the Fermi surface and universality of the metal-metal and metal-insulator transitions:
$d$-dimensional Hatsugai-Kohmoto model as an example}
\author{Gennady Y. Chitov}
\affiliation{Bogoliubov Laboratory of Theoretical Physics,
Joint Institute for Nuclear Research, Dubna 141980, Russia}
%\affiliation{D\'{e}partement de physique, Universit\'{e} de Sherbrooke,
%Sherbrooke, Qu\'{e}bec, J1K 2R1 Canada}
\date{\today}

%
%
%
%xxxxxxxxxxxxxxxxxxxxxxxxxxxxxxxxxxxxxxxxxxxxxxxxxxxxxxxxxxxxxxxxxxxxxxxxxxxxxx
%
\begin{abstract}
The earlier theory \cite{Chitov:2025} of the quantum phase transitions related to the change of the Fermi Surface Topology (FST) is advanced. For such transitions the Fermi surface as a quantum critical manifold determined by the Lee-Yang zeros, the order parameter $\mathcal{P}$ as the $d$-volume of the Fermi sea, and the special FST universality class were introduced in \cite{Chitov:2025}. The exactly solvable Hatsugai-Kohmoto (HK) $d$-dimensional ($d=1,2,3$) model of interacting fermions is analyzed. We explore the relation between the Lee-Yang zeros, the Luttinger and the plateau (Oshikawa) theorems.  The validity of the Luttinger theorem in the HK model is confirmed. It is shown that the order parameter $\mathcal{P}$ and the FST universality class describe the transitions between metal and band/Mott insulators, as well as the Lifshitz and van Hove gapless-to-gapless transitions. The gapless phases are established to be the ferromagnetic Fermi liquids (metals), where the quantum critical manifold is also a locus of ferromagnetic order with spontaneous breaking of the rotational symmetry occurring at $T=0$.  In addition to the conventional paradigm with a continuous order parameter, we apply the homology theory to analyze the FST transitions. They are critical points of the Morse function. To quantify FST we use the Euler characteristic, which is calculated for each phase of the HK model. We claim that the FST universality class is robust with respect to interactions and other model details, under the condition that the conduction band function is smooth and the critical points are non-degenerate.
\end{abstract}
\maketitle

%
%
%%%%%%%%%%%%%%%%%%%%%%%%%%%%%%%%%%%%%%%%%%%%%%%%%%%%%%%%%%%%%%%%%%%%%%%%%%%%%%
%
\section{Introduction}\label{Intro}
%
%%%%%%%%%%%%%%%%%%%%%%%%%%%%%%%%%%%%%%%%%%%%%%%%%%%%%%%%%%%%%%%%%%%%%%%%%%%%%%
%
%
%

This paper is to the great extend a sequel of our previous work \cite{Chitov:2025}, where we studied quantum criticality and quantum transitions
between metal and insulating phases, as well as gapless-to-gapless transitions within the metal state of several $d$-dimensional fermionic models ($d=1,2,3$). This was done with the analysis of zeros (or poles) of the partition function.  This rigorous theory, pioneered by Yang and Lee \cite{YangLee:1952,*LeeYang:1952}, was proven to be very successful for studies of various models at equilibrium and even out of it. For a short list of references, see, e.g., \cite{Fisher:1965,Fisher:1980,Matveev:2008,Bena:2005,TongLiu:2006,Wei:2014,Chitov:2021,Chitov:2022DL}, and more references in there. One of the great advantages of this approach is that it can detect a phase transition without any knowledge of the symmetry breaking or the order parameter involved. In case of fermionic or spin models the order accompanying a quantum transition could be quite evasive, like  for example, string or brane order parameters, and one needs an extra effort to reveal duality between branes (strings) and conventional (local) orders with broken symmetries \cite{Chitov:2018,Chitov:2019,Chitov:2021,Chitov:2022DL,Chitov:2022}.

We have shown \cite{Chitov:2025} that the set of the Lee-Yang  (LY) zeros of a $d$-dimensional model at $T=0$ defines the $(d-1)$-dimensional Fermi surface (FS) which is the manifold of quantum critical points.\footnote{\label{LYfug}We use the term ``Lee-Yang zeros'' broadly for generic partition function's zeros expressed via microscopic and thermodynamic parameters (e.g. as in Eq.~\eqref{Z} and \eqref{LYZs} below) at finite volume $V$ and/or number of particles $N$, before the thermodynamic limit is taken. In the original work, Yang and Lee \cite{YangLee:1952,*LeeYang:1952} dealt with zeros of the partition function expressed as polynomial of degree $M(V)$ in fugacity. See also \cite{Huang:1987} for a nice concise presentation and footnote \ref{LYfug} below. In the proposed terminology, e.g., the Fisher's zeros are understood as a special case (class) of the LY zeros.}
The appearance or restructuring of the FS is a quantum phase transition. In the latter case it is called the Lifshitz transition  \cite{Lifshitz:1960}. See also  \cite{Volovik:2003,Volovik:2007,*Volovik:2017,*Volovik:2018} and  \cite{Horava:2005}. Appearance of a $(d-1)$-dimensional membrane in the momentum space \cite{Horava:2005} breaks the initial symmetry of the latter, and as a consequence gapless excitations around the FS appear as well, which is a hallmark of metallic phase. The FS coincides also with the manifold of zeros of the inverse single-particle Green's function at $T=0$.

To quantify the transition accompanied change of the Fermi Surface Topology (FST) (appearance/disappearance, openning/closure, connectedness change, etc), we introduced in \cite{Chitov:2025} the order parameter $\mathcal{P}$ for the gapless metallic phase via the $d$-volume of the Fermi sea $\mathcal{S}_d$. In this paper we consider the $d$-cubic lattice with the first Brillouin zone (BZ) $\mathbf{k} \in [-\pi, \pi]^d$, so the definition \cite{Chitov:2025} of the order parameter $\mathcal{P}$ reads:
\begin{equation}
\label{OPFS}
  \mathcal{P}  \coloneqq \mathcal{S}_d~\big(\mathrm{mod} (2 \pi)^d \big)~.
\end{equation}
Such definition of $\mathcal{P}$ is consistent with the scaling properties of other observables  within the special (FST) universality class determined by the critical indices \cite{Chitov:2025}:
\begin{equation}
\label{IndAll}
 \mathrm{FST:}~~ \nu^\prime = \nu= \frac12~, ~~z=2~, ~~\beta =d  \nu^\prime~,~~\gamma=\alpha= 1-d  \nu^\prime~, ~~\eta =d~,
\end{equation}
where all notations are standard \cite{Sachdev:2011}, except the additional index $\nu^\prime$ which is introduced as the critical exponent of the incommensurate (IC) wave number \cite{Nussinov:2012}. The FS transitions in two paradigmatic interacting metals -- the Fermi and the Tomonaga-Luttinger liquids -- belong to the universality class \eqref{IndAll}.  The FST indices satisfy the Rushbrooke equality
\begin{equation}
\label{FisherSc}
  \alpha+2\beta+\gamma =2~,
\end{equation}
and the (hyper)scaling relations
\begin{equation}
\label{Hyper}
   \gamma = (2-\eta)\nu~,~~2-\alpha =(d+z) \nu~,~~~2 \beta = (d+z-2+\eta) \nu~.
\end{equation}

The Hamiltonian of interacting spin-1/2 fermions residing on the $d$-cubic lattice we study in this paper, is almost a unique example of the exactly-solvable non-trivial interacting $d$-dimensional model. It has a reach phase diagram with several metallic and insulating phases. It was introduced and solved by Hatsugai and Kohmoto in Ref.~\cite{Hatsugai:1992} (a closely related model was proposed a year before by Baskaran in Ref.~\cite{Baskaran:1991}). The original study \cite{Hatsugai:1992} was followed up by several subsequent papers at a moderate pace \cite{Continentino:1994,Nogueira:1996,Vitoriano:2000}, until recently, when the interest to the Hatsugai-Kohmoto (HK) model has grown considerably, see, e.g.,  \cite{Phillips:2018,Phillips:2020,Si:2024,Zhao:2025,Phillips:2025,Millis:2025} and more references there.

Since the exact solution yields the results for the partition function, LY zeros, Green's function, etc, via explicit equations at the difficulty level of the statistical mechanics of free fermions, the HK model is an excellent playground to test and to advance the theory \cite{Chitov:2025}.
We show in this work that the order parameter \eqref{OPFS} and the FST universality class \eqref{IndAll} describe the transitions of the HK model from the metal (gapless) phase to the phases of band insulator (BI) or the interaction-driven Mott insulator (MI). The Lifshitz \cite{Lifshitz:1960} and van Hove \cite{VanHove:1953} transitions between the gapless phases of the model belong to the same class. It appears that the Lifshitz transition which separates two gapless phases (denoted as M and M2) and its critical properties were not addressed in earlier literature on the HK model.

We show in this paper that in addition to the description of the FST transitions  within the standard paradigm of critical phenomena \eqref{OPFS} and  \eqref{IndAll}, they can be also identified as critical points of the Morse function where the cohomology group changes \cite{Milnor:1963,NovikovIII:1990}. The discrete change of the FST is quantified by the Euler characteristic which is calculated for each phase of the HK model.

Another point specifically addressed in this work is relation between the LY zeros, the filling in the gapped and gapless phases (the Luttinger theorem \cite{Luttinger:1960}, for a recent comprehensive account and more references, see \cite{Seki:2017}), and the existence of quantized plateaux \cite{Affleck:1997,Oshikawa:2000} in the gapped incompressible phases (BI/MI).
The latter result is a generalization of the Lieb-Schultz-Mattis (LSM) theorem \cite{LiebSM:1961}. A particularly relevant to the present study is Oshikawa's  version of the LSM-type theorem for the lattice fermions \cite{Oshikawa:2000}.\footlabel{LSM-IC}{According to Oshikawa \cite{Oshikawa:2000}, for a quantum many-particle fermionic system on a periodic lattice with an exactly conserved particle number, a finite excitation gap is possible only if the particle number per unit cell of the ground state is an integer, i.e., at commensurate (rational) filling, otherwise the system is in a gapless state with an IC filling.}
We introduced the formulation of the Luttinger theorem  (LT) relating the filling $\bar n$ (or the $d$-volume of the Fermi sea $\mathcal{S}_d$) to the LY zeros
which is shown to be universally working for all phases. The LT is shown to hold in the HK model.

The HK gapless phases are established to be the ferromagnetic Fermi liquids (metals), where the quantum critical manifold (LY zeros) is also a locus of ferromagnetic (FM) order with spontaneous breaking of the rotational symmetry occurring at $T=0$. At this transition the disentanglement, measured by the residual entropy ($s_d$), vanishes, and the FM order parameter $m_s \sim s_d$  takes over. It appears \emph{continuously}, together with the Fermi surface, and vanished at the BI ends.

The rest of the paper is organized as follows: In Sec.~\ref{Model} we introduce the $d$-dimensional Hatsugai-Kohmoto model and give its main exact results.
Sec.~\ref{Trans} contains the results for the phases and transitions of the HK model for $d=1,2,3$.
The technical details on the LY formalism and derivations are collected in Appendix \ref{AppLY}; results on the ground state properties (Fock space, entropy, nature of the Fermi liquid, quantum ferromagnetism) are given in Appendix \ref{GSEntGF}; applications of the Morse (homology) theory are explained in Appendix \ref{Morse}; Appendix \ref{M2CPs} contains a special derivation for the $d=2$ case. The concluding remarks and discussion are given in Sec.~\ref{Concl}.

%
%
%%%%%%%%%%%%%%%%%%%%%%%%%%%%%%%%%%%%%%%%%%%%%%%%%%%%%%%%%%%%%%%%%%%%%%%%%%%%%%%
%xxxxxxxxxxxxxxxxxxxxxxxxxxxxxxxxxxxxxxxxxxxxxxxxxxxxxxxxxxxxxxxxxxxxxxxxxxxxxx
%
\section{Model, definitions, and some exact results}\label{Model}
%
%xxxxxxxxxxxxxxxxxxxxxxxxxxxxxxxxxxxxxxxxxxxxxxxxxxxxxxxxxxxxxxxxxxxxxxxxxxxxxx
%%%%%%%%%%%%%%%%%%%%%%%%%%%%%%%%%%%%%%%%%%%%%%%%%%%%%%%%%%%%%%%%%%%%%%%%%%%%%%%
%
%
%
Most of equations we collect in this section were either obtained  explicitly or follow straightforwardly from the results of the original paper by  Hatsugai and Kohmoto \cite{Hatsugai:1992} and of the following work by Nogueira and Anda \cite{Nogueira:1996}.

The HK model treats the interacting spin-1/2 fermions hopping on the $d$-dimensional (hyper)-cubic lattice of $L^d$ sites.
We will analyze the particle-hole symmetric version of the  HK Hamiltonian \cite{Hatsugai:1992}:
\begin{eqnarray}
\label{HKsum}
  \hat H &=& \sum_{\mathbf{k}} \hat H_{\mathbf{k}}~, \\
\label{HKs}
  \hat H_{\mathbf{k}} &=& \varepsilon(\mathbf{k})(\hat n_{\mathbf{k}\uparrow}+\hat n_{\mathbf{k}\downarrow})
  +U \Big(\hat n_{\mathbf{k}\uparrow}-\frac12 \Big) \Big(\hat n_{\mathbf{k}\downarrow}-\frac12 \Big)~,  \\
\label{Spec}
  \hat n_{\mathbf{k}\sigma} &=&  c^{\dag}_{\mathbf{k}\sigma}c_{\mathbf{k}\sigma},~~\varepsilon(\mathbf{k})=-2t \sum_{i=1}^{d} \cos k_i~.
\end{eqnarray}
The wave vectors are restricted to the first Brillouin zone (BZ) $\mathbf{k} \in [-\pi, \pi]^d$. We set the units such that $\hbar=k_B=1$, the lattice
spacing $a=1$, and the energy scale $2t=1$. Thus
\begin{equation}
\label{Eband}
  \varepsilon(\mathbf{k}) \in [-d, d]~.
\end{equation}
In this paper we will analyze the most interesting case of repulsion $U >0$ and $d=1,2,3$.

\begin{figure}[h]
\centering{\includegraphics[width=12.0cm]{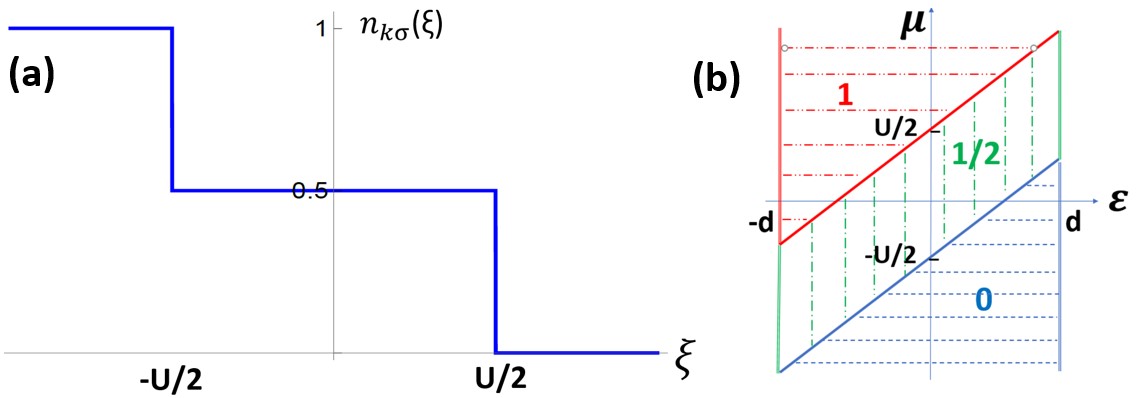}}
\caption{(a) The distribution function  $n_{\mathbf{k}\sigma}(\xi)$ at $T=0$ given by Eq.~\eqref{nksig0};
(b) The ground-state values $n_{\mathbf{k}\sigma}(\xi)=\{0,\frac12,1\}$, where $\xi=\varepsilon-\mu$, shown in the plane
$(\varepsilon,\mu)$. The bold solid lines are $\mu =\varepsilon \pm U/2$, the energy band $\varepsilon \in [-d,d]$.}
\label{Nk-AB}
\end{figure}

The  exact solvability of the HK model hinges on its fundamental property: the interacting term  $\hat H_U$  in the Hamiltonian commutes with its free
part $\hat H_0$ \cite{Hatsugai:1992}:
\begin{equation}
\label{Comm}
  [\hat H_U,\hat H_0]=0~.
\end{equation}
The eigenvalues of $\hat H_{\mathbf{k}}$ are
\begin{equation}
\label{HkEigen}
  \Bigg\{ \frac{U}{4}; ~\varepsilon(\mathbf{k})-\frac{U}{4};~\varepsilon(\mathbf{k})-\frac{U}{4};~2\varepsilon(\mathbf{k})+\frac{U}{4} \Bigg\} ~,
\end{equation}
whence the grand partition function is readily found:
\begin{eqnarray}
\label{Z}
   \mathcal{Z} &=& \mathrm{Tr} e^{- \beta (\hat H -\mu \hat N)}= \prod _{\mathbf{k}}  \mathcal{Z}_{\mathbf{k}}~, \\
 \mathcal{Z}_{\mathbf{k}} &=&  e^{- \beta U/4}+ 2 e^{- \beta(\xi_{\mathbf{k}}- U/4)}+e^{- \beta(2 \xi_{\mathbf{k}}+ U/4)}~,
\label{Zk}
\end{eqnarray}
with $\xi_{\mathbf{k}} \coloneqq \varepsilon(\mathbf{k}) - \mu$. The distribution function per spin
\begin{equation}
\label{nksigDef}
  n_{\mathbf{k}\sigma} \coloneqq \langle \hat n_{\mathbf{k}\sigma} \rangle =\frac12 n_{\mathbf{k}}
  =\frac{1}{2 \beta} \frac{\partial \ln \mathcal{Z}_{\mathbf{k}}  }{\partial \mu}
\end{equation}
(note that  $n_{\mathbf{k}\uparrow}=n_{\mathbf{k}\downarrow}$) is:
\begin{eqnarray}
\label{nksig}
  n_{\mathbf{k}\sigma} &=&
  \frac{1+ e^{ \beta(\xi_{\mathbf{k}}+ U/2)}}{1+ 2 e^{\beta(\xi_{\mathbf{k}}+ U/2)}+e^{2 \beta\xi_{\mathbf{k}} }} \\
  \xrightarrow[\beta \to \infty]{~}&~& \frac12 \big[ \Theta(U/2-\xi)+\Theta(-U/2-\xi) \big]
\label{nksig0}
\end{eqnarray}
It is plotted in Fig.~\ref{Nk-AB} at $T=0$.

%%%%%%%%%%%%%%%%%%%%%%%%%%%%%%%%%%%%%%%%%%%%%%%%%%%%%%%%%%%%%%%%%%%%%%%%%%%%%%%%%
%%%%%%%%%%%%%%%%%%%%%%%%%%%%%%%%%%%%%%%%%%%%%%%%%%%%%%%%%%%%%%%%%%%%%%%%%%%%%%%%%
%%%%%%%%%%%%%%%%%%%%%%%%%%%%%%%%%%%%%%%%%%%%%%%%%%%%%%%%%%%%%%%%%%%%%%%%%%%%%%%%%%
%%%%%%%%%%%%%%%%%%%%%%%%%%%%%%%%%%%%%%%%%%%%%%%%%%%%%%%%%%%%%%%%%%%%%%%%%%%%%%%%%%

The Matsubara single-particle Green's function for the ``particle-hole asymmetric'' Hamiltonian
$\hat H_{\mathbf{k}}^\prime = \varepsilon(\mathbf{k})(\hat n_{\mathbf{k}\uparrow}+\hat n_{\mathbf{k}\downarrow})
+U \hat n_{\mathbf{k}\uparrow} \hat n_{\mathbf{k}\downarrow}$ was found in \cite{Nogueira:1996}. For the symmetrized
Hamiltonian \eqref{HKs} the Green's function is \cite{Si:2024Let}
\begin{equation}
\label{GF}
  G_\sigma (\mathbf{k}, i \omega_n)= \frac{1- n_{\mathbf{k}\sigma}}{i \omega_n-\xi_{\mathbf{k}}+U/2}
  +\frac{n_{\mathbf{k}\sigma}}{i \omega_n-\xi_{\mathbf{k}}-U/2}
  = \frac{i \omega_n-\xi_{\mathbf{k}}+U(n_{\mathbf{k}\sigma}-1/2) }{(i \omega_n-\xi_{\mathbf{k}})^2-U^2/4}   ~,
\end{equation}
where $i \omega_n= i(2n+1)\pi/\beta$.

To better understand the physical properties of the HK model and its ground state phases, we will be analysing  zeros of the partition function \eqref{Z} in the subsequent sections. From the representation of the $\mathbf{k}$-component of the grand partition function $\mathcal{Z}_{\mathbf{k}}$
\eqref{Zk} as:
\begin{equation}
\label{ZkEq}
  \mathcal{Z}_{\mathbf{k}} =  e^{- \beta U/4} \big( e^{-\beta\xi_{\mathbf{k}}}- w_+ \big)
                             \big( e^{-\beta\xi_{\mathbf{k}}}- w_- \big)~,
\end{equation}
where
\begin{equation}
\label{wpm}
  w_\pm \coloneqq  e^{i(2n+1)\pi}  e^{\beta U/2} \big( 1 \pm \sqrt{1-e^{-\beta U}} \big)~,
\end{equation}
zeros of the partition function are found:\footnote{\label{LYfug} In terms of the fugacity $\varsigma \coloneqq e^{\beta \mu}$,  $\mathcal{Z}_{\mathbf{k}}$ is a quadratic polynomial. So the full $\mathcal{Z}$ is a polynomial $\varsigma$ of degree $2L^d$. That is exactly the original Lee-Yang setting. Its zeros are the roots of each quadratic  $\varsigma e^{- \beta \varepsilon (\mathbf{k}) } = w_\pm $, which is Eqs.~\eqref{ZkEq}–\eqref{LYZs} and Eq.~\eqref{LYZT0} in the $T \to 0$ limit.}
\begin{equation}
\label{LYZs}
  \mathcal{Z}_{\mathbf{k}}(\xi) =0~\Longrightarrow~ \xi_{\mathbf{k}}= \pm \Bigg( \frac{U}{2}+\frac{1}{\beta}
   \ln \Big(1+\sqrt{1-e^{-\beta U}} \Big) \Bigg) +i(2n+1)\pi/\beta~.
\end{equation}
As one can see from the above equation, the zeros lie in the real range ($ \xi_{\mathbf{k}} \in \mathds{R}$) only at $T=0$, that is only \textit{quantum} phase transitions can occur in the HK model in any spatial dimension.\footnote{\label{DL} Transitions of the type of disorder line can happen at finite $T$ when
the root is complex $ \xi_{\mathbf{k}} \in \mathbb{C}$  \cite{Chitov:2021}, but we will not discuss this issue in the present work.}

\begin{figure}[h]
\centering{\includegraphics[width=9.0cm]{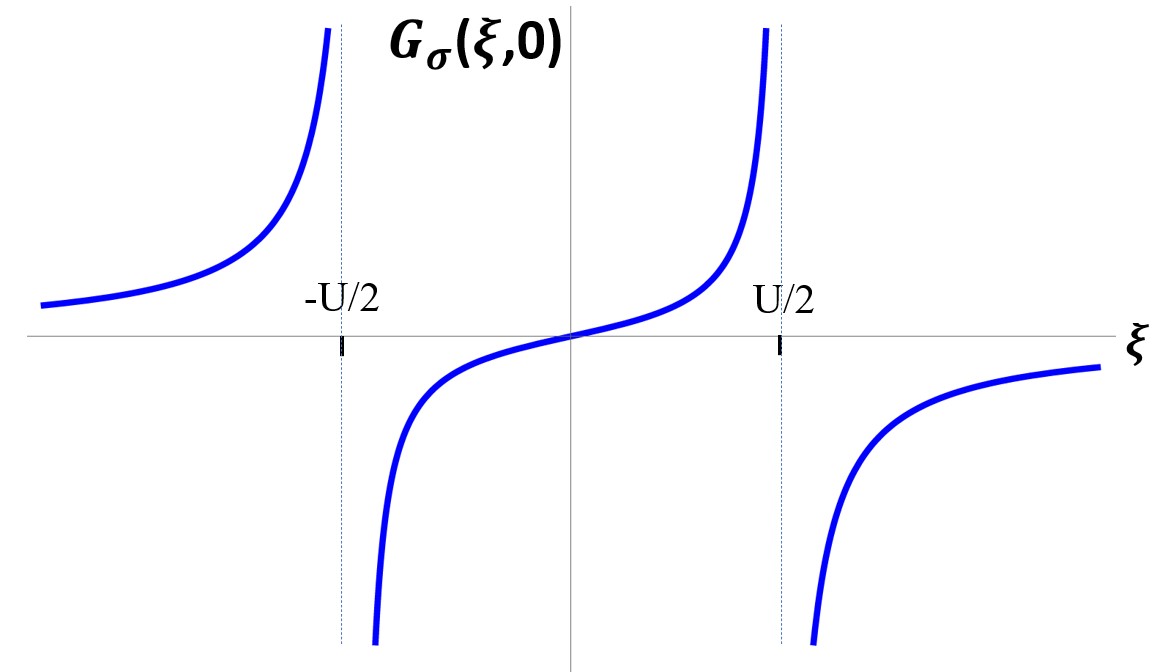}}
\caption{The Matsubara Green's function  $G_\sigma (\xi, 0)$ \eqref{GF} at $T=0$.
 Zero of $G_\sigma (\xi, 0)$ corresponds to $\xi=0$ and  $n_{\mathbf{k}\sigma}(\xi)=1/2$, see Fig.~\ref{Nk-AB}.}
\label{GFxi}
\end{figure}

We can also find easily poles of the Green's function \eqref{GF}:
\begin{equation}
\label{GFpoles}
  \xi_{\mathbf{k}}=\pm \frac{U}{2} +i(2n+1)\pi/\beta~.
\end{equation}
and its zeros:
\begin{equation}
\label{GFzeros}
   \xi_{\mathbf{k}} = U \big( n_{\mathbf{k}\sigma}- 1/2 \big) + i(2n+1)\pi/\beta~.
\end{equation}
The Matsubara Green's function $G_\sigma (\xi, 0)$  at $T=0$ is plotted in Fig.~\ref{GFxi}.

Contrary to the non-interacting case \cite{Chitov:2021},  zeros of the partition function \eqref{LYZs} and  poles of the Green's function \eqref{GFpoles} are distinct quantities when $U \neq 0$, and they coincide only in the limit $T \to 0$ yielding the single equation:
\begin{equation}
\label{LYZT0}
  \varepsilon(\mathbf{k})-\mu =\pm \frac{U}{2}~.
\end{equation}

%%%%%%%%%%%%%%%%%%%%%%%%%%%%%%%%%%%%%%%%%%%%%%%%%%%%%%%%%%%%%%%%%%%%%%%%%%%%%%%%%
%%%%%%%%%%%%%%%%%%%%%%%%%%%%%%%%%%%%%%%%%%%%%%%%%%%%%%%%%%%%%%%%%%%%%%%%%%%%%%%%%
%%%%%%%%%%%%%%%%%%%%%%%%%%%%%%%%%%%%%%%%%%%%%%%%%%%%%%%%%%%%%%%%%%%%%%%%%%%%%%%%%%
%%%%%%%%%%%%%%%%%%%%%%%%%%%%%%%%%%%%%%%%%%%%%%%%%%%%%%%%%%%%%%%%%%%%%%%%%%%%%%%%%%

Following \cite{Chitov:2025} we use the complex parametrization of the
$d$-dimensional wave numbers within the first BZ as:
\begin{equation}
\label{zi}
  z_i= e^{ \imath k_i}, ~i=1,...,d~,
\end{equation}
and express the LY zeros at $T=0$ as the roots of the following equation:
\begin{equation}
\label{LYZC}
\sum_{i=1}^{d}(z_i+z_i^{-1})/2=-\mu_\pm~,
\end{equation}
where
\begin{equation}
\label{mupm}
  \mu_\pm \coloneqq \mu \pm \frac{U}{2}~.
\end{equation}

Since there is a certain ongoing confusion in the literature regarding fulfillment of the LT in the HK model \cite{Hatsugai:1992,Phillips:2020,Zhao:2025}, we will specifically address this point in what follows. The general statement of the LT (Eq.~(33) in \cite{Luttinger:1960}) directly following from the particle conservation, is that the volume of the Fermi sea is \textit{invariant with respect to interactions}. Stated in a more general way to account for the contributions to $\bar N$ from fully filled bands $\bar n_i$, the LT  reads as:
\begin{equation}
\label{LuttGen}
 \bar n =\frac{\bar N}{V}= \sum_{i} \bar n_i + \frac{1}{(2 \pi)^d} \mathcal{S}_d(\mathrm{FS})  =~inv~,
\end{equation}
where $\mathcal{S}_d$ stands for total $d$-volume of the Fermi sea. In case of multiple connectivity of the Fermi seas, their $d$-volumes must be added on the r.h.s. of \eqref{LuttGen}.

We introduce the following formulation of the LT, relating filling $\bar n$ to the LY zeros:
\begin{equation}
\label{LuttLYZ}
 T=0:~ \mathcal{S}_d= (2 \pi)^d \bar n =\int_{A_{\s LYZ}^d} d \mathbf{k} = \Re \int_{C_{\s LYZ}^d} \prod_{i=1}^{d} \frac{d z_i}{\imath z_i}~.
\end{equation}
It holds for \textit{any} model Hamiltonian with a conserved particle number, and not only for the HK model, whose LY zeros are known exactly.
In Eq.~\eqref{LuttLYZ} the first integration is done in the interior of the $d$-dimensional region of the real $\mathbf{k}$-space within the BZ,  bounded by the
$(d-1)$-surface of the LY zeros  \eqref{LYZT0} (Fermi surface). In the second (mathematically equivalent) expression, the integration is over a real $d$-dimensional chain
$C_{\s LYZ}^d \subset \mathbb{C}_{cut}^d$,\footnote{\label{Cuts} Here we defined $\mathbb{C}_{cut}^d =\{(z_1,\dots,z_d)\in \mathbb{C}^d:\ z_i \neq 0,~ \arg z_i \in [-\pi,\pi],~ \forall ~i \}$ as a $d$-dimensional complex space with branch cuts along negative real semi-axes for all $z_i$.}
whose boundary lies on the variety of LY zeros \eqref{LYZC}. This somewhat terse definition should become clearer when we analyse the model for the cases $d=1,2,3$ in the following sections. We will explicitly show that the LT \eqref{LuttGen} in the form \eqref{LuttLYZ} is universally working for all phases, and it correctly accounts for the filled bands.

The definition \eqref{LuttLYZ} involves only more primary notion of the LY zeros, no need to deal with Green's function poles or zeros. See, e.g., \cite{Dzyalo:2003,Dzyalo:1996,Stanescu:2007,Gurarie:2011,Misawa:2022} on the latter point and Appendix~\ref{GFApp} for details.

%
%
%%%%%%%%%%%%%%%%%%%%%%%%%%%%%%%%%%%%%%%%%%%%%%%%%%%%%%%%%%%%%%%%%%%%%%%%%%%%%%
%
\section{Analysis of transitions: $d=1,2,3$}\label{Trans}
%
%%%%%%%%%%%%%%%%%%%%%%%%%%%%%%%%%%%%%%%%%%%%%%%%%%%%%%%%%%%%%%%%%%%%%%%%%%%%%%
%
%
%
%
%
%%%%%%%%%%%%%%%%%%%%%%%%%%%%%%%%%%%%%%%%%%%%%%%%%%%%%%%%%%%%%%%%%%%%%%%%%%%%%%%
%xxxxxxxxxxxxxxxxxxxxxxxxxxxxxxxxxxxxxxxxxxxxxxxxxxxxxxxxxxxxxxxxxxxxxxxxxxxxxx
%
\subsection{What can happen: Band, Lifshitz, Mott, and van Hove transitions}\label{Gen}
%
%xxxxxxxxxxxxxxxxxxxxxxxxxxxxxxxxxxxxxxxxxxxxxxxxxxxxxxxxxxxxxxxxxxxxxxxxxxxxxx
%%%%%%%%%%%%%%%%%%%%%%%%%%%%%%%%%%%%%%%%%%%%%%%%%%%%%%%%%%%%%%%%%%%%%%%%%%%%%%%
%
%
%
%

Although the HK model is interacting, its ground-state properties can be understood from analysis of the filling of two (upper/lower) free-like electron bands
(see Appendix \ref{GSEntGF} for details)
\begin{equation}
\label{EpsPM}
  \varepsilon_\pm \coloneqq \varepsilon(\mathbf{k}) \pm U/2~.
\end{equation}
Without interaction the model possesses a two-fold degenerate band $\varepsilon(\mathbf{k})$. When the noninteracting chemical potential lies in the interval $\mu_\circ \in [-d,d]$, the free model is in the gapless metallic phase. The critical values $\mu_\circ= \mp d$ correspond to the transitions into the phase of incompressible BI with empty/fully-filled bands, respectively. The interaction renormalizes these quantum critical points:
\begin{equation}
\label{muc12}
  \mu_\circ= \mp d~\longmapsto~\mu_{c,1/2}= \mp(d+U/2) \coloneqq \mp \mu_c~,
\end{equation}
while the physics is the same as for free fermions. These band metal-insulator transitions at $\mu =\mu_{c,1/2}$ with critical indices \eqref{IndAll} are discussed in \cite{Chitov:2025}, and the HK model fits that general framework.

The key role of interaction in the HK model is that it generates two additional quantum phase transitions  absent in the free model. In the following we call \textit{weak/strong
interactions} the cases $U<U_c$ and $U>U_c$, respectively, where
\begin{equation}
\label{Uc}
  U_c=2d~.
\end{equation}
1. \textit{Weak interaction:} the model stays in the gapless metal state while the bands are being continuously filled, all the way from $\mu=\mu_{c,1}$ up to the point $\mu=\mu_{c,2}$ with both bands totally filled. Splitting of the bands due to interaction \eqref{EpsPM} results in two Lifshitz metal-to-metal transitions within the gapless phase at
\begin{equation}
\label{muL12}
 \mu_{L,1/2}= \mp  (d-U/2)=\mp \frac12 (U_c-U) \coloneqq \mp \mu_L~,
\end{equation}
when appearance/dissapearance of the second Fermi surface occurs.

\begin{figure}[h]
\centering{\includegraphics[width=16.0cm]{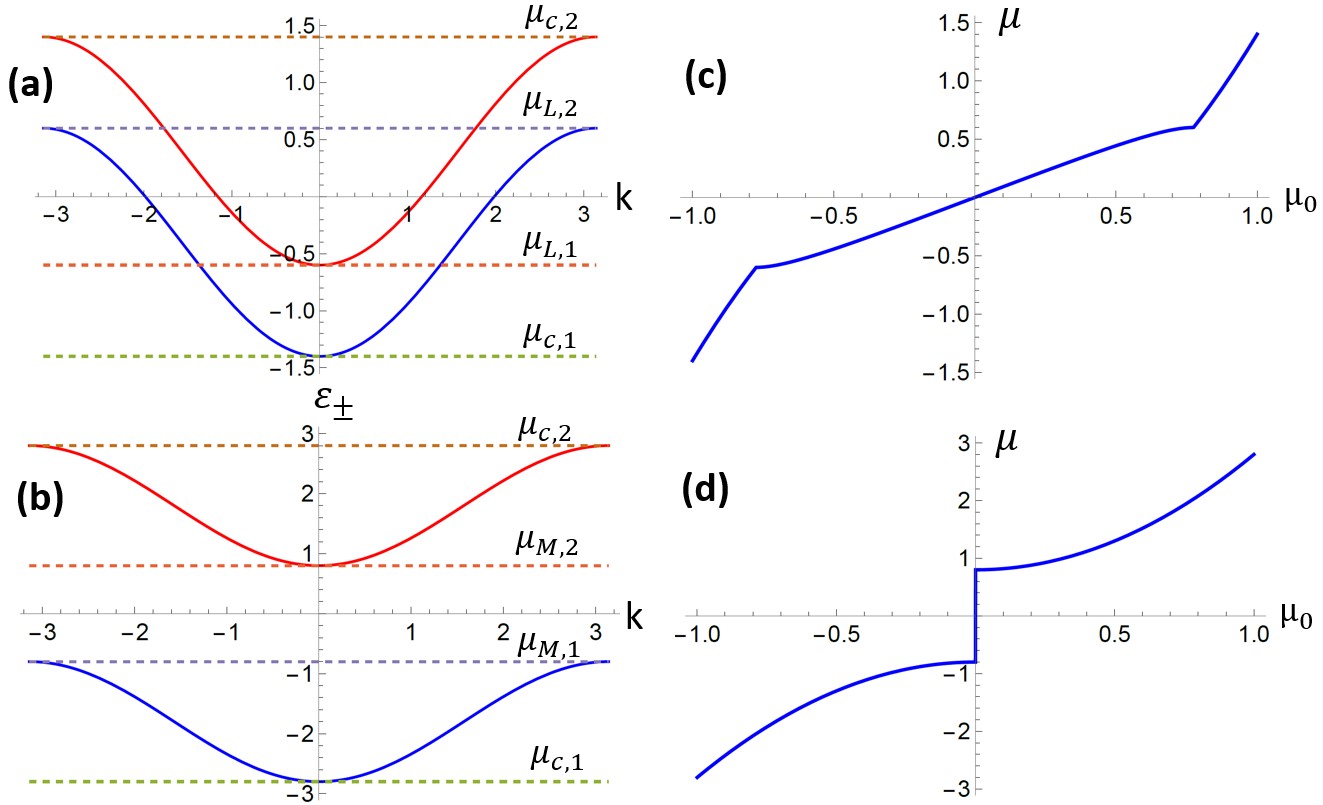}}
\caption{(a,b) One-dimensional spectra $\varepsilon_\pm(k)$  in the cases $U<U_c$ and $U>U_c$, respectively. The critical values of chemical potential:
(a): ($\mu_{c,1},\mu_{L,1},\mu_{L,2},\mu_{c,2}$), from bottom to the top;
(b):  ($\mu_{c,1},\mu_{M,1},\mu_{M,2},\mu_{c,2}$), from bottom to the top.
(c,d) The interacting chemical potential $\mu$ as a function of $\mu_\circ$ in the cases $U<U_c$ and $U>U_c$, respectively.
The cusp of $\mu$ at the Lifshitz point seen in panel (c) becomes a discontinuity at the Mott transition, panel (d).
The plots (a) and (c) are done for $U=0.8$, $\mu_{L,\sharp}=\pm 0.6$, $\mu_{c,\sharp}=\pm 1.4$ ; (b) and (d) for $U=3.6$, $\mu_{M,\sharp}=\pm 0.8$, $\mu_{c,\sharp}=\pm 2.8$.}
\label{Epm-Mu}
\end{figure}

2. \textit{Strong interaction:} with increasing $U$ splitting of the bands $\varepsilon_\pm$ leads to a gap between them,  and the Lifshitz transition becomes the Mott MI transition at $U \geq U_c$.  The latter occurs at
\begin{equation}
\label{muMott}
 \mu_{L,1/2} ~\mapsto~\mu_{M,2/1}= \pm (U/2-d)=\pm \frac12 (U-U_c) \coloneqq \pm \mu_M~.
\end{equation}
when the model is half-filled. The said above can be easily grasped qualitatively from Fig.~\ref{Epm-Mu}(a,b) where two bands $\varepsilon_\pm$ and critical values of the chemical potential in two cases $U<U_c$ and $U>U_c$ are plotted for $d=1$.\footnote{\label{12label} We used labels (1,2) in the above definitions of the critical points for convenience of the analysis presented along increasing $\mu$ (filling). When there is no risk of confusion we use the simplified notation, e.g., $\mu_c$, etc. All $\mu_\sharp \geq 0$ by definition.}

The model is \textit{gapless} if:
\begin{eqnarray}
\label{gplsCond}
  U &<& U_c~,~~ |\mu| < \mu_c~, \\
  U &>& U_c~,~~ \mu_M< |\mu| < \mu_c~.
\end{eqnarray}

Although vanishing of the FS at the transition into the band or Mott insulator can be interpreted as its restructuring, and thus, understood as a special
case of the Lifshitz transition  \cite{Lifshitz:1960,Volovik:2007,*Volovik:2017,*Volovik:2018,Chitov:2025}, in this paper we will use the latter term only for the gapless-to-gapless transitions occurring inside the metallic phase, when the FS changes its connectedness.

In general, quantum transitions correspond to the critical points of the spectrum
\begin{equation}
\label{CPs}
  \nabla_{\mathbf{k}}\varepsilon_\pm(\mathbf{k})=\nabla_{\mathbf{k}}\varepsilon(\mathbf{k})=0~.
\end{equation}
All transitions discussed above correspond to the maxima or minima of $\varepsilon(\mathbf{k})$. In $d \geq 2$ the saddle points occur among the critical points of $\varepsilon_\pm (\mathbf{k})$. At the saddle (van Hove) points $\mu_{vH}$ the topology of the FS changes, and a phase transition occurs, as to be discussed in the following sections.

The LT \eqref{LuttLYZ} for the HK model reads:\footnote{\label{LuttSimple} To appreciate the fundamental role of the LY theory in statistical mechanics, note that the LT theorem \eqref{LuttLYZ}, \eqref{LuttUD} follows immediately from the thermodynamic result \eqref{nksig0}. Since
$[\hat N ,\hat H_0]= [\hat N ,\hat H_U]=0$, so $N(\mu,U)=N(\mu_\circ,0)$, yielding straightforwardly \eqref{LuttUD} at $T=0$.}
\begin{equation}
\label{LuttUD}
   2 \mathcal{S}_d(\mu_\circ)=\mathcal{S}_d(\mu_+)+ \mathcal{S}_d(\mu_-)~.
\end{equation}
In the pertinent range $|\mu_\circ|<d$, Eq.~\eqref{LuttUD} defines the solution $\mu(\mu_\circ,U)$ as an odd function of $\mu_\circ$ for any $U \gtrless U_c$. The differential form of the LT \eqref{LuttUD} is found as:
\begin{equation}
\label{LuttDiffD}
  \frac{d \mu}{d \mu_\circ}= \frac{2 \mathcal{D}_d(\mu_\circ)}{\mathcal{D}_d(\mu_+)+\mathcal{D}_d(\mu_-)}~,
\end{equation}
where we introduced the single-band density of states (DOS) in the standard way:
\begin{equation}
\label{DOSdef}
  \mathcal{D}_d(E) \coloneqq \int \delta(E-\varepsilon(\mathbf{k})) \frac{d \mathbf{k}}{(2 \pi)^d}~.
\end{equation}
Using the result for compressibility
\begin{equation}
\label{chiD}
    \chi= \frac{\partial \bar{n}}{\partial \mu}= \frac12 \big[\mathcal{D}_d(\mu_+) +  \mathcal{D}_d(\mu_-) \big]~,
\end{equation}
Eq.~\eqref{LuttDiffD} can be also written as:
\begin{equation}
\label{LuttDiffDChi}
  \frac{d \mu}{d \mu_\circ}= \frac{\chi_\circ (\mu_\circ)}{\chi(\mu)}~,
\end{equation}
with the bare susceptibility $\chi_\circ (\mu_\circ)=\frac{\partial \bar{n}}{\partial \mu_\circ}$.

%%%%%%%%%%%%%%%%%%%%%%%%%%%%%%%%%%%%%%%%%%%%%%%%%%%%%%%%%%%%%%%%%%%%%%%%%%%%%%%%%
%%%%%%%%%%%%%%%%%%%%%%%%%%%%%%%%%%%%%%%%%%%%%%%%%%%%%%%%%%%%%%%%%%%%%%%%%%%%%%%%%
%%%%%%%%%%%%%%%%%%%%%%%%%%%%%%%%%%%%%%%%%%%%%%%%%%%%%%%%%%%%%%%%%%%%%%%%%%%%%%%%%%
%%%%%%%%%%%%%%%%%%%%%%%%%%%%%%%%%%%%%%%%%%%%%%%%%%%%%%%%%%%%%%%%%%%%%%%%%%%%%%%%%%

%
%
%%%%%%%%%%%%%%%%%%%%%%%%%%%%%%%%%%%%%%%%%%%%%%%%%%%%%%%%%%%%%%%%%%%%%%%%%%%%%%%
%xxxxxxxxxxxxxxxxxxxxxxxxxxxxxxxxxxxxxxxxxxxxxxxxxxxxxxxxxxxxxxxxxxxxxxxxxxxxxx
%
\subsection{d=1}\label{1D}
%
%xxxxxxxxxxxxxxxxxxxxxxxxxxxxxxxxxxxxxxxxxxxxxxxxxxxxxxxxxxxxxxxxxxxxxxxxxxxxxx
%%%%%%%%%%%%%%%%%%%%%%%%%%%%%%%%%%%%%%%%%%%%%%%%%%%%%%%%%%%%%%%%%%%%%%%%%%%%%%%
%
%
%
%
%
%%%%%%%%%%%%%%%%%%%%%%%%%%%%%%%%%%%%%%%%%%%%%%%%%%%%%%%%%%%%%%%%%%%%%%%%%%%%%%%
%xxxxxxxxxxxxxxxxxxxxxxxxxxxxxxxxxxxxxxxxxxxxxxxxxxxxxxxxxxxxxxxxxxxxxxxxxxxxxx
%
\subsubsection{Critical properties}\label{1DCrit}
%
%xxxxxxxxxxxxxxxxxxxxxxxxxxxxxxxxxxxxxxxxxxxxxxxxxxxxxxxxxxxxxxxxxxxxxxxxxxxxxx
%%%%%%%%%%%%%%%%%%%%%%%%%%%%%%%%%%%%%%%%%%%%%%%%%%%%%%%%%%%%%%%%%%%%%%%%%%%%%%%
%
%
%
The FST critical properties \eqref{IndAll} are most pronounced in one spatial dimension, when the phase transitions are of the second kind.
Qualitatively, the analysis below bears on some resemblance between the $1d$ HK model and the fermionic ladder dealt with earlier in \cite{Chitov:2025}.

The necessary details on the LY formalism for a single band in $1d$ are given in Sec.~\ref{LYd1}. We need to take into account
two LY zeros for each band of the  HK model at $T=0$, cf. \eqref{zkF1}:
\begin{equation}
\label{zLU}
  z_{\sharp \pm} =e^{\pm \imath k_\sharp} = -\mu_\sharp \pm  \imath \sqrt{\mu_\sharp ^2-1}~,
\end{equation}
where $\mu_\sharp$ stands for $\mu_\pm$. The LT, cf. Eqs.~\eqref{LuttLYZ} and \eqref{FillD1} yields
the occupation number (filling) per spin for all phases ($\forall~ U \geq 0$) through a single formula:
\begin{equation}
\label{Fil1D}
  \bar n(\mu) =  \Re \sum_\sharp \int_{z_{\sharp -}(\mu)}^{z_{\sharp +}(\mu)} \frac{d z }{2 \pi \imath z}=
 \frac{1}{2 \pi}  \Re \{ k_+ + k_- \} ~,
\end{equation}
where
\begin{equation}
\label{kpm}
  k_\pm = \arccos(-\mu_\mp )~.
\end{equation}
The spectrum of the HK model allows two types of solutions:
\begin{equation}
\label{Kpm}
 k_\pm=
  \left\{
  \begin{array}{lr}
  k_{F \pm} \in \mathds{R},~ &|\mu_\pm | \leq 1 \\ [0.2cm]
  \mathcal{Q}_\pm+ i \kappa_\pm \in \mathds{C},~ &|\mu_\pm | > 1 \\
  \end{array}
  \right.
\end{equation}
where $\mathcal{Q}_\pm$ admits \textit{only two values}:
\begin{equation}
\label{Qpm}
  \mathcal{Q}_\pm= \{0,\pi \}~.
\end{equation}
The case of the real IC roots $k_\pm  =k_{F \pm}$ corresponds to the quantum critical (gapless, metallic) phase.
The Fermi momenta near the critical points behave as:
\begin{equation}
\label{kFcr}
  k_{F \pm}- \mathcal{Q}_\sharp \propto \sqrt{|\mu-\mu_*|}~,
\end{equation}
where $\mu_*$ stands for the corresponding critical value of the chemical potential, thus the critical index of the IC wave number $\nu^\prime =1/2$.

\begin{figure}[h]
\centering{\includegraphics[width=16.0cm]{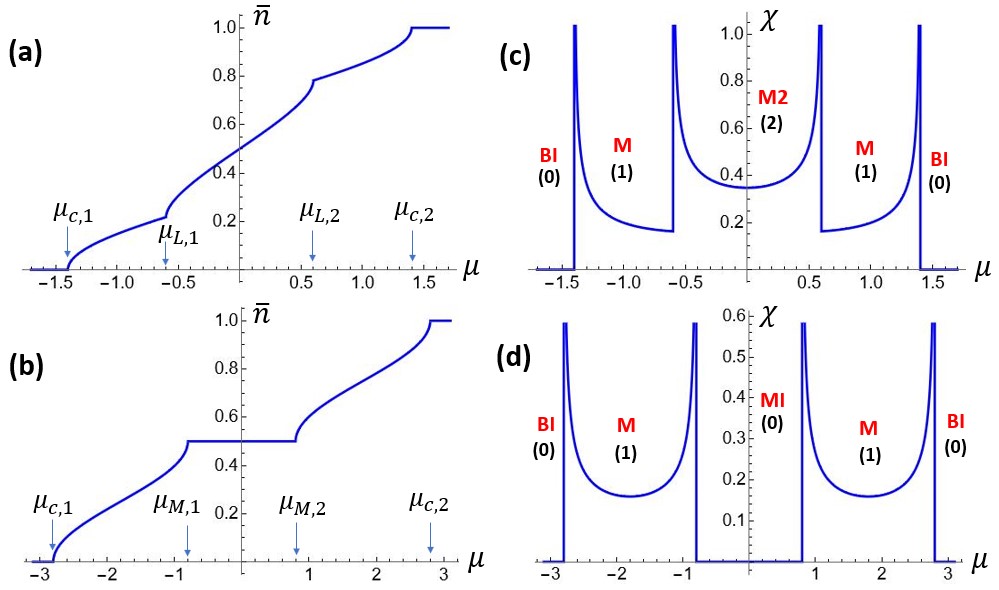}}
\caption{$d=1$ ($U_c=2$): (a,b) -- filling $\bar n(\mu)$;  (c,d) -- compressibility $\chi(\mu)$ in the cases $U<U_c$ and $U>U_c$.
The plots (a) and (c) are done for $U=0.8$ with the critical values of chemical potential: $\mu_{L,\sharp}=\pm 0.6$, $\mu_{c,\sharp}=\pm 1.4$ .
For the plots  (b) and (d): $U=3.6$, $\mu_{M,\sharp}=\pm 0.8$, $\mu_{c,\sharp}=\pm 2.8$. The Euler characteristics $\chi_E$ for each phase are shown in parentheses, see panels (c,d).}
\label{FilChi}
\end{figure}

The gapped phases (BI/MI) correspond to the range where both roots $k_\pm$ are complex. The real part of the LY root $\Re k_\pm$ gives the
commensurate (0,1/2) contribution to the filling, while $\kappa=\min \Im k_\pm$ yields the inverse correlation length. Since near the critical points
in the gapped phases $\kappa \propto \sqrt{|\mu-\mu_*|}$, the critical index $\nu =1/2$, and the equality  $\nu^\prime =\nu$ holds. The (additive) order parameter \eqref{OPFS} of the metallic phase behaves as $\mathcal{P} \propto \sqrt{|\mu-\mu_*|}$ near the critical points, yielding $\beta =1/2$.  The compressibility
\begin{equation}
\label{Comp1D}
    \chi=  \frac{1}{2 \pi }
    \Re \Bigg\{ \frac{1}{\sqrt{1-\mu_+^2}}+ \frac{1}{\sqrt{1-\mu_-^2}} \Bigg\}
\end{equation}
behaves near the critical points as $\chi \propto |\mu-\mu_*|^{-1/2}$, yielding $\gamma =1/2$.
The parameters $\bar n$ and $\chi$ are shown in Fig.~\ref{FilChi}. The Mott insulator is incompressible, see Figs.~\ref{FilChi}(d), and the critical properties of the metallic phase near the Mott transition are determined by the FST universality class \eqref{IndAll} with $d=1$.

Upon approaching the critical interaction from both sides $U \to U_c \pm 0$, two Lifshitz points $\mu_{L,\sharp}$ or the Mott points $\mu_{M,\sharp}$ merge
into a single critical point at $\mu_{L/M,1} \to \mu_{L/M,2} \to 0$, cf. Eqs.~\eqref{muL12} and \eqref{muMott}. The symmetric (with respect to $\mu \leftrightarrow -\mu$) features seen in Fig.~\ref{FilChi} at the Lifshitz and Mott critical points merge at $\mu=0$ when  $U \to U_c$.

%%%%%%%%%%%%%%%%%%%%%%%%%%%%%%%%%%%%%%%%%%%%%%%%%%%%%%%%%%%%%%%%%%%%%%%%%%%%%%%%%
%%%%%%%%%%%%%%%%%%%%%%%%%%%%%%%%%%%%%%%%%%%%%%%%%%%%%%%%%%%%%%%%%%%%%%%%%%%%%%%%%
%%%%%%%%%%%%%%%%%%%%%%%%%%%%%%%%%%%%%%%%%%%%%%%%%%%%%%%%%%%%%%%%%%%%%%%%%%%%%%%%%%
%%%%%%%%%%%%%%%%%%%%%%%%%%%%%%%%%%%%%%%%%%%%%%%%%%%%%%%%%%%%%%%%%%%%%%%%%%%%%%%%%%

%
%
%%%%%%%%%%%%%%%%%%%%%%%%%%%%%%%%%%%%%%%%%%%%%%%%%%%%%%%%%%%%%%%%%%%%%%%%%%%%%%%
%xxxxxxxxxxxxxxxxxxxxxxxxxxxxxxxxxxxxxxxxxxxxxxxxxxxxxxxxxxxxxxxxxxxxxxxxxxxxxx
%
\subsubsection{Luttinger and plateau theorems}\label{1DLutt}
%
%xxxxxxxxxxxxxxxxxxxxxxxxxxxxxxxxxxxxxxxxxxxxxxxxxxxxxxxxxxxxxxxxxxxxxxxxxxxxxx
%%%%%%%%%%%%%%%%%%%%%%%%%%%%%%%%%%%%%%%%%%%%%%%%%%%%%%%%%%%%%%%%%%%%%%%%%%%%%%%
%
%
%

The $1d$ LT \eqref{LuttUD} amounts to the following equation:
\begin{equation}
\label{LutTheo}
  2 \arccos(-\mu_\circ)= \bar n(\mu) ~,
\end{equation}
where $\mu_\circ$ is the chemical potential of the non-interacting model and $\bar n(\mu)$ is given explicitly by the r.h.s. of Eq.~\eqref{Fil1D}.
This equation we need to solve in order to find $\mu(\mu_\circ, U)$. The Fermi energy is understood as
\begin{equation}
\label{eF}
  \varepsilon_F=\mu(T=0)= \varepsilon(k_{F \pm}) \pm U/2~.
\end{equation}

$\underline{Weak~interaction:}$ when the bare chemical potential lies in the interval $\sqrt{1-U/2} < |\mu_\circ| <1$, $\mu$ is bound within the range $\mu_{L}< |\mu| < \mu_{c}$. In this case a simply connected Fermi sea is localized between two Fermi points $\pm k_{F\sharp}$.
See Fig.~\ref{Epm-Mu} (a) for visualization. When $|\mu_\circ| < \sqrt{1-U/2}$, ($|\mu| <\mu_{L}$) the Fermi sea is disconnected with four Fermi points
$\pm k_{F \pm}$. Explicitly we find:
\begin{eqnarray}
 \mathrm{At~~} \sqrt{1-U/2} < |\mu_\circ| <1 ~\Leftrightarrow ~ \mu_{L}< |\mu| < \mu_{c}: \nonumber \\
 k_{F\sharp} = 2\arccos(-\mu_\circ),~~ \mu = \pm(1-U/2-2 \mu_\circ^2)~,  \\
\label{M1}
\end{eqnarray}
and
\begin{eqnarray}
\mathrm{At~~} |\mu_\circ| < \sqrt{1-U/2}~ \Leftrightarrow ~ |\mu| <\mu_{L}: \nonumber \\
k_{F \pm} = \arccos(-\mu \pm U/2),~~ \mu = \mu_\circ \sqrt{1-\frac{U^2}{4(1-\mu_\circ^2)}}~.
\label{M2}
\end{eqnarray}
The Lifshitz transition at $\mu= \mu_{L,\sharp}=\pm (1-U/2)$, where the simply connected Fermi sea splits into two disconnected patches, corresponds to
$\mu_\circ = \pm  \sqrt{1-U/2}$. A representative curve $\mu(\mu_\circ)$ is  shown in Fig.~\ref{Epm-Mu} (c).

$\underline{Strong~interaction:}$ the bands $\varepsilon_\pm$ are separated by a gap, see Fig.~\ref{Epm-Mu}(b).
When $\mu$ varies within the range  $\mu_{M}< |\mu| < \mu_{c}$, the HK model is in the metal phase with a simply connected Fermi sea localized between $\pm k_F= \pm 2\arccos(-\mu_\circ)$. From invariance of the volume of the Fermi sea  we obtain:
\begin{equation}
\label{muRMott}
  \mathrm{At}~~0<|\mu_\circ|<1~\mathrm{(gapless~phase)}:~\mu = (U/2- 1+2 \mu_\circ^2) ~\mathrm{sign} (\mu_\circ)  ~,
\end{equation}
see Fig.~\ref{Epm-Mu}(d). The gap opens at half-filling when $\mu_\circ=0$. The interacting chemical potential \eqref{muRMott} is unambiguous when approaching the Mott phase from the metal sides $\mu \to \mp \mu_{M} \mp 0$. However $\mu(\mu_\circ)$  is an ill-defined function in the Mott phase corresponding to the range $[-\mu_{M},\mu_{M}]$ with the plateau $\bar n=1/2$, seen in  Fig.~\ref{FilChi} (b). Opening of the gap in the Mott phase appears as a discontinuity of the function  $\mu(\mu_\circ)$, compare Figs.~\ref{Epm-Mu} (c) and (d). We redefine
$\mu(\mu_\circ)$ such that
\begin{equation}
\label{mu0}
  \mu(0) \coloneqq 0~.
\end{equation}

As discussed in Appendix~\ref{LYd1}, the satisfaction of the Oshikawa theorem\footref{LSM-IC} is guaranteed by the  result \eqref{FillD1} following from the
LY version of the LT. Another way to put it: it is guaranteed by the only two types of solutions \eqref{Kpm} for the LY zeros admitted by the model's spectrum.
The gapless metallic phase corresponds to one or two real solutions $k_\pm =k_{F \pm}$, cf.  Eq.~\eqref{Kpm} ($\kappa=0$, infinite correlation length). The complex LY roots $k_\pm \in \mathds{C}$ corresponding to the gapped phases  ($\kappa \neq 0$ $\Leftrightarrow$ finite correlation length), occur with $\Re k_\pm = \mathcal{Q}_\pm \propto\bar n $. Since  $\mathcal{Q}_\pm$ only admits the values 0 or $\pi$,  the filling in the BI/MI phase demonstrates plateaux $\bar n=0,1/2,1$ when $(\mathcal{Q}_-,\mathcal{Q}_+)=(0,0)$,  $(\pi,0)$, $(\pi,\pi)$, respectively.  See Figs.~\ref{FilChi} (a,b).

Along with the order parameter in the metal phase $\mathcal{P}$ given by the measure of the Fermi sea, on can readily calculate the Euler characteristic which signals the topological transformations of the FS (change of homotopy classes, cohomology groups). See Appendix \ref{Morse} for definitions and details. The $1d$ case with two bands can be easily incorporated into the Morse theory outlined in App.~\ref{Morse}, if we deal with the problem of filling two circles placed at the different ground levels $\varepsilon=-1\mp U/2$. At weak interaction we find using Eqs.~\eqref{ChiECircle},\eqref{FSSum}, and \eqref{EulSum}:
\begin{equation}
\label{ChiEd1HK}
 U<2:~\chi_E =
  \left\{
  \begin{array}{lr}
  0,~ &|\mu|>1+U/2~~ \mathrm{(BI)}\\ [0.2cm]
  1,~ & 1-U/2< |\mu|< 1+U/2~~ \mathrm{(M)}\\ [0.2cm]
  2, ~& |\mu|< 1-U/2~~ \mathrm{(M2)}\\
  \end{array}
  \right.
\end{equation}
where M(M2) denotes metallic phase with a simply connected (disconnected) FS, respectively. $\chi_E$ changes at the critical points of band transitions where
FS appears/dissapears, as well at the Lifshitz points where the FS changes its connectedness.  At strong interaction the Euler numbers are:
\begin{equation}
\label{ChiEd1MottHK}
 U>2:~\chi_E =
  \left\{
  \begin{array}{lr}
  0,~ &|\mu|>U/2+1~~ \mathrm{(BI)}\\ [0.2cm]
  1,~ & U/2-1< |\mu|< U/2+1~~ \mathrm{(M)}\\ [0.2cm]
  0, ~& |\mu|< U/2-1~~ \mathrm{(MI)}\\
  \end{array}
  \right.
\end{equation}
For reader's convenience we have shown $\chi_E$ in different phases directly in Figs.~\ref{FilChi} (c,d).

%
%
%%%%%%%%%%%%%%%%%%%%%%%%%%%%%%%%%%%%%%%%%%%%%%%%%%%%%%%%%%%%%%%%%%%%%%%%%%%%%%%
%xxxxxxxxxxxxxxxxxxxxxxxxxxxxxxxxxxxxxxxxxxxxxxxxxxxxxxxxxxxxxxxxxxxxxxxxxxxxxx
%
\subsection{d=2}\label{2D}
%
%xxxxxxxxxxxxxxxxxxxxxxxxxxxxxxxxxxxxxxxxxxxxxxxxxxxxxxxxxxxxxxxxxxxxxxxxxxxxxx
%%%%%%%%%%%%%%%%%%%%%%%%%%%%%%%%%%%%%%%%%%%%%%%%%%%%%%%%%%%%%%%%%%%%%%%%%%%%%%%
%
%
%
%
%%%%%%%%%%%%%%%%%%%%%%%%%%%%%%%%%%%%%%%%%%%%%%%%%%%%%%%%%%%%%%%%%%%%%%%%%%%%%%%
%xxxxxxxxxxxxxxxxxxxxxxxxxxxxxxxxxxxxxxxxxxxxxxxxxxxxxxxxxxxxxxxxxxxxxxxxxxxxxx
%
\subsubsection{Critical properties}\label{2DCrit}
%
%xxxxxxxxxxxxxxxxxxxxxxxxxxxxxxxxxxxxxxxxxxxxxxxxxxxxxxxxxxxxxxxxxxxxxxxxxxxxxx
%%%%%%%%%%%%%%%%%%%%%%%%%%%%%%%%%%%%%%%%%%%%%%%%%%%%%%%%%%%%%%%%%%%%%%%%%%%%%%%
%
%
%
The key points of the analysis of this case will be essentially the same as for $d=1$, so we shall be more brief from now. The $2d$ FST transitions are weaker, cf. Eqs.~\eqref{IndAll}. All the technicalities on the LY formalism and the topological properties of critical points are collected in the Appendices \ref{LYd2} and \ref{Morse}. The main difference of $2d$ is the additional transitions at special fillings (van Hove singularities) coming from the saddle points at a half-filled single band. It is insightful to analyze the BI,  MI, Lifshitz, and the van Hove transitions of the $2d$ model in the framework of the Morse theory (see Appendix~\ref{Morse}).  These transitions are the critical points of the Morse (height) function which controls the filling of two ``collapsed" tori $\mathds{T}^2$ with removed central points (see Fig.~\ref{Donuts}). The \textit{communicating} tori are placed at the different ground levels $\varepsilon=-2 \mp U/2$. The transitions correspond to the critical points when $n$-cells are consecutively glued to the filled manifold (as explained for a single torus in  Appendix~\ref{Morse}) and the manifold's topology changes, until both tori are fully filled.

\begin{figure}[h]
\centering{\includegraphics[width=16.0cm]{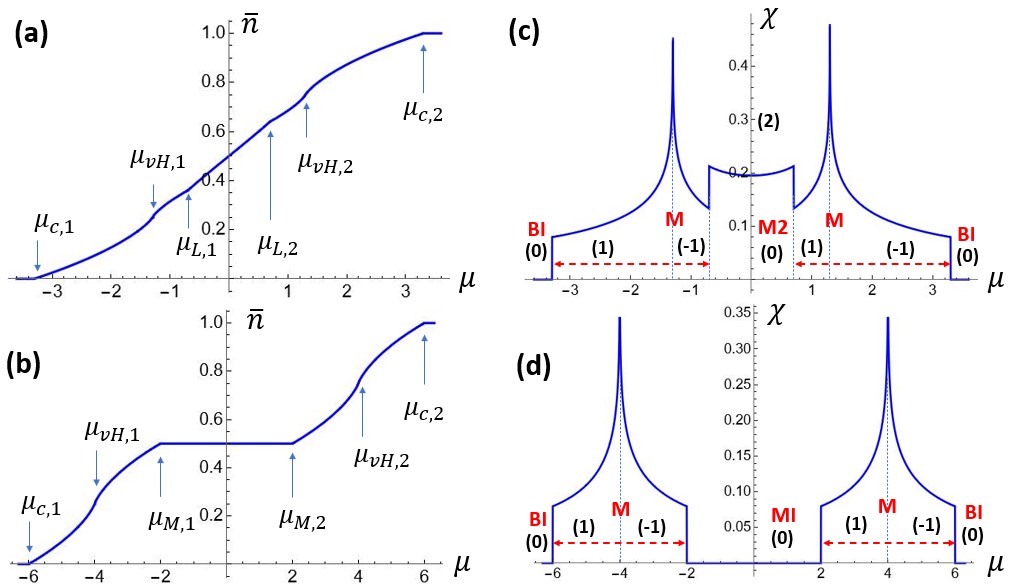}}
\caption{$d=2$ ($U_c=4$): (a,b) -- filling $\bar n(\mu)$;  (c,d) -- compressibility $\chi(\mu)$ in the cases $U<U_c$ and $U>U_c$.
The plots (a) and (c) are done for $U=2.6$ with the critical values of chemical potential: $\mu_{L,\sharp}=\pm 0.7$, $\mu_{c,\sharp}=\pm 3.3$.
The van Hove critical points correspond to  $\mu_{vH,\sharp}=\pm 1.3$ and $\bar n=1/4,3/4$.
For the plots  (b) and (d): $U=8$, $\mu_{M,\sharp}=\pm 2$, $\mu_{c,\sharp}=\pm 6$, $\mu_{vH,\sharp}=\pm 4$.
The Euler characteristics $\chi_E$ for each phase are shown in parentheses, see panels (c,d).}
\label{2DFilChi}
\end{figure}

We find the filling analytically as:
\begin{equation}
\label{Fill2D}
  \bar n = \frac{1}{2 } \sum_{\pm} \int_{\max\{ -\mu_\pm,-2\}}^{2}  \mathcal{D}_2(\varepsilon) d \varepsilon~,
\end{equation}
where the integration of the DOS \eqref{DOS2} yields the Meijer G-function \eqref{S2D2}.
The $2d$ compressibility is found from \eqref{chiD}.
The representative plots of $\bar n$ and $\chi$ are shown in Fig.~\ref{2DFilChi}.
Qualitatively, the behavior of these parameters at the critical points of the band  or  Mott/Lifshitz  transitions ($\mu_\sharp  =\pm(2 \pm U/2)$) is similar to the $1d$ case discussed in great detail above. The critical indices \eqref{IndAll} signal weaker singularities, e.g., the compressibility demonstrates only the cusps at those points. The new $2d$-feature is the van Hove transition at
\begin{equation}
\label{muVHove2D}
 \pm \mu_{vH}= \pm U/2~.
\end{equation}
with logarithmic  divergencies in $ \chi \propto D_2(\epsilon) \propto -\ln |\mu -\mu_{vH}|$, which however belong to the same FST universality class \eqref{IndAll} with $\gamma=0$ as the other $2d$ critical points.

%%%%%%%%%%%%%%%%%%%%%%%%%%%%%%%%%%%%%%%%%%%%%%%%%%%%%%%%%%%%%%%%%%%%%%%%%%%%%%%%%
%%%%%%%%%%%%%%%%%%%%%%%%%%%%%%%%%%%%%%%%%%%%%%%%%%%%%%%%%%%%%%%%%%%%%%%%%%%%%%%%%
%%%%%%%%%%%%%%%%%%%%%%%%%%%%%%%%%%%%%%%%%%%%%%%%%%%%%%%%%%%%%%%%%%%%%%%%%%%%%%%%%%
%%%%%%%%%%%%%%%%%%%%%%%%%%%%%%%%%%%%%%%%%%%%%%%%%%%%%%%%%%%%%%%%%%%%%%%%%%%%%%%%%%

The rest of the results we present for two cases separately:

\textit{1). Weak interaction:} we infer from Eqs.~\eqref{Fill2D}, \eqref{chiD}, and equations given  in Appendix \ref{LYd2} that in the metal phase
$\mu_{c,1} <\mu <\mu_{c,2}$ the model undergoes two Lifshitz and two van Hove transitions, corresponding to the points where the FS changes its topology:
simply connected $\leftrightarrow$ disconnected (Lifshitz) or closed (particle-like)  $\leftrightarrow$   open (hole-like) (van Hove). Those transitions are clearly
seen in Fig.~\ref{2DFilChi} c, plotted for $U=2.6$.  The additive order parameter  measuring contributions to the Fermi sea from both bands
\begin{equation}
  \mathcal{P} = \mathcal{P}_+ + \mathcal{P}_- ~,
\label{Psum}
\end{equation}
where $\mathcal{P}_\pm = \mathcal{S}_2(\mu_\pm)$,
is continuous and $\mathcal{P}>0$  through the gapless-to-gapless transitions, and $\mathcal{P}$ vanishes linearly upon approaching the BI phases (cf. \eqref{gBI}), in agreement with \eqref{IndAll}. Appearance/disappearance of the second Fermi sea near the Lifshitz transition where $\mathcal{P}_\pm \propto \epsilon$,   or the particle$\leftrightarrow$hole change of the FS near the van Hove critical point, where $\mathcal{P}_\pm \propto -\epsilon \ln\epsilon^2$, lead to the singularities
in the slope  of $\mathcal{P}(\mu)$ at the critical points, clearly seen in $\chi(\mu)$. Compare Figs.~\ref{2DFilChi} (a) and (c), and recall that $\mathcal{P}\propto \bar n$.

Another clean-cut indication of the transitions where the FS undergoes topological transformations is provided by the Euler characteristic $\chi_E$.  It changes at the critical points. $\chi_E$ is calculated as $\chi_E=\chi_E^+ + \chi_E^-$ (cf. Eq.~\eqref{EulSum}), where the contributions $\chi_E^\pm$ of the upper/lower bands are found from  \eqref{ChiETorus}. For each phase $\chi_E$ is shown in Fig.~\ref{2DFilChi} (c).

Note that the order of appearance of the Lifshitz and van Hove transitions changes at weak interaction $U<2$:  upon increasing filling ($\mu$) the overlap of the lower and upper bands at $\mu=\mu_{L,1}= -2+U/2$  occurs before the lower band becomes half-filled at $\mu=\mu_{vH,1}=-U/2$, cf. \eqref{muL12}, similarly on the positive side $\mu>0$. A simple exercise allows to check that the physics does not change, so below we present the results for the case when the van Hove transition occurs in the M1-phase with a single FS ($U>2$). The opposite case is analysed in Appendix \ref{M2CPs}.

\textit{2). Strong interaction:} the bands $\varepsilon_\pm$ are separated by the gap, their filling occurs consecutively, there is only one FS for
each metal phase, and no Lifshitz transitions exist in this case. The MI phase appears at $\bar n=1/2$ when $-\mu_{M}< \mu <\mu_{M}$. The critical properties of $\mathcal{P}$ are the same as in above case. Near the boundaries with the band/Mott insulator $\mathcal{P}_\pm \propto \epsilon$, and $\mathcal{P}_\pm \propto -\epsilon \ln\epsilon^2$ near the van Hove critical points. Representative plots are given in  Figs.~\ref{2DFilChi} (b) and (d). Qualitatively, the gapless phase M2 with two FS is replaced by the incompressible Mott phase. The Euler characteristics for this case are shown in Fig.~\ref{2DFilChi} (d).

%
%
%%%%%%%%%%%%%%%%%%%%%%%%%%%%%%%%%%%%%%%%%%%%%%%%%%%%%%%%%%%%%%%%%%%%%%%%%%%%%%%
%xxxxxxxxxxxxxxxxxxxxxxxxxxxxxxxxxxxxxxxxxxxxxxxxxxxxxxxxxxxxxxxxxxxxxxxxxxxxxx
%
\subsubsection{Luttinger theorem}\label{2DLutt}
%
%xxxxxxxxxxxxxxxxxxxxxxxxxxxxxxxxxxxxxxxxxxxxxxxxxxxxxxxxxxxxxxxxxxxxxxxxxxxxxx
%%%%%%%%%%%%%%%%%%%%%%%%%%%%%%%%%%%%%%%%%%%%%%%%%%%%%%%%%%%%%%%%%%%%%%%%%%%%%%%
%
%
%
The LT \eqref{LuttUD} in the pertinent range $|\mu_\circ|<2$  (cf. Eqs.~\eqref{gRdef} and \eqref{S2R}) amounts to:
\begin{equation}
\label{LuttG}
   2 g(\mu_\circ)= g_R(\mu_+)+ g_R(\mu_-)~,
\end{equation}
with the solution $\mu(\mu_\circ,U)$ as an odd function of $\mu_\circ$ for any $U \gtrless U_c$.

\textit{Weak interaction:} Eq.~\eqref{LuttG} yields a unique continuous solution $\mu(\mu_\circ,U)$ for each value of $\mu_\circ$ in the whole range $-2<\mu_\circ < 2$.
The function $\mu(\mu_\circ,U)$ found numerically, is qualitatively similar to the one shown in  Fig.~\ref{Epm-Mu}(c) for the $1d$ case, even smoother, because of the weaker $2d$ singularities. Although the function $\mu(\mu_\circ,U)$ is available only numerically, it is possible to find $\mu$ analytically near special points to leading order, using the results collected in Appendix \ref{LYd2}. The potential $\mu(\mu_\circ)$ vanishes at zero $\mu_\circ$ as
\begin{equation}
\label{mu1}
  \mu(\mu_\circ,U) \approx -\frac{2}{g^\prime(U/2)} \mu_\circ  \ln \frac{\mu_\circ^2}{4}, ~~|\mu_\circ| \ll 1~.
\end{equation}
In the vicinity of the BI we find from \eqref{gBI}:
\begin{equation}
\label{NearBI}
  \mu_\circ \to \pm 2 \mp 0:~ \mu = \mp(2-U/2) +2 \mu_\circ + \mathcal{O}(U^2)~,
\end{equation}
valid for $U \gtrless U_c$.

\textit{Strong interaction:} the situation is similar to the $1d$ case shown in  Fig.~\ref{Epm-Mu}(d): a unique continuous solution of Eq.~\eqref{LuttG} $\mu(\mu_\circ,U)$ exists in the range $0<|\mu_\circ| < 2$, but the interacting chemical potential is discontinuous at half-filling:
\begin{equation}
\label{MuJump2D}
  \mu_\circ \to \pm 0:~~ \mu \to \pm \mu_M~.
\end{equation}
To lowest order we find
\begin{equation}
\label{muMott1}
  \mu  \approx \pm (U/2-2)-\frac{2}{\pi} \mu_\circ  \ln \frac{\mu_\circ^2}{4}~,~~ |\mu_\circ| \ll 1
\end{equation}
As follows from the above equation, when $\mu_\circ \to \pm 0$, $\mu$ tends to its critical values $\pm \mu_M$ with an infinite slope, in agreement
with the exact result \eqref{DerM1}.

\textit{The two van Hove critical points} at $\mu = \mp U/2$ correspond to the values \eqref{134} of the bare chemical potential  $\mu \to \mp \mu_{\circ,s}= \mp 0.7198$, when the single non-interacting band is $1/4$ and $3/4$-filled. Then, correspondingly, the lower or upper band of the interacting model is $1/2$-filled. Using in Eq.~\eqref{LuttG} the expansions \eqref{gTaylor} and \eqref{gHalf}, we get in lowest order the transcendental equation for $\mu$ valid for $U>2$:
\begin{equation}
\label{MuTrans}
  g_1(\mu_\circ \pm \mu_{\circ,s}) = -\mu_\pm  \ln \frac{\mu_\pm^2}{4}
\end{equation}
Closely to the critical point when $|\mu_\pm|<2/e$ the approximate solution is:
\begin{equation}
\label{muVH2D}
  \mu \to + \mu_{vH}:~ \mu \approx  U/2 - g_1 (\mu_\circ - \mu_{\circ,s})/ \ln \Big( \frac{g_1 (\mu_\circ - \mu_{\circ,s})}{4} \Big)^2~.
\end{equation}
The approximation \eqref{muVH2D} retains the exact properties following from the LT \eqref{LuttG} and \eqref{DerM1}. At the van Hove points $\mu_\circ \to \pm \mu_{\circ,s}$ ($\mu \to \pm U/2$), the interacting chemical potential $\mu(\mu_\circ)$ is a continuous function without cusps, its slope
smoothly vanishes $\mu^\prime(\pm \mu_{\circ,s})=0$, and the higher-order derivatives  $\mu^{(n)}(\mu_{\circ,s}),~n \geq 2$, are divergent, in agreement with the exact result \eqref{DerM1}.\footnote{\label{DerSum} The exact Eq.~\eqref{DerM1} is valid for the M-phase with a single FS, where only one of two terms
on the r.h.s. of Eq.~\eqref{LuttG} contributes. In the M2-phase with two FSs, both terms contribute to $\frac{d \mu}{d \mu_\circ}$ in the region(s) where $[\mu^2_-<4] \cap [\mu^2_+<4] \neq \emptyset$, so the denominator on the r.h.s. of \eqref{DerM1} becomes $\mathbf{K}(1-\mu_+^2/4)+\mathbf{K}(1-\mu_-^2/4)$. In particular, this recovers the noninteracting limit  $\frac{d \mu}{d \mu_\circ}=1$ at $U \to 0$. See Eqs.~\eqref{DerM2} and \eqref{DerM1} in Appendix \ref{M2CPs}.}
In Appendix \ref{M2CPs} we show that the above results hold with inessential numerical differences for the case $U<2$, when the van Hove transition occurs inside the M2 phase, as well as in the limit $U \to 2$, when the Lifshitz and van Hove critical points merge.

\textit{Near the Lifshitz transition} we find:
\begin{equation}
\label{muLif}
 \mu \to + \mu_L:~\mu \approx \mu_L +2 (\mu_\circ- \mu_{\circ,\s L}) \times
  \left\{
  \begin{array}{lr}
   \frac{g^\prime(\mu_{\circ,\s L})}{g^\prime(2-U)},~ &\mu_{\circ} >\mu_{\circ,\s L}~~ \mathrm{(M)}\\ [0.2cm]
    \frac{g^\prime(\mu_{\circ,\s L})}{2 \pi +g^\prime(2-U)},~ &\mu_{\circ} <\mu_{\circ,\s L}~~ \mathrm{(M2),}\\
  \end{array}
  \right.
\end{equation}
where the bare potential at the Lifshitz point is determined from the following equation:
\begin{equation}
\label{mu0L}
   g(\mu_{\circ,\s L})= \pi^2+\frac12 g(2-U)~.
\end{equation}

%
%
%%%%%%%%%%%%%%%%%%%%%%%%%%%%%%%%%%%%%%%%%%%%%%%%%%%%%%%%%%%%%%%%%%%%%%%%%%%%%%%
%xxxxxxxxxxxxxxxxxxxxxxxxxxxxxxxxxxxxxxxxxxxxxxxxxxxxxxxxxxxxxxxxxxxxxxxxxxxxxx
%
\subsection{d=3}\label{3D}
%
%xxxxxxxxxxxxxxxxxxxxxxxxxxxxxxxxxxxxxxxxxxxxxxxxxxxxxxxxxxxxxxxxxxxxxxxxxxxxxx
%%%%%%%%%%%%%%%%%%%%%%%%%%%%%%%%%%%%%%%%%%%%%%%%%%%%%%%%%%%%%%%%%%%%%%%%%%%%%%%
%
%
%
This case is not essentially very different from $d=2$ of the previous subsection. Avoiding being too repetitive, we highlight below the key qualitative points without going into technical details. The $3d$ spectrum yields four the van Hove critical points, see sections \ref{LYd3} and \ref{Morse} in Appendices, and not just two as for $d=2$. We introduce an extra label to distinguish the type of the saddle point (S1 and S2, cf. Eq.~\eqref{CPsD3}) and define
\begin{equation}
\label{muVHS12}
  \mu_{\s vH}^{\s S1}=-1 \pm U/2~,~~\mu_{\s vH}^{\s S2}=1 \pm U/2~.
\end{equation}
The band filling given by the following equation
\begin{equation}
\label{n3DHK}
   \bar n(\mu) = \frac12  \int_{-3}^3 d x  \mathcal{D}_3(x) \big( \Theta(\mu_+ -x)+ \Theta(\mu_- -x) \big)~,
\end{equation}
cf. Eqs.~\eqref{Fill3D} and \eqref{DOS3D}, behaves similarly to that shown in Fig.~\ref{2DFilChi} a,b. For BI ($|\mu|>\mu_c$), $\bar n(\mu) =0,1$.\\
\textit{Weak interaction:} $\bar n(\mu)$ is a continuous monotonous function, as in Fig.~\ref{2DFilChi} a. \\
\textit{Strong interaction:} $\bar n$ has a plateau
$\bar n(\mu)=1/2$ in the Mott phase ($|\mu|<\mu_M$), as in Fig.~\ref{2DFilChi} b.

\begin{figure}[h]
\centering{\includegraphics[width=17.0cm]{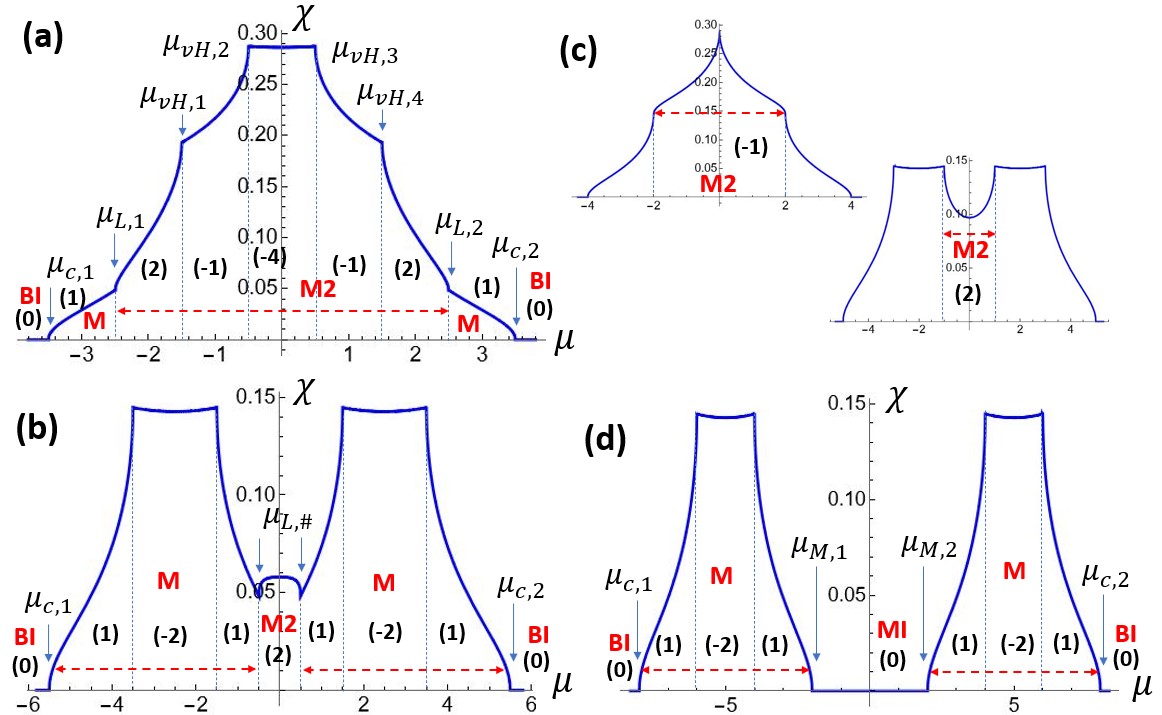}}
\caption{$d=3$ ($U_c=6$): Compressibility  $\chi(\mu)$ for different interaction strengths.
(a) $U=1$, critical values of chemical potential are: $\mu_{c,\sharp}=\pm 3.5$, $\mu_{L,\sharp}=\pm 2.5$, $\mu_{vH,\sharp}=\pm 1.5, \pm 0.5$.
(b) $U=5$, $\mu_{c,\sharp}=\pm 5.5$, $\mu_{L,\sharp}=\pm 0.5$, $\mu_{vH,\sharp}=\pm 3.5, \pm 1.5$.
(c) Two insets for the special cases when the Lifshitz and van Hove points merge. Upper left: $U=2$, $\mu_{c,\sharp}=\pm 4$, $\mu_{L,\sharp}=\mu_{vH,1/4}=\pm 2$,
$\mu_{vH,2/3}=0$. Lower right: $U=4$, $\mu_{c,\sharp}=\pm 5$, $\mu_{L,\sharp}=\mu_{vH,2/3}=\pm 1$, $\mu_{vH,1/4}=\pm 3$.
(d) $U=10$, $\mu_{c,\sharp}=\pm 8$, $\mu_{M,\sharp}=\pm 2$, $\mu_{vH,\sharp}=\pm 6, \pm 4$.
The Euler characteristics $\chi_E$ for each phase are shown in parentheses.}
\label{3DComp}
\end{figure}

The Lifshitz or van Hove points are not pronounced in $\bar n(\mu)$, while they are well noticeable in the behavior of the compressibility or DOS, cf. Eq.~\eqref{chiD}.
Several characteristic cases of $\chi(\mu)$ at different $U$ are presented in Fig.~\ref{3DComp}. Similarly to the $2d$ case, the order of appearance of the van Hove and Lifshitz points in the gapless regime can change, depending on the interaction strength, cf. Eqs.~\eqref{muL12} and \eqref{muVHS12}. At two particular values $U=2,4$ two of the van Hove points and the Lifshitz points merge. According to \eqref{DOS3DAs3}, the order parameter \eqref{OPFS} vanishes at the critical points $\mu_c,\mu_M$ of transitions into the BI or MI phases with exponent $\beta =3/2$. $\mathcal{P}$ is non-vanishing and continuous; it has an additive contribution with the
same exponent at the metal-to-metal (Lifshitz, van Hove) transitions. The compressibility clearly demonstrates the non-analyticities with $\gamma =-1/2$ at the critical points. The topology of the FS surface changes at each critical point, which is signalled by change of the Euler characteristic $\chi_E$. The latter is calculated using Eqs.~\eqref{muVHS12},\eqref{CPsD3},\eqref{ChiED3}, and \eqref{EulSum}. $\chi_E$ is shown in  Fig.~\ref{3DComp}.

%%%%%%%%%%%%%%%%%%%%%%%%%%%%%%%%%%%%%%%%%%%%%%%%%%%%%%%%%%%%%%%%%%%%%%%%%%%%%%%%%
%%%%%%%%%%%%%%%%%%%%%%%%%%%%%%%%%%%%%%%%%%%%%%%%%%%%%%%%%%%%%%%%%%%%%%%%%%%%%%%%%
%%%%%%%%%%%%%%%%%%%%%%%%%%%%%%%%%%%%%%%%%%%%%%%%%%%%%%%%%%%%%%%%%%%%%%%%%%%%%%%%%%
%%%%%%%%%%%%%%%%%%%%%%%%%%%%%%%%%%%%%%%%%%%%%%%%%%%%%%%%%%%%%%%%%%%%%%%%%%%%%%%%%%

The interacting chemical potential  $\mu(\mu_\circ,U)$ is found numerically from the LT \eqref{LuttUD} and \eqref{Fill3D}.
It agrees with \eqref{muc12}, i.e., $\mu(\pm 3,U)=\pm(3+U/2)$. At $U<U_c$, $\mu(\mu_\circ)$ is a monotonous odd function,
similar to the one shown in Fig.~\ref{Epm-Mu} (c), with $\mu(0,U)=0$. At $U>U_c$, $\mu(\mu_\circ)$ undergoes a discontinuity
in the Mott phase at half-filling ($\mu_\circ =0$), similarly to Fig.~\ref{Epm-Mu} (d):
\begin{equation}
\label{muMott3D}
  \mu_\circ \to \pm 0:~ \mu \to \pm(U/2-3)~.
\end{equation}

%
%
%
%xxxxxxxxxxxxxxxxxxxxxxxxxxxxxxxxxxxxxxxxxxxxxxxxxxxxxxxxxxxxxxxxxxxxxxxxxxxxxx
%
\section{Conclusion and discussion}\label{Concl}
%
%xxxxxxxxxxxxxxxxxxxxxxxxxxxxxxxxxxxxxxxxxxxxxxxxxxxxxxxxxxxxxxxxxxxxxxxxxxxxxx
%
%

The earlier theory of the FS-transitions \cite{Chitov:2025} is corroborated and advanced further. The exactly solvable $d$-dimensional HK model allows in particular to elucidate the role of interactions. The model demonstrates transitions accompanied by the appearance/disappearance of the FS between metal and the band or interaction-driven Mott insulators. It also undergoes the Lifshitz and van Hove gapless-to-gapless transitions in the metal phases, related to the topological changes of the FS (connectedness, openness, closure). All these transitions are described by the non-local topological order parameter \eqref{OPFS} and the FST universality class \eqref{IndAll}. The critical exponents $\nu^\prime,~\beta,~z,~\gamma,$ and $\alpha$ are independently determined from the exact results, and are shown to satisfy \eqref{IndAll} for $d=1,2,3$. At the transitions into the band/Mott insulator the order parameter  \eqref{OPFS} vanishes; and it demonstrates cusps  at the metal-metal (Lifshitz, van Hove) transitions.

In addition to the standard paradigm of critical phenomena \eqref{OPFS} and \eqref{IndAll}, we apply the homology (Morse) theory to analyze the metal-metal and metal-insulator transitions. They are identified as critical points of the Morse function where the cohomology group, the Betti numbers, and some other topological quantities of the Fermi sea change. To quantify discrete changes of the FS topology we use the Euler characteristic, which is calculated for each phase.

We build up the theory with the LY zeros as the primary fundamental point of the analysis, since the partition function gives directly the grand potential and other thermodynamic parameters. From the exact results of the HK model we explore the relation between the LY zeros, the Luttinger and the plateau (Oshikawa) theorems. In particular, we introduced the general (model-independent) formulation of the LT yielding the filling $\bar n$ via the integration in the region of $\mathds{C}^d$ complex space bounded by the LY zeros \eqref{LuttLYZ}. This formulation relating $\bar n$ (or the $d$-volume of the Fermi sea $\mathcal{S}_d$) to the LY zeros is shown to be working in the metal (gapless) phases, it correctly accounts for the filled bands, and it applies continuously to the gapped phases as well.

The general version of the LT \eqref{LuttGen} is shown to hold in the HK model via explicit calculations for $d=1,2,3$. The analytic expressions \eqref{LuttUD}, \eqref{LuttDiffD} of this theorem for the HK model determine the interacting chemical potential $\mu(\mu_\circ,U)$ as an odd function of $\mu_\circ$. The latter is the bare chemical potential of the non-interacting model. In the gapless regime  $\mu(\mu_\circ,U)$ is a continuous function; several explicit formulas for are calculated near the critical points (band, Lifshitz, Mott, van Hove). At strong interaction the Mott transition occurs at half-filling, when $\mu_\circ=0$, and a discontinuity is found in $\mu(\mu_\circ \to \pm 0) \to \pm \mu_M$.

The gapless phases (M, M2) of the HK model are characterized by: \\
  \textit{(i)} interaction-independent $d$-volume of the Fermi sea;\\
   \textit{(ii)} one (M) or two (M2) Fermi surfaces (quantum critical manifolds) defined by $T=0$ LY zeros at $\mu(\mu_\circ,U)$;
   \footnote{\label{Luk}  The low-energy sector of the gapless M2-phase preserves scaling, as of M-phase, but no conformal symmetry because of two distinct Fermi velocities $v_{\s F\pm}$. I am thankful to S.L. Lukyanov for discussions on related issues.} \\
   \textit{(iii)} simple poles \eqref{GFpoles} of the Green's function which control the low-energy fermionic excitations around the Fermi surface(s).\\
    \textit{(iv)} quantum ferromagnetism with spontaneously broken $SU(2)$ symmetry and gapless rigid rotational modes, occurring on the same quantum critical manifold of the LY zeros.\\
These phases are  identified as the \textit{ferromagnetic Fermi liquids} (ferromagnetic metals).

The important question to address: how robust and universal the FST universality class \eqref{IndAll} is?
The short answer: whatever is the model Hamiltonian of a metal,\footnote{\label{LL} It is not even necessary to be a normal metal, i.e., the Fermi liquid.
In the XXZ chain in the external field $h$, the Fermi sea (order parameter) in the gapless Tomonaga-Luttinger phase vanishes near the critical point $h_c$
as $\sqrt{h_c-h}$  \cite{Korepin:1986}, in agreement with \eqref{IndAll}.}
\textit{it suffices to have a ``good''  conduction band available for filling.} \\
To put it more rigorously:\\
1). Assume a model with the Hamiltonian $\hat H$ is defined on a Bravais lattice, such that $\hat H$ meets  the requirements of the \textit{Bloch theorem} (see, e.g. Ch.~8 in \cite{Ashcroft:1976}). Then the spectrum of electron eigenstates splits into the bands $\mathcal{E}_n(\mathbf{k})$ which are periodic functions of $\mathbf{k}$ on the reciprocal lattice. If in addition we require $[\hat H,\hat N]=0$, then the conjugate variable $\mu$ (chemical potential) is meaningfully defined and  controls the filling.\\
2). Assume for one (or several) band energy function $\mathcal{E}_i(\mathbf{k})$ the solution of
\begin{equation}
\label{GPless}
  \mathcal{E}_i(\mathbf{k})=\mu
\end{equation}
with real $\mathbf{k} \in \mathbb{R}^d$ exists, i.e., a conduction band(s) exists and the Fermi energy can be defined as $\mathcal{E}_F=\mu$.  With the requirements that the band energy function $\mathcal{E}_i(\mathbf{k})$ to be bounded and differentiable at least three times, no degenerate critical points, the \textit{Morse theory} can be applied.

\textit{Then}, the function $\mathcal{E}_i(\mathbf{k})$ (it can also be a sum of several $\mathcal{E}_i$ solving \eqref{GPless}) defined on a compactified BZ, can be identified with the Morse function with the height  $\mu$. For different models their Morse functions, the number of critical points, their order of appearance and multiplicity are different. However \textit{there are only maxima, minima, and saddle points} (for $d \geq 2$) among the critical points of any Morse function. These points are specified by $(d,s)$: for a given $d$ the index $s$ can admit values in the range $0 \leq s \leq d$. The maxima and minima correspond to the band, Mott, or Lifshitz transitions, while the saddle points correspond to the van Hove transitions. Since the expansion of  $\mathcal{E}_i(\mathbf{k})$ starts from the second order near a non-degenerate critical point, one always gets $z=2$. Only the signature of the expansion of the Morse function near the critical point (determined solely by $d$ and $s$) matters for the values of the critical exponents $\mathcal{P} \propto |\mu-\mu_c|^\beta$ and $\partial \mathcal{P}/\partial \mu \propto |\mu-\mu_c|^{-\gamma}$, \textit{which are always yielding the FST universality class \eqref{IndAll}}.

Along with the critical properties of the order parameter, compressibility (DOS), residual entropy, etc, characterized by the exponents  \eqref{IndAll}, topology of the FS undergoes discrete transformations at the transition points. The type and sequence of those transformations are not universal and depend on the model's details. For instance for the model with $d$-cubic lattice the Morse function is defined on the $d$-dimensional torus $\mathds{T}^d$. In this work we considered
several examples of how the topology of the FS undergoes consecutive transformation when $n$-cell (or cells) are added at each critical point, which is signalled by a change of the Euler characteristic, until the band gets fully filled and $\mathrm{FS} \mapsto \mathds{T}^d$.

At the degenerate critical points where the Hessian of the band energy $\mathcal{E}(\mathbf{k})$ vanishes, the Morse theory is not applicable.
Such high-order critical points occur at some particular (fine-tuned) values of the microscopic parameters, and we assume the situation with non-degenerate critical points of the Morse function as generic. The critical exponents at the high-order points are different from \eqref{IndAll}, see, e.g., \cite{Yuan:2020,Betouras:2020,Betouras:2025} and need to be studied on the case-by-case basis.

The very interesting aspects of the role of infinitely-ranged HK interaction, which allows to work out the exact $d$-dimensional FLT and  
to connect it to the properties of generic Fermi liquids, will be presented in a separate paper.

%
%xxxxxxxxxxxxxxxxxxxxxxxxxxxxxxxxxxxxxxxxxxxxxxxxxxxxxxxxxxxxxxxxxxxxxxxxxxxxxx
\begin{acknowledgments}
I thank A.M. Povolotsky for useful discussions and  A.-M. Tremblay for bringing important references to my attention.
I am very grateful to V.A. Timorin for reading the manuscript and illuminating discussions of the Morse theory.
I also thank an anonymous Referee \#5 whose insistence on several issued helped me to obtain more results and improve the presentation.  
The author acknowledges financial support from the Russian Science Foundation (RSF),
grant No.~24-22-00075.

\end{acknowledgments}
%xxxxxxxxxxxxxxxxxxxxxxxxxxxxxxxxxxxxxxxxxxxxxxxxxxxxxxxxxxxxxxxxxxxxxxxxxxxxxx

%xxxxxxxxxxxxxxxxxxxxxxxxxxxxxxxxxxxxxxxxxxxxxxxxxxxxxxxxxxxxxxxxxxxxxxxxxxxxxx

%%%%%%%%%%%%%%%%%%%%%%%%%%%%%%%%%%%%%%%%%%%%%%%%%%%%%%%%%%%%%%%%%%%%%%%%%%%%%%
\begin{appendix}
\section{Details on the Lee-Yang formalism and other derivations}\label{AppLY}
%%%%%%%%%%%%%%%%%%%%%%%%%%%%%%%%%%%%%%%%%%%%%%%%%%%%%%%%%%%%%%%%%%%%%%%%%%%%%%
%
%

To give necessary technical details on the results used in the main text, we consider a
single noninteracting band. We analyze the solutions of
\begin{equation}
\label{Eps0}
  -\sum_{i=1}^{d} \cos k_i = \varepsilon~,
\end{equation}
in the complex parametrization \eqref{zi}  with  $\varepsilon \in \mathds{R}$, $- \infty < \varepsilon < \infty$ for the three cases $d=1,2,3$.
The critical points of the spectrum \eqref{CPs} correspond to the following equations:
\begin{equation}
\label{LYCrP}
\frac{\partial \varepsilon }{\partial \ln z_i}=-\frac12 (z_i- z_i^{-1})=0~,~~i=1,...,d.
\end{equation}

%%
%%%%%%%%%%%%%%%%%%%%%%%%%%%%%%%%%%%%%%%%%%%%%%%%%%%%%%%%%%%%%%%%%%%%%%%%%%%%%%
\subsection{d=1}\label{LYd1}
%%%%%%%%%%%%%%%%%%%%%%%%%%%%%%%%%%%%%%%%%%%%%%%%%%%%%%%%%%%%%%%%%%%%%%%%%%%%%%
%%

\begin{figure}[h]
\centering{\includegraphics[width=7.0cm]{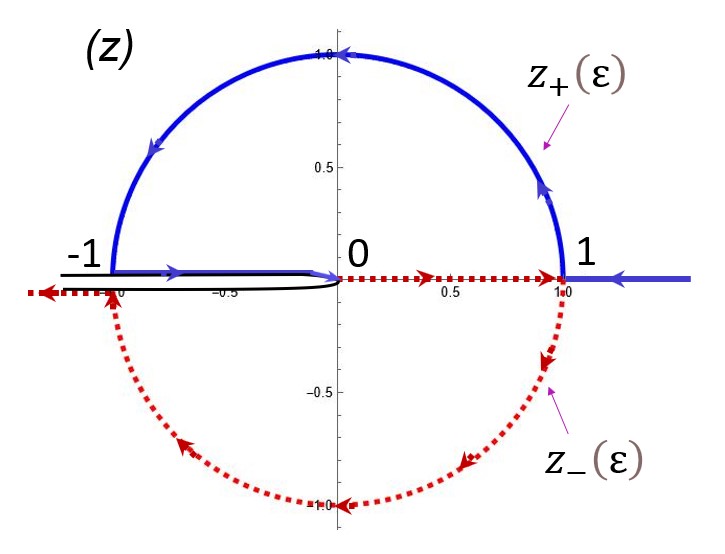}}
\caption{Path of the LY zeros $z_\pm (\varepsilon)$ on the complex plane $z$ as $\varepsilon$ varies from $- \infty$ to  $+ \infty$. The roots $z_\pm$
on the upper/lower edges of the branch cut along the negative real semi-axis are defined such that $\arg z_\pm = \pm \pi$.}
\label{ZpmD1path}
\end{figure}

In this case the roots of Eqs.~\eqref{Eps0} and \eqref{zi} are:
\begin{equation}
\label{zpmD1}
  z_\pm = -\varepsilon \pm \sqrt{\varepsilon^2-1}~,~~ z_+ z_-=1~.
\end{equation}
We trace their path on the complex plane $z$ as $\varepsilon$ varies from $- \infty$ to  $+ \infty$. The plane has
a branch cut along the negative real semi-axis, see Fig.~\ref{ZpmD1path}. $z_\pm$ are both real positive/negative for $\varepsilon<-1$ ($\varepsilon>1$),
respectively, and they are unimodular complex conjugate numbers at $|\varepsilon| \leq 1$.
\begin{equation}
\label{zkF1}
  z_+=z_-^\ast~,~~   z_+ =e^{\imath k_F} = -\varepsilon +  \imath \sqrt{1-\varepsilon^2}~.
\end{equation}

%%%%%%%%%%%%%%%%%%%%%%%%%%%%%%%%%%%%%%%%%%%%%%%%%%%%%%%%%%%%%%%%%%%%%%%%%%%%%%%%%
%%%%%%%%%%%%%%%%%%%%%%%%%%%%%%%%%%%%%%%%%%%%%%%%%%%%%%%%%%%%%%%%%%%%%%%%%%%%%%%%%
%%%%%%%%%%%%%%%%%%%%%%%%%%%%%%%%%%%%%%%%%%%%%%%%%%%%%%%%%%%%%%%%%%%%%%%%%%%%%%%%%%
%%%%%%%%%%%%%%%%%%%%%%%%%%%%%%%%%%%%%%%%%%%%%%%%%%%%%%%%%%%%%%%%%%%%%%%%%%%%%%%%%%

The complex version of the LT  \eqref{LuttLYZ} yields:
\begin{equation}
\label{FillD1}
  \bar n(\varepsilon)  = \Re \int_{z_-(\varepsilon)}^{z_+(\varepsilon)} \frac{d z }{2 \pi \imath z}=
  \Re \frac{1 }{2 \pi \imath } [\ln z_+(\varepsilon) - \ln z_-(\varepsilon)]= \frac{\arg z_+(\varepsilon)}{\pi}
\end{equation}
As one can see from the above result, the formulation of the LT via the LY zeros works smoothly for both gapped and gapless phases.
Eq.~\eqref{FillD1} connects $\bar n$ to the measure of the Fermi sea $2 \pi \bar n= 2 k_F = 2 \arccos(-\varepsilon)$  for the partially filled band ($|\varepsilon| <1$), and it works for the empty and fully filled bands as well, $\bar n =0,1$ ($|\varepsilon| >1$).

The LT  \eqref{FillD1} immediately leads to the LSM-type plateau theorem by Oshikawa \cite{Oshikawa:2000}: in the gapped phases when $|z_\pm| \neq 1$, the spectrum yields $\arg z_\pm = 0, \pm \pi$, so only commensurate fillings $\bar n =0,1$ can occur.  For the gapless phase the condition of criticality $|z_\pm|=1$ must hold, and $\arg z_+ =k_F$ yields the incommensurate filling.

Two extrema (min/max) of the spectrum \eqref{LYCrP} at $z_\pm =- \varepsilon=\mp 1$
correspond to the quantum phase transitions between the metal and BI. At $\varepsilon=-1$ the two roots merge at the real axis  with $z_+ =z_-=1$, see Fig.~\ref{ZpmD1path}. At $\varepsilon=1$ the roots tend to $z_+ =z_-=-1$ from the opposite edges of the branch cut: $z_\pm =\exp(\pm \imath \pi)$.
The critical points $\varepsilon=\pm 1$ bound two unit semicircles on the complex plane, which are the arcs of quantum criticality corresponding to $-1<\varepsilon<1$, see Fig.~\ref{ZpmD1path}.

%%
%%%%%%%%%%%%%%%%%%%%%%%%%%%%%%%%%%%%%%%%%%%%%%%%%%%%%%%%%%%%%%%%%%%%%%%%%%%%%%
\subsection{d=2}\label{LYd2}
%%%%%%%%%%%%%%%%%%%%%%%%%%%%%%%%%%%%%%%%%%%%%%%%%%%%%%%%%%%%%%%%%%%%%%%%%%%%%%
%%

The equations \eqref{LYCrP} for the critical points yield the minimum $\varepsilon=-2$ at $z_x=z_y=1$, the maximum  $\varepsilon=2$ at $z_x=z_y=-1$, and the saddle
(van Hove) points $\varepsilon=0$ at $z_x=-z_y= \pm 1$ when the band is half-filled. In the gapless range $-1<\varepsilon<1$, the FS is determined by the LY roots where $|z_x|=|z_y|=1$.
From the LT \eqref{LuttLYZ} one can get the filling as a single integral over a real parameter:
\begin{equation}
\label{FillD2}
  (2 \pi)^2 \bar n(\varepsilon)  = 2 \Im \int_0^\pi dk  [\ln z_+ (\varepsilon,k) - \ln z_-(\varepsilon,k)]
  = \mathcal{S}_2(\varepsilon) ~,
\end{equation}
where
\begin{equation}
\label{zpmD2}
  z_\pm (\varepsilon,k)= -a \pm \sqrt{a^2-1}~,~~a \coloneqq \varepsilon + \cos k~,
\end{equation}
and $\mathcal{S}_2(\varepsilon)$ denotes the area of the reciprocal space filled by the levels $\varepsilon(\mathbf{k})=-\cos k_x -\cos k_y \leq \varepsilon$.
Note that $z_+ z_-=1$. We chose a branch cut along the negative real semi-axis of $z$, and the roots $z_\pm$ on the upper/lower edges of the branch cut are defined such that $\arg z_\pm = \pm \pi$.

As in the $1d$ case, analysis of the positions of $z_\pm$ on the complex plane as $k$ varies from $0$ to $\pi$ at different $\varepsilon \in (- \infty,+ \infty)$, allows us to get a better insight on the connection between the LY zeros, Luttinger's  and Oshikawa's theorems. In the gapped states when $|\varepsilon| >  2$, both roots $z_\sharp (\varepsilon,k) \in \mathds{R}$. For $\varepsilon \lessgtr \mp 2$, the roots $z_\sharp (\varepsilon,k) \gtrless 0$, $\forall k \in [0,\pi]$. Then the integration in \eqref{FillD2} is trivial, yielding
\begin{equation}
\label{FillBI2D}
 \bar n =
  \left\{
  \begin{array}{lr}
  1,~ &\varepsilon >  2 \\ [0.2cm]
  0,~ &\varepsilon <  -2.  \\
  \end{array}
  \right.
\end{equation}
In the gapless region $|\varepsilon|<2$, $z_\pm$ are unimodular complex conjugate numbers $z_+=z_-^\ast$ in certain ranges of $k$.
At $-2< \varepsilon<0$, only $z_\pm$ sitting on the arcs of the unit circle $|z|=1$ contribute to \eqref{FillD2}, resulting in an IC filling. At the saddle critical point $\varepsilon=0$, when the diamond-shaped FS touches the edges of the BZ \cite{Economou:2006}, the roots $z_\pm$ run across
whole upper/lower semicircles during the integration in \eqref{FillD2}. At $\varepsilon>0$, $\bar n (\varepsilon)$ acquires additional contributions from the parts
where $z_\pm$ are located on the negative real semi-axis.

It is convenient to bring  \eqref{FillD2} to the following form:
\begin{equation}
\label{S2Gen2D}
  \mathcal{S}_2(\varepsilon)=4 \int_{\max\{-\varepsilon,-2\}}^2 \mathbf{K}(1-x^2/4) dx~,
\end{equation}
where the complete elliptic integral of the first kind $\mathbf{K}$ in the integrand comes from the DOS.
The latter is known from the literature, see, e.g., \cite{Economou:2006}:
\begin{equation}
\label{DOS2}
  \mathcal{D}_2(\varepsilon) =   \int \delta(\varepsilon -\varepsilon(\mathbf{k})) \frac{d \mathbf{k}}{(2 \pi)^2}
  =\frac{1}{\pi^2} \mathbf{K}(1-\varepsilon^2/4) \Theta(4-\varepsilon^2)~,
\end{equation}
where $\Theta(x)$ is the Heaviside step function.
In the gapless phase we find
\begin{equation}
\label{S2D2}
  \mathcal{S}_2(\varepsilon)=4 \int_{-\varepsilon}^2 \mathbf{K}(1-x^2/4) dx   =
  2 \pi^2 +\frac{\varepsilon}{ \pi} \mathcal{G}(\varepsilon^2/4) ~,~~|\varepsilon| \leq 2~,
\end{equation}
where $\mathcal{G}(x)$ is Meijer G-function.\footnote{\label{Meijer} This is the Mathematica built-in function
$MeijerG[\{ \{1/2, 1/2, 1/2\}, \{\} \}, \{\{0, 0\}, \{-1/2\}\}, x^2/4]$.}
It is convenient to introduce an odd auxiliary function
\begin{equation}
\label{gx}
  g(x) \coloneqq \frac{x}{ \pi} \mathcal{G}(x^2/4)~.
\end{equation}
Near the band end the Taylor expansion of $\mathcal{G}$ yields:
\begin{equation}
\label{gBI}
  g(x)= 2 \pi^2 - 2 \pi(2-x) - \frac{\pi}{4}(2-x )^2 -\frac{5 \pi}{96}(2-x)^3
  + \mathcal{O}((2-x)^4),~~ x \sim  2^-~,
\end{equation}
and near half-filling:
\begin{equation}
\label{gHalf}
  g(x) = - 2 x \ln \frac{x^2}{4}+\mathcal{O}(x),~x \sim  0.
\end{equation}
For the analysis of the interacting case we will need the expansion of $g(x)$ near the special points $\mp \zeta_\circ$ where
the band is $1/4-$ and $3/4$-filled:
\begin{equation}
\label{134}
  \zeta_\circ =0.71984482...:~~g(\pm \zeta_\circ)= \pm \pi^2~,
\end{equation}
and $g(x)$ is analytical:
\begin{equation}
\label{gTaylor}
  g(x)=\pi^2+g_1 (x - \zeta_\circ) - g_2 (x - \zeta_\circ)^2 +g_3 (x - \zeta_\circ)^3
  + \mathcal{O}((x - \zeta_\circ)^4)~.
\end{equation}
Numerically, $g_1=4 \mathbf{K}(1-\zeta_\circ^2/4)=9.82789$, $g_2=2.58482$, and  $g_3=1.31111$.

%We will also use analytical expansions  \eqref{gTaylor} of $g$ around other points $x_0$, using notations $g_n(x_0)$.

With the function $g_R(x)$ defined as:
\begin{equation}
\label{gRdef}
 g_R(x) \coloneqq
  \left\{
  \begin{array}{lr}
  2 \pi^2,~   &x >2 \\ [0.2cm]
   g(x),~     &|x|<2   \\ [0.2cm]
  -2 \pi^2 ,~ &x<-2,  \\
  \end{array}
  \right.
\end{equation}
the filled area is
\begin{equation}
\label{S2R}
  \mathcal{S}_2(\varepsilon)=2 \pi^2+ g_R(\varepsilon)~,~~\forall~ \varepsilon \in (-\infty, \infty)~.
\end{equation}
Note also that
\begin{equation}
\label{dDOS}
  g_R^\prime(\varepsilon)= 4 \pi^2 \mathcal{D}_2(\varepsilon)~,
\end{equation}
as follows from  definitions \eqref{gRdef} and \eqref{DOS2}.

%%
%%%%%%%%%%%%%%%%%%%%%%%%%%%%%%%%%%%%%%%%%%%%%%%%%%%%%%%%%%%%%%%%%%%%%%%%%%%%%%
\subsection{d=3}\label{LYd3}
%%%%%%%%%%%%%%%%%%%%%%%%%%%%%%%%%%%%%%%%%%%%%%%%%%%%%%%%%%%%%%%%%%%%%%%%%%%%%%
%%

The equations \eqref{LYCrP} for the critical points yield the minimum $\varepsilon=-3$ at $(z_x,z_y,z_z)=(1,1,1)$, the maximum  $\varepsilon=3$ at
$(z_x,z_y,z_z)=(-1,-1,-1)$, and two saddle (van Hove) points with $\varepsilon=-1$ at $(z_x,z_y,z_z)=(1,1,-1)$; $(1,-1,1)$; $(-1,1,1)$, and
with $\varepsilon=1$ at $(z_x,z_y,z_z)=(1,-1,-1)$; $(-1,1,-1)$; $(-1,-1,1)$. Cf. also Eq.~\eqref{CPsD3}. In the gapless range $-3<\varepsilon<3$, the FS is determined by the LY roots where $|z_x|=|z_y|=|z_z|=1$.
The Lee-Yang formulation of the LT  \eqref{LuttLYZ} can be brought to the form similar to \eqref{FillD2} with an extra integration.
The key steps of analysis and the outcomes are analogous to those of the case $d=2$. The filled volume is found as follows:
\begin{equation}
\label{Fill3D}
  \mathcal{S}_3(\varepsilon)= (2 \pi)^3  \bar n(\varepsilon) = (2 \pi)^3  \int_{-3}^3 d x  \mathcal{D}_3(x) \Theta(\varepsilon-x)~,
\end{equation}
where the single-band DOS is
\begin{equation}
\label{DOS3D}
  \mathcal{D}_3(\varepsilon) =
  \frac{1}{\pi^3} \int_{\varepsilon-1}^{\varepsilon+1}  \frac{d z}{\sqrt{1-(z-\varepsilon)^2}}  \mathbf{K}(1-z^2/4) \Theta(4-z^2)~.
\end{equation}
It vanishes near the band ends $\varepsilon= \pm 3$ as:
\begin{equation}
\label{DOS3DAs1}
  \mathcal{D}_3(\varepsilon) \propto \sqrt{|\varepsilon \mp 3|}, ~~\varepsilon \to \pm 3^\mp~,
\end{equation}
and in the vicinity of the van Hove (saddle) points $\varepsilon= \pm 1$ it behaves as:
\begin{equation}
\label{DOS3DAs2}
  \mathcal{D}_3(\varepsilon) - \mathcal{D}_3(\pm 1)
  \propto \sqrt{|\varepsilon \mp 1|} \Theta(-1 \pm \varepsilon) + \mathcal{O}(\varepsilon \mp 1), ~~\varepsilon \to \pm 1~,
\end{equation}
where $\mathcal{D}_3(\pm 1) =0.2893$. Note that the non-analytic square-root term appears only from one side of the van Hove points.
All this leads to the non-analytic contributions to the Fermi sea volume near the critical points \eqref{CPsD3}:
\begin{equation}
\label{DOS3DAs3}
  \delta \mathcal{S}_3(\varepsilon) \propto |\varepsilon -\varepsilon_c|^{3/2}~,
\end{equation}
so the exponents  $\gamma$ and $\beta$ agree with the general result \eqref{IndAll}. Note that for the $3d$ spectrum \eqref{Eps0}, the van Hove
critical point occurs at a generic IC filling $\bar n(\mp 1)=0.21327/0.78222$, while the point of half-filling $\varepsilon=0$ is not critical.

%xxxxxxxxxxxxxxxxxxxxxxxxxxxxxxxxxxxxxxxxxxxxxxxxxxxxxxxxxxxxxxxxxxxxxxxxxxxxxx
%xxxxxxxxxxxxxxxxxxxxxxxxxxxxxxxxxxxxxxxxxxxxxxxxxxxxxxxxxxxxxxxxxxxxxxxxxxxxxx

%%%%%%%%%%%%%%%%%%%%%%%%%%%%%%%%%%%%%%%%%%%%%%%%%%%%%%%%%%%%%%%%%%%%%%%%%%%%%%
\section{Ground state, residual entropy, and the nature of the gapless phase}\label{GSEntGF}
%%%%%%%%%%%%%%%%%%%%%%%%%%%%%%%%%%%%%%%%%%%%%%%%%%%%%%%%%%%%%%%%%%%%%%%%%%%%%%
%
%
In the earlier literature on the HK model, see, e.g.,  \cite{Hatsugai:1992,Vitoriano:2000}, the macroscopic degeneracy of the ground state of the HK model was found, leading to non-zero residual specific entropy $s_d$. The existence of $s_d \neq 0$ is sometimes linked in the literature to the alleged violation of the LT, which altogether feed speculations about some exotic non-Fermi liquid occurring  in the gapless phase of the HK model. In this Appendix we will specifically address these points.

%%
%%%%%%%%%%%%%%%%%%%%%%%%%%%%%%%%%%%%%%%%%%%%%%%%%%%%%%%%%%%%%%%%%%%%%%%%%%%%%%
\subsection{Ground state}\label{GSApp}
%%%%%%%%%%%%%%%%%%%%%%%%%%%%%%%%%%%%%%%%%%%%%%%%%%%%%%%%%%%%%%%%%%%%%%%%%%%%%%
%%

Since the total HK Hamiltonian is separable in the $\mathbf{k}$-space \eqref{HKsum}, its ground state can be constructed from eigenvectors of  $\hat H_{\mathbf{k}}$. The interaction does not generate the net magnetization in the HK model \cite{Hatsugai:1992}, and even more,
the number of particles is conserved for each spin separately  for each $\mathbf{k}$,
\begin{equation}
\label{nkudCom}
  [\hat n_{\mathbf{k}\uparrow},\hat H_{\mathbf{k}}]=[\hat n_{\mathbf{k}\downarrow},\hat H_{\mathbf{k}}]=0~,
\end{equation}
and so does the magnetization:
\begin{equation}
\label{Mk}
 \hat M_\mathbf{k} \coloneqq \hat n_{\mathbf{k}\uparrow}- \hat n_{\mathbf{k}\downarrow}~,~~   [\hat M_\mathbf{k},\hat H_{\mathbf{k}}]=0~.
\end{equation}
Note that since
\begin{equation}
\label{HuMk}
  \hat H_{\mathbf{k}}^{\s U}- \frac14 U= \frac12 U \hat n_\mathbf{k}  +U \hat n_{\mathbf{k}\uparrow} \hat n_{\mathbf{k}\downarrow}
  =- \frac12 U \hat M_\mathbf{k}^2~,
\end{equation}
the interacting Hamiltonian $\hat H_{\mathbf{k}}\neq H_{\mathbf{k}\uparrow}+\hat H_{\mathbf{k}\downarrow}$.
For a given $\mathbf{k}$ we will specify the quantum states by the eigenvalues $(H_\mathbf{k}^\prime,n_\mathbf{k},M_\mathbf{k})$ of commuting operators  $\hat H_{\mathbf{k}}$, $\hat n_{\mathbf{k}}$, and $\hat M_{\mathbf{k}}$, where  $\hat H_\mathbf{k}^\prime \coloneqq  \hat H_\mathbf{k}-U/4$.
The orthonormal set of four eigenvectors of $\hat H_\mathbf{k}^\prime$ with the eigenvalues  \eqref{HkEigen} is found as:
\begin{eqnarray}
  (H_\mathbf{k}^\prime,n_\mathbf{k},M_\mathbf{k}&)&,~~\mathrm{eigenvector}: \nonumber\\
  (0,0,0&)&,~~|0 \rangle; \nonumber\\
  (\varepsilon_-,1,\pm 1&)&,~~ |\Phi_{1 \sigma} \rangle =c^{\dag}_{\sigma}(\mathbf{k}) |0 \rangle;  \nonumber \\
  (2\varepsilon,2,0&)&,~~ |\Phi_2 \rangle = c^{\dag}_{\uparrow}(\mathbf{k}) c^{\dag}_{\downarrow}(\mathbf{k}) |0 \rangle~.
\end{eqnarray}
Note that $\varepsilon_+$ is not an eigenvalue of $\hat H_{\mathbf{k}}$, these excitations appear only in
the two-particle state $|\Phi_2 \rangle$ as $2 \varepsilon=\varepsilon_+ +\varepsilon_-$. The thermodynamic average in the grand canonical ensemble at $T=0$
is equivalent to the quantum-mechanical average with the ground state vector $|GS \rangle$. The latter is the eigenvector of
$\hat H_\mathbf{k}^\prime-\mu \hat n_\mathbf{k}$ with the lowest eigenvalue \cite{LandauV9}. Thus we find:
\begin{equation}
\label{GS}
 U>0:~ |GS \rangle =
  \left\{
  \begin{array}{lr}
  |0 \rangle,~ &\xi_- >0,~  H_\mathbf{k}^\prime-\mu n_\mathbf{k}=0, \\ [0.2cm]
  | \Phi_{1 \pm} \rangle,~ &(\xi_- <0) \cup (\xi_+ >0),~ H_\mathbf{k}^\prime-\mu n_\mathbf{k}=\xi_-,\\ [0.2cm]
  |\Phi_2 \rangle,~ &(\xi_- <0) \cup (\xi_+ <0),~  H_\mathbf{k}^\prime-\mu n_\mathbf{k}=\xi_+ + \xi_- ~,\\
  \end{array}
  \right.
\end{equation}
where we introduced the single-particle states
\begin{equation}
\label{Phi1}
  | \Phi_{1 \pm} \rangle = \frac{1}{\sqrt{2}} \big(  |\Phi_{1 \uparrow} \rangle \pm | \Phi_{1 \downarrow} \rangle \big)
\end{equation}
with $\langle \Phi_{1 \pm} | \hat M_\mathbf{k}| \Phi_{1 \pm} \rangle=0$  and $\xi_\pm \coloneqq \varepsilon_\pm - \mu$. With the above equations we get
\begin{equation}
\label{nkGS}
 U>0:~ n_\mathbf{k}=\langle GS | \hat n_\mathbf{k}|GS \rangle =
  \left\{
  \begin{array}{lr}
  0,~ &\xi_- >0, \\ [0.2cm]
  1,~ &(\xi_- <0) \cup (\xi_+ >0), \\ [0.2cm]
  2,~ &(\xi_- <0) \cup (\xi_+ <0),\\
  \end{array}
  \right.
\end{equation}
which is exactly the result \eqref{nksig0} obtained from thermodynamics in the limit  $T \to 0$.

\begin{figure}[h]
\centering{\includegraphics[width=12.0cm]{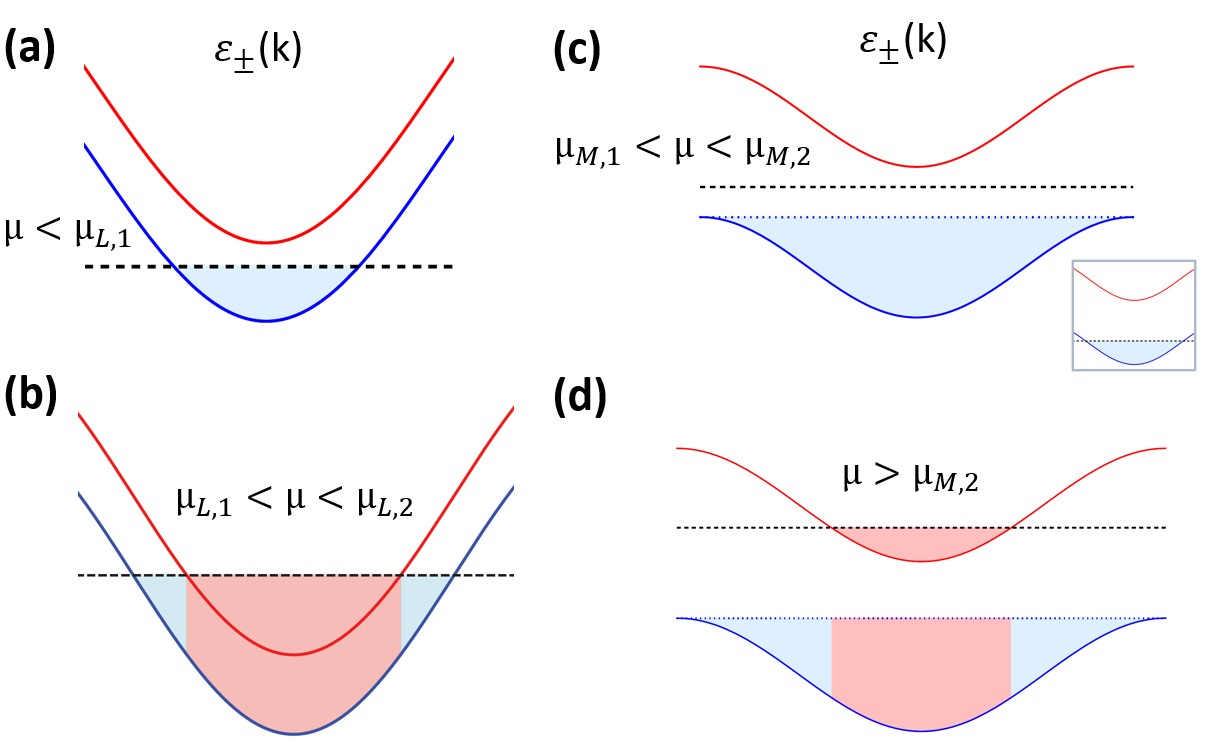}}
\caption{Color-coded distribution (BZ) of the degenerate singly occupied states $| \Phi_{1 \pm}(k) \rangle$ (light blue) and the doubly occupied states
$| \Phi_{2}(k) \rangle$ (light red), $d=1$. The upper/lower bands $\varepsilon_\pm(k)$ -- red/blue solid lines, $\mu$ -- black dashed line. (a-b): gapless regime ($U<U_c$), partially filled band(s);  (a): $\mu<\mu_{L,1}$ , (b): $\mu_{L,1}<\mu<\mu_{L,2}$.
(c-d): Mott regime ($U>U_c$). The inset: $\mu<\mu_{M,1}$, case similar to (a). (c): MI phase, fully filled lower band; (d): gapless phase
$\mu_{M,2}<\mu<\mu_{c,2}$, partially filled upper band.}
\label{FSPhi12}
\end{figure}

Having the ground-state vectors \eqref{GS} and the occupation numbers \eqref{nkGS}  for each given $\mathbf{k}$, we can analyze the evolution of the GS  with growing $\mu$. At weak coupling when $\mu < \mu_{L,1}$ and at strong coupling when $\mu < \mu_{M,1}$, the ground state is the Fermi sea
of the fermions filling up the lower band $\varepsilon_-$ in the singly occupied ($n_\mathbf{k}=1$) states $| \Phi_{1 \pm}(\mathbf{k}) \rangle$, the wave vectors are bounded by the chemical potential  $\varepsilon_-(\mathbf{k})<\mu$. Cf. Fig.~\ref{FSPhi12} (a) and inset for visualization. (The $1d$ spectrum is taken for simplicity
only, the results and arguments work for all $d$ considered.)
%%%%%%%%%%%%%%%%%%%%%%%%%%%%%%%%%%%%%%%%%%%%%%%%%%%%%%%%%%%%%%%%%%%%%%%%%%%%%%%%%
%%%%%%%%%%%%%%%%%%%%%%%%%%%%%%%%%%%%%%%%%%%%%%%%%%%%%%%%%%%%%%%%%%%%%%%%%%%%%%%%%
%%%%%%%%%%%%%%%%%%%%%%%%%%%%%%%%%%%%%%%%%%%%%%%%%%%%%%%%%%%%%%%%%%%%%%%%%%%%%%%%%%
%%%%%%%%%%%%%%%%%%%%%%%%%%%%%%%%%%%%%%%%%%%%%%%%%%%%%%%%%%%%%%%%%%%%%%%%%%%%%%%%%%

When $\mu > \mu_{L,1}$ at $U<U_c$ (or $\mu > \mu_{M,2}$ at $U>U_c$), for each $\mathbf{k}$ within the range
$\varepsilon_+(\mathbf{k})<\mu$ the (second) single-particle state in the upper band becomes available. Filling up each $\mathbf{k}$ in this region with two fermions
lifts the two-fold degeneracy and results in two-particle ground state $|\Phi_2 \rangle$:
\begin{equation}
\label{GSpmProd}
  | \Phi_{1 +}(\mathbf{k}) \rangle \otimes | \Phi_{1 -}(\mathbf{k}) \rangle = - | \Phi_{2}(\mathbf{k}) \rangle
\end{equation}
with $n_\mathbf{k}=2$. The fraction of the band
\begin{equation}
\label{DegFrac}
  \varepsilon_-(\mathbf{k}) < \mu < \varepsilon_+(\mathbf{k})
\end{equation}
stays singly occupied ($| \Phi_{1 \pm}(\mathbf{k}) \rangle$, $n_\mathbf{k}=1$) in the ground state, which is two-fold degenerate  for each $\mathbf{k}$.

The partition of the momentum space between the fractions of the degenerate singly occupied and non-degenerate doubly occupied states is shown in Fig.~\ref{FSPhi12} for several cases to visualize the following:

For any $d$ the appearance of the filled  $| \Phi_{2}(\mathbf{k}) \rangle$ states takes place at the critical point ($\mu_L$ or $\mu_M$), the fraction of the degenerate states shrinks with growing $\mu>\mu_{L/M}$, and it vanishes at the BI boundary $\mu=\mu_{c,2}$, when the two bands become fully filled.

In the non-interacting limit the band split vanishes, and the single band $\varepsilon(\mathbf{k})=\varepsilon_-(\mathbf{k})=\varepsilon_+(\mathbf{k})$ can accommodate two electrons for each $\mathbf{k}$.  To maintain zero net magnetization, one can fill up the Fermi sea with two electrons for a given $\mathbf{k}$ either one by one with $| \Phi_{1 \pm}(\mathbf{k}) \rangle$ states, or, equivalently (cf. \eqref{GSpmProd}), by pairs in the state $| \Phi_{2}(\mathbf{k}) \rangle$. Either way one ends up with the non-degenerate ground state of the free gas.

%%
%%%%%%%%%%%%%%%%%%%%%%%%%%%%%%%%%%%%%%%%%%%%%%%%%%%%%%%%%%%%%%%%%%%%%%%%%%%%%%
\subsection{Residual entropy}\label{ResEntApp}
%%%%%%%%%%%%%%%%%%%%%%%%%%%%%%%%%%%%%%%%%%%%%%%%%%%%%%%%%%%%%%%%%%%%%%%%%%%%%%
%%
We calculate the entropy defined via the grand canonical potential $\Omega$ as:
\begin{equation}
\label{STDdef}
S=- \Big(\frac{\partial \Omega  }{\partial T} \Big)_{\mu,V}~,
\end{equation}
so the entropy density is:
\begin{eqnarray}
\label{EntrDen}
  s(\mu,T) &=& \int_{BZ} \frac{d \mathbf{k}}{(2 \pi)^d} S_\mathbf{k}~,\\
  S_\mathbf{k}(\mu,T) &=& \ln \mathcal{Z}_{\mathbf{k}}+ T \Big( \frac{\partial \ln \mathcal{Z}_{\mathbf{k}}}{\partial T} \Big)_{\mu}~.
\label{Sk}
\end{eqnarray}
The partial entropy density $S_\mathbf{k}$ is found as
\begin{equation}
\label{SkRez}
  S_\mathbf{k}(x,u)=\ln \big(1+2 e^{u-x}+e^{-2x} \big)+ 2 \frac{(x-u)e^{u-x}+x e^{-2x}}{1+2 e^{u-x}+e^{-2x}}~,
\end{equation}
where
\begin{equation}
\label{ux}
  x \coloneqq \beta \xi, ~~u \coloneqq  \beta U/2~.
\end{equation}
Since $S_\mathbf{k}(x,u)$ is an even function of $x$
\begin{equation}
\label{Seven}
  S_\mathbf{k}(x,u)=S_\mathbf{k}(-x,u)~,
\end{equation}
it suffices to analyze the range $x \geq 0$ only.

In the non-interacting case we recover the Fermi gas result:
\begin{equation}
\label{SkFree}
  S_\mathbf{k}(x,0)=2 \ln \big(1+e^{-x} \big)+  \frac{2x}{1+e^x}
\end{equation}
with the zero-temperature limit $\beta \to \infty$:
\begin{equation}
\label{SkUT0}
 T = 0, ~U=0:~ S_\mathbf{k}(x,0)=
  \left\{
  \begin{array}{lr}
  2 \ln 2,~ &x=0 ~~(\xi_\mathbf{k} =0) \\ [0.2cm]
  0,~ & x > 0 ~~(\xi_\mathbf{k } > 0) \\
  \end{array}
  \right.
\end{equation}
Since the integrand $S_\mathbf{k}$ in the integral \eqref{EntrDen} \textit{over a finite} $d$-cube $[-\pi,\pi]^d$ is zero, except at a subspace of zero integration measure where $S_\mathbf{k}$ is \textit{finite}, the integral is zero. So, the entropy of free fermions $S(T=0)=0$, in agreement with the textbooks.

For the interacting HK model we find  in the zero-temperature limit:
\begin{equation}
\label{SkU}
 T = 0, ~U>0:~ S_\mathbf{k}(x,u)=
  \left\{
  \begin{array}{lr}
  \ln 2,~ &x<u ~~(\xi_\mathbf{k} <U/2) \\ [0.2cm]
   \ln 3,~ &x=u ~~(\xi_\mathbf{k} =U/2) \\ [0.2cm]
  0,~   &x >u ~~(\xi_\mathbf{k} >U/2)\\
  \end{array}
  \right.
\end{equation}
The plateau $S_\mathbf{k}(x,u)=\ln 2$ at $x<u$ yields the residual ($T=0$) entropy of the HK model, vanishing in the non-interacting limit. From
Eqs.~\eqref{EntrDen} and  \eqref{SkU} we find the specific residual $d$-dimensional entropy $s_d$ as
\begin{equation}
\label{sd}
  s_d= \ln 2 \int_{-d}^d \mathcal{D}_d (\varepsilon) \Theta(U^2/4-(\varepsilon-\mu)^2) d \varepsilon~.
\end{equation}
In the $d=1$ case we obtain a particularly simple result:
\begin{equation}
\label{s1}
  s_1= \frac{\ln 2}{\pi} \Re \big\{ \arcsin(\mu+U/2)-\arcsin(\mu-U/2) \big\}~,
\end{equation}
In cases $d=2,3$ the integral for $s_d$ involves special functions, with the single-band DOS given by Eqs.~\eqref{DOS2} and \eqref{DOS3D}.

%%%%%%%%%%%%%%%%%%%%%%%%%%%%%%%%%%%%%%%%%%%%%%%%%%%%%%%%%%%%%%%%%%%%%%%%%%%%%%%%%
%%%%%%%%%%%%%%%%%%%%%%%%%%%%%%%%%%%%%%%%%%%%%%%%%%%%%%%%%%%%%%%%%%%%%%%%%%%%%%%%%
\begin{figure}[h]
\centering{\includegraphics[width=14.0cm]{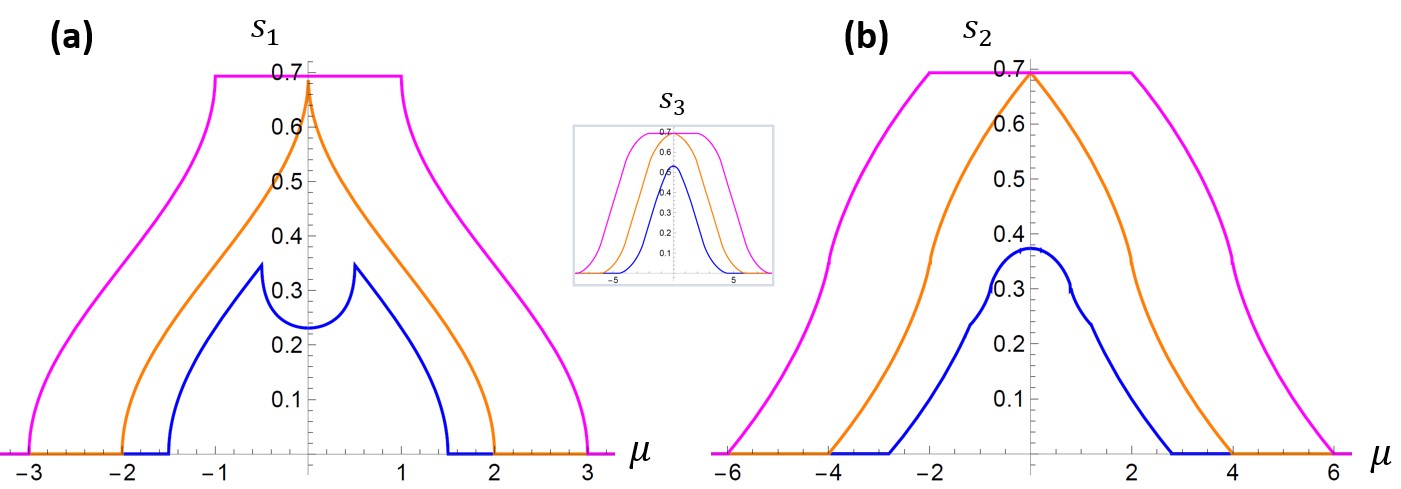}}
\caption{The ground state entropy $s_d(\mu)$ for $U <U_c$ (blue), $U=U_c$ (orange), and $U >U_c$ (magenta). (a): $d=1$, $U=1,2,4$; (b): $d=2$, $U=1.6,4,8$;
inset: $d=3$, $U=3,6,10$. The $s_d(\mu)$ features \eqref{CapAlpha} are most pronounced in $1d$, and are barely seen at $\mu_L$ and $\mu_{vH}$, $d=2$, $U<U_c$. The plateau $s_d(\mu)= \ln 2$ at $U > U_c$ and $\mu_{M,1}<\mu<\mu_{M,2}$ corresponds to the half-filled MI phase. }
\label{PlGSEnt}
\end{figure}
%%%%%%%%%%%%%%%%%%%%%%%%%%%%%%%%%%%%%%%%%%%%%%%%%%%%%%%%%%%%%%%%%%%%%%%%%%%%%%%%%
%%%%%%%%%%%%%%%%%%%%%%%%%%%%%%%%%%%%%%%%%%%%%%%%%%%%%%%%%%%%%%%%%%%%%%%%%%%%%%%%%

Since the plateau boundary of the integrand  $S_\mathbf{k}$ in Eq.~\eqref{SkU} coincides with the boundary of the degenerate singly occupied states \eqref{DegFrac}, Eq.~\eqref{sd} for the residual entropy amounts to a simple ``geometric'' formula:
\begin{equation}
\label{sdGeom}
  s_d= \ln 2 \times \{\mathrm{fraction ~of~ BZ~ filled ~with~ degenerate~states} ~ | \Phi_{1 \pm} \rangle~ \} =\ln 2 \times \mathcal{F}_1~.
\end{equation}
For the case $d=1$ one can elementarily calculate the geometric factor in the above equation using the results given in Sec.~\ref{1D}. The real parts of LY zeros $k_\pm$, cf. Eqs.~\eqref{kpm} and \eqref{Kpm}, yield $0/\pi/k_{F \pm}$ for emplty/filled/partially filled $\varepsilon_\pm$ bands. With some help for intuition from Figs.~\ref{FSPhi12} and \ref{Epm-Mu} (a,b), one can find
\begin{equation}
\label{Geom1D}
   \{\mathrm{geometric~factor,}~ d=1 \} =\frac{1}{ \pi}  \Re \{ k_- - k_+ \}
\end{equation}
and easily establish equivalence between  \eqref{sd} and \eqref{sdGeom}.

The plots of the ground state entropy $s_d$ for different interactions ($U \gtrless U_c$ and $U=U_c$) are shown in Figs.~\ref{PlGSEnt} for $d=1,2,3$.
The features of $s_d(\mu)$ at the critical points are most pronounced in one dimension, while the are barely seen at the Lifshitz and van Hove points in $d=2$, $U<U_c$.
When $U \leq U_c$ ($\forall d$) the entropy can reach its maximal value $s_d(\mu)= \ln 2$ only at the critical interaction $U=U_c$, when $\mu \to \mu_L=\mu_M =0$: the model is half-filled with the lower band fully occupied by the degenerate states and the empty upper band. When $U >  U_c$ the entropy has a plateau at $s_d(\mu)= \ln 2$
in the half-filled MI phase, corresponding to the range $\mu_{M,1}<\mu<\mu_{M,2}$.

From analysis of several models in the earlier work \cite{Chitov:2025}, it was conjectured that the critical exponent $\alpha$ in the universality class
\eqref{IndAll} can be retrieved from the global entanglement (or the von Neumann entropy) scaling \textit{in the gapless phase} near the FST critical point(s) $\mu_\sharp$ as $\propto  |\mu -\mu_\sharp|^{1-\alpha}$, or more directly from the entanglement capacity $\mathfrak{C}$, introduced in that work (cf. Eqs.~[22] and [40] in Ref.~\cite{Chitov:2025}).

The critical scaling of the entanglement is usually explored for the models without residual entropy. In case of the HK model with $s_d \neq 0$, the (entanglement) capacity $\mathfrak{C}$ is a natural quantum counterpart of the (specific) heat capacity near $T_c$ at a thermal phase transition. To link $s_d$ to the entanglement note:\\
\textit{(a)} the ground state of the model is always ($n_\mathbf{k}=1,2$) a factorized state over all available $\mathbf{k}$-s (i.e. it is a tensor product of \eqref{GS} for each $\mathbf{k}$); \\
\textit{(b)} the free-electron-like  states $|\Phi_2 (\mathbf{k}) \rangle$  ($n_\mathbf{k}=2$) are also factorized over the single particle eigenstates of the magnetization $\hat M_\mathbf{k}$: $| \Phi_{2}(\mathbf{k}) \rangle = | \Phi_{1 \uparrow}(\mathbf{k}) \rangle \otimes | \Phi_{1 \downarrow}(\mathbf{k}) \rangle$. This fully factorized (non-degenerate) state is disentangled \cite{Chitov:2022DL,Chitov:2021}; \\
\textit{(c)} the degenerate state $|\Phi_{1 \pm} (\mathbf{k}) \rangle$ ($n_\mathbf{k}=1$) is not an eigenstate of  $\hat M_\mathbf{k}$, it is entangled over $|\Phi_{1 \uparrow \downarrow} (\mathbf{k}) \rangle$, and adding a second electron ($n_\mathbf{k}=1 \mapsto n_\mathbf{k}=2$)
removes \textit{both} degeneracy and entanglement, cf. Eq.~\eqref{GSpmProd}.\\
It is natural to conclude from these arguments and Eq.~\eqref{sdGeom} that $s_d$ is a measure of the degeneracy and of the entanglement as well, so it should demonstrate the FST critical scaling as found for the latter in several solvable cases \cite{Chitov:2025,WeiGold:2011}.

From the analytical expression for  the entanglement capacity
\begin{equation}
\label{EnCapD}
  \mathfrak{C}_d \coloneqq \mu \frac{\partial s_d }{\partial \mu}= \mu \ln 2 \Big[ \mathcal{D}_d (\mu +U/2)- \mathcal{D}_d (\mu -U/2) \Big]~,
\end{equation}
the scaling near the BI, MI, Lifshitz, and van Hove critical points is easily found as
\begin{equation}
\label{CapAlpha}
 \mathfrak{C}_d \propto  |\mu -\mu_\sharp|^{-\alpha}~
\end{equation}
for $d=1,2,3$ with the FST universality class values of $\alpha=\gamma$, as expected.

To understand the physical nature of the GS of the HK model and further implications, let us now add a Zeeman term $-h(\hat n_{k\uparrow}-\hat n_{k\downarrow})$ into the Hamiltonian, with $\mu_B=1$ and $m=n_\uparrow-n_\downarrow$ per site. With
\begin{equation}
\label{Zk-h}
  \mathcal{Z}_\mathbf{k}(h)=e^{-\beta U/4}+2\cosh(\beta h)\,e^{-\beta(\xi-U/4)}+e^{-\beta(2\xi+U/4)}
\end{equation}
we recover Eq.~~\eqref{sdGeom} in the limit $\lim_{T \to 0} s_d(h=0,T)$, while in the regime $T \ll h$ we get from the asymptotic of $s_d(h,T)$ that even an infinitesimal field kills the residual entropy and restores the third law:
\begin{equation}
\label{3law}
  s_d(0^\pm,0)=0~.
\end{equation}
Note that the thermodynamical derivation of Eq.~\eqref{sdGeom} was done under the condition $h=0$ held exactly. 
And \eqref{sdGeom} is in agreement with the explicit construction of the Fock's space of the entangled GS in Sec.~\ref{GSApp}, where net zero magnetization was built in at each $\mathbf{k}$. The physical implications of \eqref{3law} is that at zero temperature even a field of a single electron (say number of electrons is changed from even to odd) suffices to remove the residual entropy, i.e., to bring the system into the disentangled polarized ground state.  We analyze further implications of the latter in the following subsection.

%%
%%%%%%%%%%%%%%%%%%%%%%%%%%%%%%%%%%%%%%%%%%%%%%%%%%%%%%%%%%%%%%%%%%%%%%%%%%%%%%
\subsection{Nature of the fermionic liquid}\label{GFApp}
%%%%%%%%%%%%%%%%%%%%%%%%%%%%%%%%%%%%%%%%%%%%%%%%%%%%%%%%%%%%%%%%%%%%%%%%%%%%%%
%%
%%
%%%%%%%%%%%%%%%%%%%%%%%%%%%%%%%%%%%%%%%%%%%%%%%%%%%%%%%%%%%%%%%%%%%%%%%%%%%%%%
\subsubsection{Green's function}\label{GFonly}
%%%%%%%%%%%%%%%%%%%%%%%%%%%%%%%%%%%%%%%%%%%%%%%%%%%%%%%%%%%%%%%%%%%%%%%%%%%%%%
%%
%%
%%

Let us comment now on the nature of the phase in gapless states of the HK model in the connection to the properties of the Green's function. For any $U \gtrless U_c$ the particle distribution $n_\mathbf{k} \equiv 2 n_{\mathbf{k} \sigma}$ is a step-like Fermi-Dirac function  with one or two FSs, cf. Eq.~\eqref{nksig0}  and Fig.~\ref{Nk-AB}. The total volume bounded by the FS(s), i.e., the particle number, is shown to be preserved in agreement with the LT \eqref{LuttGen}, \eqref{LuttUD}. Since the analysis of Sec.~\ref{Trans} is fully based on the available exact partition function, i.e., thermodynamics, some additional input is warranted to address the behavior of $G_\sigma(\mathbf{k},i \omega_n)$ in different phases. (Consistency check: summation of the Green's function \eqref{GF} over the Matsubara frequencies yields Eq.~\eqref{nksig0} in the limit $T \to 0$, as it must.) $G_\sigma(\mathbf{k},i \omega_n)$ has only poles \eqref{GFpoles} and zeros \eqref{GFzeros}.  The locus of its zeros in the ground state (a.k.a. the Luttinger surface \cite{Dzyalo:2003}) is
\begin{equation}
\label{GFZerosGS}
  \varepsilon(\mathbf{k})-\mu=0~,~~ n_{\mathbf{k} \sigma}=1/2~,
\end{equation}
see Fig.~\ref{GFxi}. Note that this is not a non-interacting FS, since $\mu \neq \mu_\circ$.
In particular,
\begin{equation}
\label{LTGfZeros}
  \mathcal{S}_d(FS)   \centernot{\Longleftrightarrow} \int_{G_\sigma(\mathbf{k},0)>0} d \mathbf{k} ~.
\end{equation}
In the gapless phases (M, M2) the low-energy single-particle excitations correspond to the real poles of $G_\sigma(\mathbf{k},0)$,  which are also the LY zeros. The real solutions of \eqref{GFZerosGS} occur in the depth of the Fermi sea (compare Eqs.~\eqref{LYZT0} and \eqref{GFZerosGS}), and have no bearing on the low-energy sector of the model. (To visualize distributions of zeros and poles of $G_\sigma$ within the BZ, we collected characteristic plots for different phases ($d=1$) in the regimes of strong and weak couplings in a separate file with Supplementary Materials.) The \textit{exact} spectral function
\begin{equation}
\label{SpectrFun}
  A_\sigma(\mathbf{k}, \omega)=(1-n_{\mathbf{k} \sigma}) \delta(\omega-\xi_-)+ n_{\mathbf{k} \sigma} \delta(\omega-\xi_+)
\end{equation}
demonstrates two delta-function peaks corresponding to the ideal Fermi gas-like quasi-particles with energies $\xi_\pm$ and infinite lifetimes (no damping). The effects of interaction on the single-particle properties manifest themselves in:\\
1) the weights of the peaks $\Longleftrightarrow$ residues of Green's function  are as in the ideal gas due to extra plateau $n_{\mathbf{k} \sigma}=1/2$;\\
2) excitations - while $\xi_-$ is the eigenvalue of $H_{\mathbf{k}}^\prime$, $\xi_+$ is truly \textit{a quasi-particle} created due to interaction-generated band splitting, and it is not an eigenvalue of $H_{\mathbf{k}}^\prime$.  \\
\textit{So the gapless phases,  qualify as a Fermi liquid.}

At strong coupling the Green's function  near the Mott transition and inside the MI phase warrants a special comment in the context of active research on this topic, see e.g., \cite{Volovik:2007,*Volovik:2017,*Volovik:2018,Si:2024,Dzyalo:2003,Dzyalo:1996,Stanescu:2007,Gurarie:2011,Misawa:2022}.
When the HK model approaches the gapped MI phase ($\mu \to \pm \mu_M \pm 0$, the FS with its gapless excitations (that is  FL) vanishes, along with the real parts of the LY zeros/ poles of $G_\sigma(\mathbf{k},0)$.  In the Mott phase ($|\mu| < \mu_M$) the thermodynamics yields plateau of the filling $\bar n(\mu)=1/2$ and discontinuity of $\mu(\mu_\circ)$ at $\mu_\circ=0$, see Figs.~\ref{Epm-Mu}, \ref{FilChi}, and \ref{2DFilChi}. $G_\sigma(\mathbf{k},0)$ has only real zeros and no poles in this range of $\mu$. One can check from the limits of the solutions of Eqs.~\eqref{LYZT0} and \eqref{GFZerosGS} near the ends of plateau $\mu \to \pm \mu_M \pm 0$, that redefinition \eqref{mu0} of the chemical potential inside the Mott phase results in the following relation  between the Luttinger and Fermi surfaces: at $\mu=0$  Eq.~\eqref{GFZerosGS} for zeros coincides with that for the FS of the non-interacting model at half-filling. Since the latter has two un-split bands, twice the non-interacting FS volume is equal to the  FS-bounded volume of the fully-filled lower band of the interacting model:
\begin{equation}
\label{FLS-MI}
  \mathrm{MI,}~\bar n=\frac12,~\mu \equiv 0 =\mu_\circ:~2 \mathcal{S}_d(LS)=2 \mathcal{S}_d(FS|_{\s U=0})=\mathcal{S}_d(FS)=(2 \pi)^d
\end{equation}
In other phases, in principle, one can derive relations between the $d$-volumes within the locus of zeros and Fermi surfaces from the available results, but
they are not as simply related as in Eq.~\eqref{FLS-MI} when $\mu \neq 0$, and this does not seem to be a useful exercise.  As we have shown in Sec.~\ref{Trans}, the LT \eqref{LuttUD} works smoothly across all phases, including the case \eqref{FLS-MI}.  In addition, as follows from the equation for the partial grand canonical potential (calculated with the shifted Hamiltonian $\hat H_\mathbf{k}^\prime$):
\begin{equation}
\label{OmK}
\Omega_\mathbf{k}^\prime =
  \left\{
  \begin{array}{lr}
  0,~ &\xi_\mathbf{k} >U/2, \\ [0.2cm]
  \xi_\mathbf{k} -U/2,~ &-U/2 < \xi_\mathbf{k} < U/2, \\ [0.2cm]
  2 \xi_\mathbf{k},~ &\xi_\mathbf{k} <-U/2,\\
  \end{array}
  \right.
\end{equation}
the locus of zeros of the Green's function is not a critical point of the HK model, since neither $\Omega_\mathbf{k}^\prime$, nor its derivatives demonstrate any singularities at $\xi_\mathbf{k}=0$, $U \neq 0$, while the cusps of $\Omega_\mathbf{k}^\prime$ at  $\xi_\mathbf{k}=\pm U/2$ correspond to the FST transitions.

At weak coupling $\mu(\mu_\circ)$ is smooth, and in the gapless phase at half-filling $\mu=\mu_\circ=0$. See Figs.~\ref{Epm-Mu}, \ref{FilChi}, and \ref{2DFilChi}.
In this case \eqref{FLS-MI} holds automatically (only at $\mu=0$!), where $\mathcal{S}_d(FS)$ is the sum of two contributions from partially filled lower and upper bands.

%%
%%%%%%%%%%%%%%%%%%%%%%%%%%%%%%%%%%%%%%%%%%%%%%%%%%%%%%%%%%%%%%%%%%%%%%%%%%%%%%
\subsubsection{Quantum ferromagnetism}\label{FM}
%%%%%%%%%%%%%%%%%%%%%%%%%%%%%%%%%%%%%%%%%%%%%%%%%%%%%%%%%%%%%%%%%%%%%%%%%%%%%%
%%
The phenomenon we will discuss in this subsection is not universal as the main focus of this paper -- topological transitions related to the FS transformations.
It is \textit{specific for the HK model} and related to the its particular degenerate (entangled) GS induced by the infinite-ranged interaction. From the viewpoint of general theory of phase transitions, it is a neat example of the \textit{secondary order parameter}, occurring due to disentanglement (polarization) of the GS.

So here I will rather briefly give basic results, leaving more details and analysis for a separate comprehensive paper, where the exact FLT for the HK model and connections to the properties of generic Fermi liquids, especially those predicted by the RG FLT \cite{Chitov:1998}, will be presented.\footnote{\label{ExFLT} G.Y. Chitov, paper in preparation.}

Introducing $\mathbf{k}$-components of the magnetization and spin susceptibility as
\begin{equation}
\label{mchi}
  m_\mathbf{k}=\frac1\beta\frac{\partial\ln \mathcal{Z}_\mathbf{k}(h)}{\partial h}\Big|_\mu, \qquad  \chi_\mathbf{k}=\frac{\partial m_\mathbf{k}}{\partial h}\Big|_\mu~,
\end{equation}
we can find the magnetization per site
\begin{equation}
\label{m-approx}
  m(h,T)=\tanh\frac hT\,\Big[\mathcal{F}_1+T\ln\!\Big(2\cosh\frac hT\Big)\,\big(\mathcal{D}(\mu_+)+\mathcal{D}(\mu_-)\big)\Big]+\mathcal{O}(T^2,h^2),
\end{equation}
where $\mathcal{F}_1$ is the fraction of singly-occupied BZ defined in Eq.~\eqref{sdGeom}. 

As follows from the above expression at $T=0$, even an infinitesimal field polarizes all electrons in the singly-occupied states and lifts completely the degeneracy of those states (blue states shown in Fig.~\ref{FSPhi12}):
\begin{equation}
\label{mGS}
 m(h \to 0^\pm,T=0)=\pm \mathcal{F}_1~,
\end{equation}
which signals a quantum phase transition. Indeed, using the grand canonical potential at exactly zero temperature $\Omega_0(h,\mu)$, assuming $h>0$, we obtain
\begin{equation}
\label{mhmu}
   m(\mu,h)=-\frac{\partial\Omega_0}{\partial h}=  \frac{1}{(2\pi)^d} \Big(\mathcal{S}_d(\mu+U/2+h)-\mathcal{S}_d(\mu-U/2-h) \Big)~,
\end{equation}
and  $\bar n(\mu,h)$ given by the r.h.s. of Eq.~\eqref{LuttUD} (LT) with \eqref{h-shift}. 
The spontaneous order parameter is therefore
\begin{equation}
\label{mSSB}
  m_s(\mu) \equiv \lim_{h\to0^+}m= \frac{1}{(2\pi)^d} \Big(\mathcal{S}_d(\mu_+)-\mathcal{S}_d(\mu_-) \Big)  =\mathcal{F}_1(\mu) 
\end{equation}
with the spin susceptibility
\begin{equation}
\label{chiSSB}
  \chi_s(\mu)=-\frac{\partial^2\Omega_0}{\partial h^2}\Big|_{h\to0^+}=\mathcal{D}(\mu_+)+\mathcal{D}(\mu_-)\ \ge 0.  
\end{equation}
Up to 1/2, $\chi_s$  is equal to the compressibility \eqref{chiD} analysed in the main text for $d=1,2,3$. $\chi_s$ is strictly positive FL-like in the metal phases, it vanishes in the band insulators, where $m_s=0$, and in the Mott phase, where $m_s$ has a plateau (lower band is fully filled). 

Eqs.~\eqref{mSSB} and \eqref{chiSSB} establish a quantum phase transition with the SSB and the order parameter $m_s$ of the stable FM phase.
It is also possible to make the Legendre transformation  to the Gibbs potential $G_0(\mu,m)=\Omega_0(\mu,h)+hm$ and to demonstrate that  $\partial G/\partial m =0$  and  $\partial^2G/\partial m^2 =1/\chi_s>0$ at $m=\pm m_s\neq 0$.

The FM order parameter  $m_s(\mu)=f_1(\mu)=s_d(\mu)/\ln2$  is a fraction of the BZ filled with degenerate states \eqref{sdGeom}, which becomes the spontaneous moment, and qualitatively, is the residual entropy release, up to a factor.  It appears \emph{continuously}, together with the Fermi surface, and vanished at the BI ends $\mu =\mu_{c,1/2}$. From the analytical expressions \eqref{mSSB} and \eqref{chiSSB} for $m_s$ and $\chi_s$, it is clear that their critical exponents $\beta,\gamma$ are of the FST universality class  \eqref{IndAll}. Since $m_s(\mu)=s_d(\mu)/\ln2$, its behavior is actually visualized in Fig.~\ref{PlGSEnt}. The plateau in the Mott phase
is the spin counterpart of the Oshikawa filling $\bar n(\mu)$ plateau, and an instance of the Oshikawa--Yamanaka--Affleck magnetization-plateau condition \cite{Affleck:1997}.
 
In the context of the LY formalism consistently applied in our analysis, it is important to point out: \textit{the locus of the FM order is the Fermi surface, the same quantum critical manifold determined by the $T=0$ LY zeros}. Indeed, from \eqref{Zk-h} we easily find the LY zeros at $h>0$. They are given by Eq.\eqref{wpm}  with the change $e^{-\beta U/2} \mapsto e^{-\beta U/2}\cosh(\beta h)$, yielding at $T=0$
Eq.~\eqref{LYZT0} with
\begin{equation}
\label{h-shift}
  U/2 \longmapsto U/2+h~.
\end{equation}
So, \textit{for the HK model}  the condition for the real LY roots where a spontaneous order could occur ($h=0$) coincides with the condition for appearance of the FS. In the Mott phase where the roots are complex, we get plateau $m_s=1$, as already explained in the complex formalism of Appendix~\ref{AppLY}. \textit{In general}  there are no reasons for the FS and any other type or order to coexist/overlap: this is clear from examples of different models considered in Refs.~\cite{Chitov:2025,Chitov:2022DL}.
Another point to stress is that the HK quantum ferromagnetism is a particular phenomenon occurring due to the interaction-induced macroscopically degenerate GS of the HK model, $m_s \sim s_d$. In the non-interacting limit the magnetic field induces the split of free bands, but $m_s(h)$ trivially vanishes as $h\to0$, while the FST order parameter $\mathcal{P}$ stays.

Since the interactions in the HK model is of infinite range, the Mermin-Wagner and Goldstone theorems are not applicable, and  there is no contradiction with its spin rotational $SU(2)$ symmetry spontaneous breaking in any $d$ at $T=0$. However, the gapless modes due to rigid rotations of $\vec{m}$ do exist: adding an infinitesimal
transverse component to a given field   $\vec{h}$ ($\vec{m} \upuparrows \vec{h}$) leads to  a divergent transverse susceptibility 
\begin{equation}
\label{trans}         
\chi_s^\perp=m_s/h \ \xrightarrow{h\to0^+}\ \infty,
\end{equation}
while the longitudinal component given by \eqref{chiSSB} is finite. 

With the results of this and previous subsections, we identify the gapless phase of the HK model as the \textit{ferromagnetic Fermi liquid}.

%%%%%%%%%%%%%%%%%%%%%%%%%%%%%%%%%%%%%%%%%%%%%%%%%%%%%%%%%%%%%%%%%%%%%%%%%%%%%%%%%%
%%%%%%%%%%%%%%%%%%%%%%%%%%%%%%%%%%%%%%%%%%%%%%%%%%%%%%%%%%%%%%%%%%%%%%%%%%%%%%%%%%

%xxxxxxxxxxxxxxxxxxxxxxxxxxxxxxxxxxxxxxxxxxxxxxxxxxxxxxxxxxxxxxxxxxxxxxxxxxxxxx

%%%%%%%%%%%%%%%%%%%%%%%%%%%%%%%%%%%%%%%%%%%%%%%%%%%%%%%%%%%%%%%%%%%%%%%%%%%%%%
\section{FST quantum transitions and Morse theory}\label{Morse}
%%%%%%%%%%%%%%%%%%%%%%%%%%%%%%%%%%%%%%%%%%%%%%%%%%%%%%%%%%%%%%%%%%%%%%%%%%%%%%
%
%
In this Appendix we explore the topological nature of the metal-metal and metal-BI quantum phase transitions, advancing the
pioneering ideas of the classical van Hove paper  \cite{VanHove:1953}.
The key idea is to apply the Morse theory \cite{Milnor:1963} and more generally, the homology theory \cite{NovikovIII:1990}
for the analysis of the band filling. Due to its periodicity, the spectrum $\varepsilon(\mathbf{k})$ can be defined
on a $d$-dimensional torus $\mathds{T}^d$  and it satisfies the conditions for the Morse function \cite{Milnor:1963}.
The mathematical definition of the critical point in the Morse theory $\nabla_{\mathbf{k}}\varepsilon(\mathbf{k})=0$
coincides with the physical condition for the transition \eqref{CPs}. The universal nature of the critical points corresponding
to the band, Mott, Lifshitz and van Hove transitions can be understood within a single framework of the cohomology theory outlined below.

%%
%%%%%%%%%%%%%%%%%%%%%%%%%%%%%%%%%%%%%%%%%%%%%%%%%%%%%%%%%%%%%%%%%%%%%%%%%%%%%%
\subsection{Plain vanilla spectrum \eqref{Eps0}}\label{MorseSimple}
%%%%%%%%%%%%%%%%%%%%%%%%%%%%%%%%%%%%%%%%%%%%%%%%%%%%%%%%%%%%%%%%%%%%%%%%%%%%%%
%%

\begin{figure}[h]
\centering{\includegraphics[width=10.0cm]{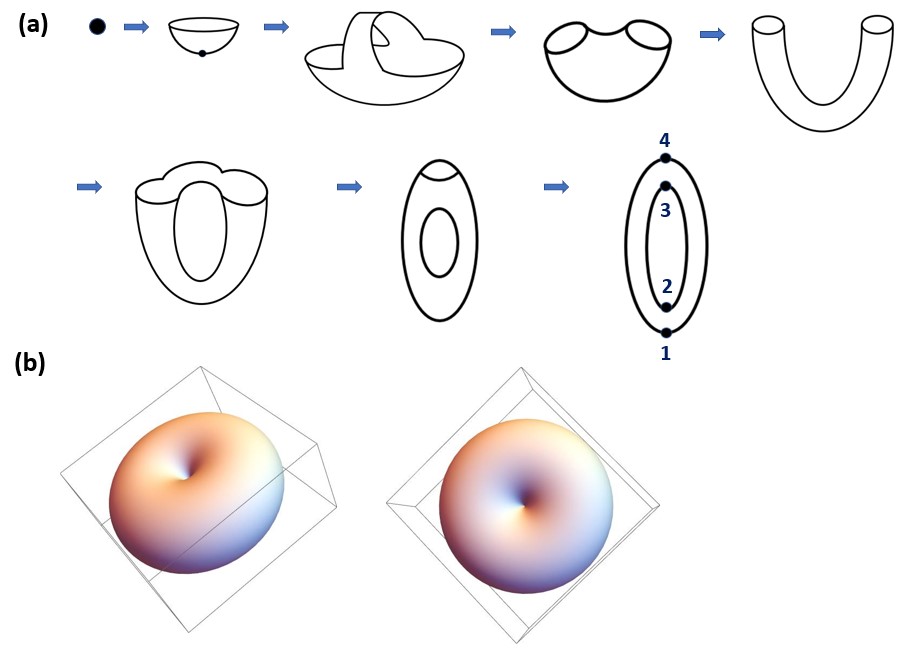}}
\caption{(a) Consecutive building up the torus $\mathds{T}^2$ by attaching $n$-cells, as discussed in the text.
Adapted from \cite{NovikovIII:1990}. Indices of the critical points (1-4): min (1) $s=0$, saddle points (2,3) $s=1$, and max (4) $s=2$.
(b) The collapsed doughnut with a punctured cental point (its inner radius $r_{inn} \to 0$) shown from different angles.}
\label{Donuts}
\end{figure}

We start with the \textbf{$\mathbf{2d}$ case} of \eqref{Eps0} which is non-trivial and easy to visualize. The topologically distinct cases of band filling amount to the standard problem of the Morse theory: analysis of the height function for the torus, see Fig.~\ref{Donuts} a. It can be simply formulated as a problem
of, say, filling with the water a glassy doughnut touching the ground at single point \cite{Milnor:1963}.

We define $M^\varepsilon$ as a manifold which is the ``filled" part of the torus $\mathds{T}^d$ at a given $\varepsilon$:
\begin{equation}
\label{Me}
 \forall ~\mathbf{k}  \in M^\varepsilon  \subseteq \mathds{T}^d: ~\varepsilon(\mathbf{k})\leq \varepsilon
\end{equation}
We will be particularly interested in the $d$-area $\mathcal{S}_d(M^\varepsilon)$ which measures the filled part of the BZ (FS).

%%%%%%%%%%%%%%%%%%%%%%%%%%%%%%%%%%%%%%%%%%%%%%%%%%%%%%%%%%%%%%%%%%%%%%%%%%%%%%%%%
%%%%%%%%%%%%%%%%%%%%%%%%%%%%%%%%%%%%%%%%%%%%%%%%%%%%%%%%%%%%%%%%%%%%%%%%%%%%%%%%%
%%%%%%%%%%%%%%%%%%%%%%%%%%%%%%%%%%%%%%%%%%%%%%%%%%%%%%%%%%%%%%%%%%%%%%%%%%%%%%%%%%
%%%%%%%%%%%%%%%%%%%%%%%%%%%%%%%%%%%%%%%%%%%%%%%%%%%%%%%%%%%%%%%%%%%%%%%%%%%%%%%%%%

A generic  $2d$ torus yields four critical points (1-4 in Fig.~\ref{Donuts} a). Points 1 and 4 correspond, respectively,  to the minimum and maximum of the Morse function;
and points 2 and 3 are the saddle points. At each critical point the topology of the filled part of the torus surface changes.
The point 1 corresponds to attaching $0$-cell to the empty space; at the saddle point 2, one $1$-cell (handle) is added, and the second handle is added at the saddle point 3. The 2-cell (membrane) is added at the maximum 4. The homotopies of the manifolds between the critical points are different, and so are their cohomology groups \cite{Milnor:1963,NovikovIII:1990}.

To numerically account for the change of topology at the critical points one can use the Euler characteristic $\chi_E$.
For the  manifold $M$ it can be calculated in two ways  \cite{NovikovIII:1990}:
\begin{eqnarray}
\label{ChiEf}
  \chi_E(M)  &\overset{\forall f}{=}&  \sum_{s \geq 0} (-1)^s C_s(f)  \\
  &=& \sum_{n \geq 0} (-1)^n b_n~,
\label{ChiEb}
\end{eqnarray}
where $C_s$ is the number of the critical points with index $s$,\footnote{\label{s} The index $s$ is defined as the number of negative quadratic terms in the expansion of the Morse function $f$ in the vicinity of the critical point. In our case $f=\varepsilon(\mathbf{k})$.} and $b_n$ are the Betty numbers. For $\mathds{T}^2:$ $b_0=1,b_1=2,b_2=1$. It is important to emphasize that Eq.~\eqref{ChiEf} holds for any Morse function, and it is determined solely by the parameters of the cohomology group of $M$ as expressed by Eq.~\eqref{ChiEb}.

Special symmetries can lead to merging of the saddle points. In our case the $x \leftrightarrow y$ symmetry of the spectrum leads to a single saddle point in the center of the ``collapsed'' doughnut (i.e. the doughnut with a punctured cental point when its inner radius $r_{inn} \to 0$, see  Fig.~\ref{Donuts} b). The saddle point is non-degenerate since the Hessian at $(k_x,k_y)=(\pi, 0)$ and $(0,\pi)$ is equal to $-1$.
The critical points are:
\begin{eqnarray}
  \mathrm{Min}&:&~~ \varepsilon(0,0) = -2,~~s=0,~ C_0=1,  \nonumber  \\
  \mathrm{S}&:&~~ \varepsilon(0,\pi) =\varepsilon(\pi,0) =0,~~s=1,~ C_1=2,  \nonumber  \\
  \mathrm{Max}&:&~~ \varepsilon(\pi,\pi) = 2,~~s=2,~ C_2=1.
\label{CPsD2}
\end{eqnarray}
Using the above numbers $C_s$ (or equivalently, $b_n$) which change when $M^\varepsilon$ goes through the critical point \cite{Milnor:1963,NovikovIII:1990},
we find:
\begin{equation}
\label{ChiETorus}
 d=2,~M^\varepsilon \subseteq \mathds{T}^2:~~\chi_E(M^\varepsilon) =
  \left\{
  \begin{array}{lr}
  0,~ &\varepsilon <-2 \\ [0.2cm]
  1,~ &-2 < \varepsilon <0 \\ [0.2cm]
  -1, ~ &0 <\varepsilon <2 \\ [0.2cm]
  0, ~ & \varepsilon > 2. \\
  \end{array}
  \right.
\end{equation}
The minimum/maximum  ($\varepsilon=\mp 2$) correspond to the appearance/dissapearance of the FS, respectively.
The saddle point at $\varepsilon=0$ (called the van Hove singularity in the literature) corresponds to the half-filled band with the diamond-shaped FS which touches the BZ edges by its corners, see, e.g. \cite{Economou:2006}. After crossing this point the closed particle-like FS becomes an open hole-like one at $\varepsilon >0$. However a shift by the reciprocal lattice vector $\mathbf{Q}=(\pi,\pi)$ with $\varepsilon(\mathbf{k}+\mathbf{Q})=-\varepsilon(\mathbf{k})$ allows us to switch to the hole representation within the new BZ with a closed hole-like FS. The area of the hole Fermi sea (hole order parameter) can be easily found from equation
$\mathcal{S}_{2,h}(\varepsilon)|_{\varepsilon>0} =4 \pi^2- \mathcal{S}_2(\varepsilon)|_{\varepsilon<0}$. At the BI critical points $\mathcal{S}_2(-2)=\mathcal{S}_{2,h}(2)=0$, and at the van Hove point $\mathcal{S}_2(0)=\mathcal{S}_{2,h}(0)=2 \pi^2$. From the topological point of view, this is a manifestation of the \textit{Poincar\'{e} duality}. Upon changing the sign of the Morse function $f$, the critical point with index $s$ of $f$ becomes the critical point of index $d-s$ of $-f$, and $C_s \mapsto C_{d-s}$ (and $b_n \mapsto b_{d-n}$). Thus, $max \leftrightarrow min$, while the saddle points stay.

\begin{figure}[h]
\centering{\includegraphics[width=10.0cm]{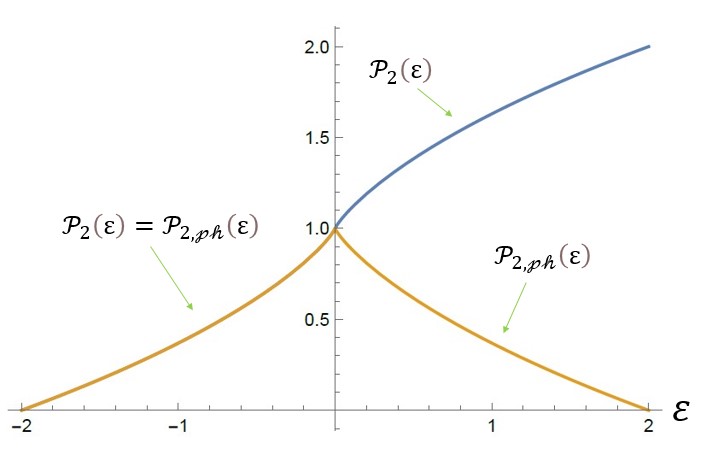}}
\caption{Two normalized order parameters for the gapless phase, $d=2$: the particle parameter $\mathcal{P}_2(\varepsilon) \coloneqq \mathcal{S}_2(\varepsilon)/\mathcal{S}_2(0)$, where  $\mathcal{S}_2(\varepsilon)$ is defined by Eq.~\eqref{S2R}, and the particle-hole parameter  $\mathcal{P}_{2,ph}(\varepsilon)$ \eqref{phOPD2}. By definition $\mathcal{P}_{2,ph}(\varepsilon)=\mathcal{P}_2(\varepsilon)$ at $\varepsilon <0$. }
\label{OPsD2}
\end{figure}

Since the order parameter \eqref{OPFS} is defined modulo the fully filled band, it could be more natural to use
the combined particle-hole order parameter, normalized by the area of the half-filled band, which appears to be more informative for the general analysis:
\begin{equation}
\label{phOPD2}
      \mathcal{P}_{2,ph}(\varepsilon)  \coloneqq 1-   \frac{g_R(\varepsilon)}{2 \pi^2}\mathrm{sign}(\varepsilon)~.
\end{equation}
It vanishes at the boundaries of the metal phase where the FS disappears, and it signals the topological transformations of the FS in the metallic phase by a cusp.\footnote{\label{ph1d} Introducing the particle-hole parameter in $1d$ is not useful, since the point of half-filling $\varepsilon=0$ is not critical.}
See  Fig.~\ref{OPsD2}.
%%%%%%%%%%%%%%%%%%%%%%%%%%%%%%%%%%%%%%%%%%%%%%%%%%%%%%%%%%%%%%%%%%%%%%%%%%%%%%%%%
%%%%%%%%%%%%%%%%%%%%%%%%%%%%%%%%%%%%%%%%%%%%%%%%%%%%%%%%%%%%%%%%%%%%%%%%%%%%%%%%%
%%%%%%%%%%%%%%%%%%%%%%%%%%%%%%%%%%%%%%%%%%%%%%%%%%%%%%%%%%%%%%%%%%%%%%%%%%%%%%%%%%
%%%%%%%%%%%%%%%%%%%%%%%%%%%%%%%%%%%%%%%%%%%%%%%%%%%%%%%%%%%%%%%%%%%%%%%%%%%%%%%%%%

The \textbf{$\mathbf{1d}$ case} of the circle $\mathds{T}^1=\mathds{S}^1$ easily fits in the above analysis with a single minimum/maximum at the ends of the interval $\varepsilon= \mp 1$:
\begin{eqnarray}
  \mathrm{Min}&:&~~ \varepsilon(0) = -1,~~s=0,~ C_0=1,  \nonumber  \\
  \mathrm{Max}&:&~~ \varepsilon(\pi) = 1,~~s=1,~ C_1=1.
\label{CPsD1}
\end{eqnarray}
The Euler characteristic is:
\begin{equation}
\label{ChiECircle}
 d=1:~~ \chi_E(M^\varepsilon) =
  \left\{
  \begin{array}{lr}
  0,~ &|\varepsilon| > 1 \\ [0.2cm]
  1,~ &|\varepsilon| <1 \\
  \end{array}
  \right.
\end{equation}

\begin{figure}[h]
\centering{\includegraphics[width=14.0cm]{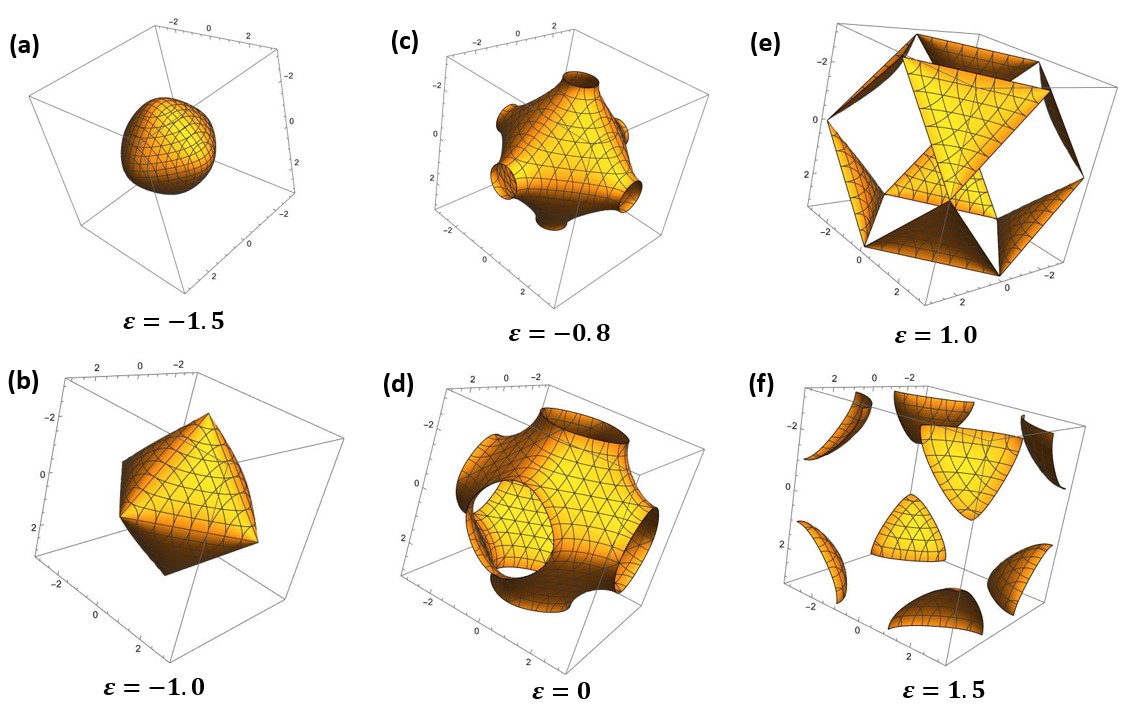}}
\caption{The Fermi surfaces of the 3D band \eqref{Eps0} at different fillings, shown within the cube of the first BZ $[-\pi,\pi]^3$.
The first saddle point ($S1$) at $\varepsilon=-1$ (b), when the vertices of the FS touch the faces of the BZ, precedes appearance of a neck, i.e., closed FS on the faces of the BZ, (c,d). After crossing the second saddle point ($S2$) at $\varepsilon=1$ (e), the necks on the faces become open surfaces, (f).}
\label{3DFS}
\end{figure}

In the \textbf{case $\mathbf{d=3}$} the Morse function  $\varepsilon(k_x,k_y,k_z)$ is defined on the $3d$ torus  $\mathds{T}^3$. We will not attempt to visualize filling it, like we did for $\mathds{T}^2$ in Fig.~\ref{Donuts}, but proceed analytically getting some help for intuition from Fig.~\ref{3DFS}. The Morse function on $\mathds{T}^3$ has 8 critical points: 1 minimum, 1 maximum, and 2 saddle points (S1, S2) of multiplicity 3 each. $S1:(\pi,0,0)+2$ permutations and $S2:(\pi,\pi,0)+2$ permutations \cite{VanHove:1953}:
\begin{eqnarray}
  \mathrm{Min}&:&~~ \varepsilon(0,0,0) = -3,~~s=0,~ C_0=1,  \nonumber  \\
  \mathrm{S1}&:&~~ \varepsilon(S1) = -1,~~s=1,~ C_1=3,  \nonumber  \\
  \mathrm{S2}&:&~~ \varepsilon(S2) = 1,~~s=2,~ C_2=3,   \nonumber  \\
  \mathrm{Max}&:&~~ \varepsilon(\pi,\pi,\pi) = 3,~~s=3,~ C_3=1.
\label{CPsD3}
\end{eqnarray}
The Betty numbers are $b_0=1$, $b_1=3$, $b_2=3$, $b_3=1$.  Similarly to the $2d$ case, at the critical points $n$-cells are glued to $M^\varepsilon$, so its topology (cohomology group) changes. We quantify this change with the Euler characteristic, which is found as follows:
\begin{equation}
\label{ChiED3}
 d=3,~M^\varepsilon \subseteq \mathds{T}^3:~~ \chi_E(M^\varepsilon) =
  \left\{
  \begin{array}{lr}
  0,~ &\varepsilon <-3 \\ [0.2cm]
  1,~ &-3 < \varepsilon <-1 \\ [0.2cm]
  -2, ~ &-1 <\varepsilon <1 \\ [0.2cm]
  1, ~ &1 <\varepsilon <3 \\ [0.2cm]
  0, ~ & \varepsilon > 3 \\
  \end{array}
  \right.
\end{equation}
As one can see from Fig.~\ref{3DFS}, the first saddle point ($S1$) at $\varepsilon=-1$ corresponds to the appearance of the FS necks, while at the second point ($S2$) at $\varepsilon=1$ the necks on the faces of the BZ become open.

%%%%%%%%%%%%%%%%%%%%%%%%%%%%%%%%%%%%%%%%%%%%%%%%%%%%%%%%%%%%%%%%%%%%%%%%%%%%%%%%%
%%%%%%%%%%%%%%%%%%%%%%%%%%%%%%%%%%%%%%%%%%%%%%%%%%%%%%%%%%%%%%%%%%%%%%%%%%%%%%%%%
%%%%%%%%%%%%%%%%%%%%%%%%%%%%%%%%%%%%%%%%%%%%%%%%%%%%%%%%%%%%%%%%%%%%%%%%%%%%%%%%%%
%%%%%%%%%%%%%%%%%%%%%%%%%%%%%%%%%%%%%%%%%%%%%%%%%%%%%%%%%%%%%%%%%%%%%%%%%%%%%%%%%%

The class of universality of all topological transitions at the critical points discussed in this subsection is determined by the critical exponents  \eqref{IndAll}. As was first done by van Hove for $d=2,3$ \cite{VanHove:1953}, the index $\gamma$, for instance, can be extracted from the DOS calculated in the vicinity of critical points. It is determined \textit{solely by the topology of $\mathds{T}^d$}: only the signature of the quadratic expansion of the Morse function defined on $\mathds{T}^d$  (i.e., the index $s$) and the spatial dimension $d$  matter to get the result \eqref{IndAll}.

The $1d$ case is obvious. In $2d$ two extrema at $\varepsilon= \pm 2$ yield the discontinuities of $D_2(\varepsilon)$, while
at the saddle point $\varepsilon=0$ the DOS is logarithmically divergent $D_2(\varepsilon) \propto -\ln \varepsilon^2$ \cite{VanHove:1953}. The both types of singularities fall into the case $\gamma=0$, as understood in the theory of phase transitions. Similarly, calculating the area $\mathcal{S}_2(\varepsilon)$ from the quadratic expansion of the spectrum $\varepsilon(\mathbf{k})$ near $\varepsilon= \mp 2$, ($s=0,2$, two terms of the same sign) is sufficient to recover the exact result \eqref{gBI} to leading order: $\mathcal{P}_2 \propto |\varepsilon \pm 2|$ near the band ends. The quadratic expansion of $\varepsilon(\mathbf{q})$ near the saddle (van Hove) point ($s=1$, two terms of the opposite sign) agrees to leading order with the exact result \eqref{gHalf}: $\mathcal{P}_2-1  \propto -\varepsilon \ln \varepsilon^2$. So, we infer $\beta =1$ and the dynamical critical index $z=2$. Similar exercise with the expansions of $\varepsilon(\mathbf{q})$ near the critical points yields the $3d$ FST indices \eqref{IndAll}, in agreement with those obtained from the exact results, see Sec.~\ref{LYd3}.  Thus, we proved that the whole class universality FST \eqref{IndAll} is determined uniquely by $s$ and $d$,  where $d=1,2,3$.

To apply the Morse theory to the HK model,  we need to make an extension of the above analysis to account for the occurrence of several bands. For our purposes it suffices to deal with two bands defined on two tori, say left/right (L/R). We extend the definition \eqref{Me} as follows:
\begin{equation}
\label{MeLR}
 \forall ~\mathbf{k}  \in M_i^\varepsilon  \subseteq \mathds{T}_i^d: ~\varepsilon_i(\mathbf{k})\leq \varepsilon,~~i=L,R~.
\end{equation}
(Since in the case of the HK model the bands are distinguished by the constant energy shift only, one can visualize the problem as filling of two
communicating tori placed at different ground levels). The resulting manifold:
\begin{equation}
\label{MeSum}
  M^\varepsilon  \coloneqq  M_L^\varepsilon \cup  M_R^\varepsilon  \subseteq \mathds{T}_L^d \cup \mathds{T}_R^d~.
\end{equation}
 As follows from the above definition:
 \begin{equation}
\label{FSSum}
  \mathcal{S}_d(M^\varepsilon)  = \mathcal{S}_d(M_L^\varepsilon)+ \mathcal{S}_d(M_R^\varepsilon)~,
\end{equation}
and
\begin{equation}
\label{EulSum}
  \chi_E(M^\varepsilon)=   \chi_E(M_L^\varepsilon)+  \chi_E(M_R^\varepsilon)~.
\end{equation}

%
%
%%
%%%%%%%%%%%%%%%%%%%%%%%%%%%%%%%%%%%%%%%%%%%%%%%%%%%%%%%%%%%%%%%%%%%%%%%%%%%%%%
\subsection{Generalizations and universality}\label{MorsePerturb}
\begin{figure}[h]
\centering{\includegraphics[width=7.0cm]{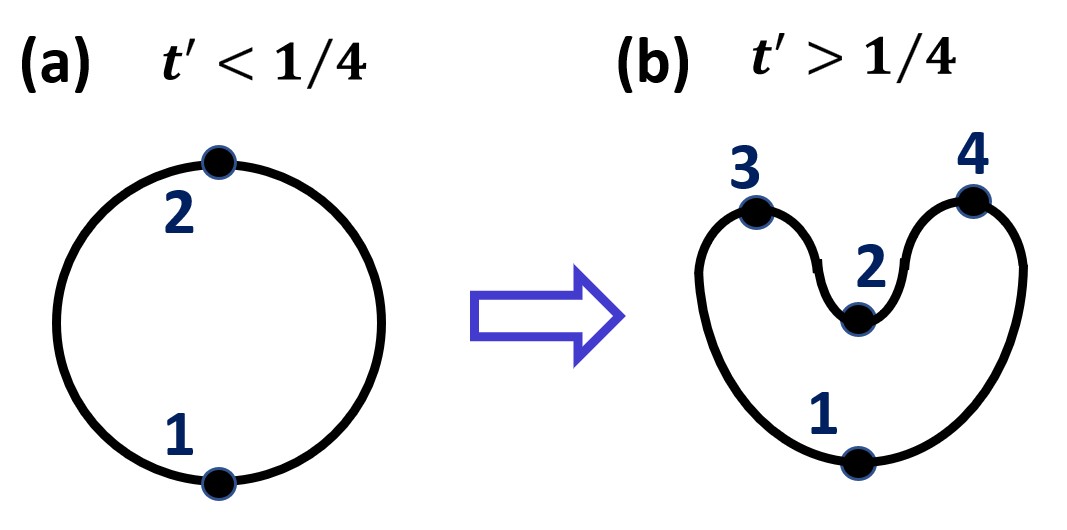}}
\caption{(a): circle $\mathds{T}^1=\mathds{S}^1$  at $t^\prime <1/4$, two critical points with indices \eqref{CPsD1}; (b): Deformed circle
at $t^\prime >1/4$ with four critical points: min (1), $s=0$, $C_0=1$; min (2), $s=0$, $C_0=1$; two maxima (3,4), $s=1$, $C_1=2$.}
\label{1Dmodif}
\end{figure}

%%%%%%%%%%%%%%%%%%%%%%%%%%%%%%%%%%%%%%%%%%%%%%%%%%%%%%%%%%%%%%%%%%%%%%%%%%%%%%
%%
Now we want to check how robust the results of the previous subsection are with respect to modifications of the spectrum (Morse function). To compare apples to apples, we preserve periodicity of the spectrum $\varepsilon(\mathbf{k})$ defined on $\mathds{T}^d$.
For instance we can include the next-nearest hopping $t^\prime>0$ along main axes. The spectrum
\begin{equation}
\label{epsNNN}
  \varepsilon(\mathbf{k})=- \sum_{i=1}^{d} (\cos k_i + t^\prime \cos(2 k_i) )
\end{equation}
at $t^\prime<1/4$ yields the same critical points and the critical behavior of the case $t^\prime=0$, up to irrelevant change of some numbers. At $t^\prime>1/4$ an extra IC critical point (or points, depending on $d$) appears, and additional analysis is warranted. In case of $d=1$ the maximum $\varepsilon(\pi)=1-t^\prime$ becomes minimum  when $t^\prime>1/4$. It is a new Lifshitz point where the Fermi sea becomes disconnected. See Fig.~\ref{1Dmodif}. The band gets fully filled at the energy $\varepsilon(x_{\s IC})=t^\prime+1/8t^\prime$ corresponding to two maxima at $x_{\s IC}= \pm \arccos(-1/4t^\prime)$.\footnote{\label{ICedge} There are certain subtleties when the band gets filled at the IC points, and not at the edge or at the center. For details, see analysis of the modulated ladder in \cite{Chitov:2025}.}

%%%%%%%%%%%%%%%%%%%%%%%%%%%%%%%%%%%%%%%%%%%%%%%%%%%%%%%%%%%%%%%%%%%%%%%%%%%%%%%%%
%%%%%%%%%%%%%%%%%%%%%%%%%%%%%%%%%%%%%%%%%%%%%%%%%%%%%%%%%%%%%%%%%%%%%%%%%%%%%%%%%
%%%%%%%%%%%%%%%%%%%%%%%%%%%%%%%%%%%%%%%%%%%%%%%%%%%%%%%%%%%%%%%%%%%%%%%%%%%%%%%%%%
%%%%%%%%%%%%%%%%%%%%%%%%%%%%%%%%%%%%%%%%%%%%%%%%%%%%%%%%%%%%%%%%%%%%%%%%%%%%%%%%%%

The Euler characteristic (compare to \eqref{ChiECircle}) at $t^\prime >1/4$ is found as follows:
\begin{equation}
\label{ChiED1mod}
 t^\prime >1/4:~~ \chi_E(M^\varepsilon) =
  \left\{
  \begin{array}{lr}
  0,~ &\varepsilon <-1 -t^\prime \\ [0.2cm]
  1,~ &-1 -t^\prime < \varepsilon <1-t^\prime \\ [0.2cm]
  2, ~ &1-t^\prime <\varepsilon <t^\prime+1/8t^\prime \\ [0.2cm]
  0, ~ & \varepsilon > t^\prime+1/8t^\prime \\
  \end{array}
  \right.
\end{equation}
The cases $d=2,3$ can be readily analyzed along these lines, revealing additional minima, maxima,  and saddle points at $t^\prime >1/4$.

\begin{figure}[h]
\centering{\includegraphics[width=12.0cm]{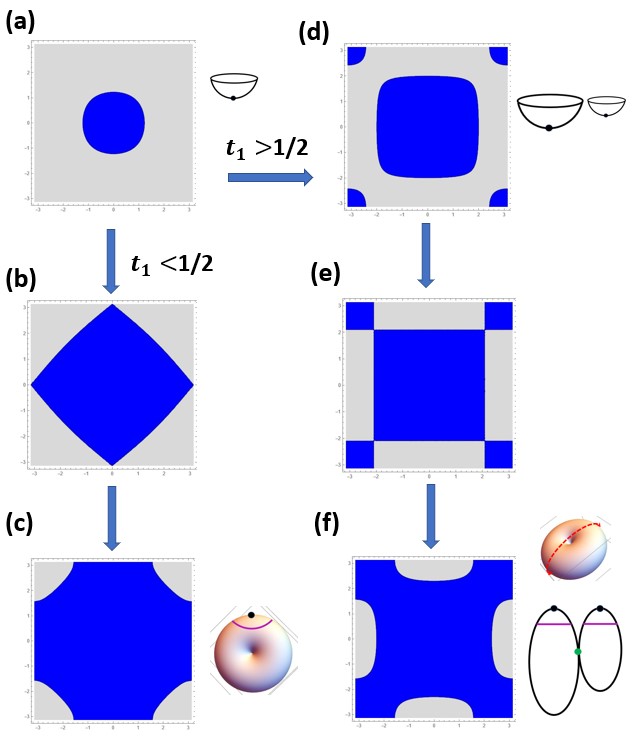}}
\caption{(a-c): Evolution of the FS at $t^\prime <1/2$ with increasing filling. The manifold $M^\varepsilon$ follows the steps shown in Fig.~\ref{Donuts}.
(d-f): Evolution of the FS at  $t^\prime >1/2$.  (a)$\to$ (d): At the second minimum (Lifshitz point) $\varepsilon=\varepsilon(\pi,\pi) = 2-2 t_1$, the FS becomes disconnected. $M^\varepsilon$  is a unity of two detached cups at $\varepsilon> 2-2 t_1$. (e): FS at the van Hove critical point, four IC saddle points at $\varepsilon=\varepsilon(\pm x_{\s IC},\pm y_{\s IC})=1/2t_1$.
(f): FS at $1/2t_1<\varepsilon<2t_1$. Passing though the van Hove point, four handles are attached to the two cups shown in (d). $M^\varepsilon$  is (connected) ``distorted'' umbilical torus with two holes. The deformation of the torus can be visualized as if the torus was tied with a tight loop drawn by red dashed line
and distorted then. Its cross-section is drawn in (f), bottom right. Two holes at the energy level $1/2t_1<\varepsilon<2t_1$ are shown in magenta. Two membranes are added at the maxima (black dots) when the band is fully filled, $M^\varepsilon \mapsto \mathds{T}^2$.
}
\label{tt1Evol}
\end{figure}

It is quite pedagogical to demonstrate in detail what happens to the $2d$ punctured doughnut (see Fig.~\ref{Donuts}) with modifications of the Morse function \eqref{Eps0}. Leaving the case \eqref{epsNNN} as an exercise, we present analysis of another interesting extension of \eqref{Eps0} in $2d$ with additional hopping ($t_1>0$) along diagonals of the elementary plaquette included. The spectrum is  \cite{Chitov:2005}:
\begin{equation}
\label{epst1}
  \varepsilon(k_x,k_y)=- \cos k_x - \cos k_y - 2t_1 \cos k_x  \cos k_y~.
\end{equation}
When $t_1<1/2$ the critical behavior is essentially the same as in the case $t^\prime=0$. At different energy levels the filled area of the torus $M^\varepsilon$ follows the path $a \rightarrow b \rightarrow c$ in Fig.~\ref{tt1Evol}, i.e., the same transformations as discussed in the context of Fig.~\ref{Donuts}.
At low filling above the minimum $\varepsilon(0,0)=-2 -2 t_1$, $M^\varepsilon$ is a cup, see Fig.~\ref{tt1Evol} a. At the saddle (van Hove) critical
point $\varepsilon(0,\pi)=\varepsilon(\pi,0)=2 t_1$ the cup gets two handles attached, and at $\varepsilon> 2 t_1$, $M^\varepsilon$ is a punctured
(umbilical) doughnut with a hole on the top (see Fig.~\ref{tt1Evol} c), until $M^\varepsilon \to \mathds{T}^2$ at the maximum  $\varepsilon=\varepsilon(\pi,\pi)=2 -2t_1$. While going through these critical points, the Euler characteristic changes exactly as in Eq.~\eqref{ChiETorus}.

When  $t_1>1/2$ the number of the critical points and their locations within the BZ change as follows:
 \begin{eqnarray}
  \mathrm{Min~1}&:&~~ \varepsilon(0,0) = -2-2 t_1,~~s=0,~ C_0=1,  \nonumber  \\
  \mathrm{Min~2}&:&~~ \varepsilon(\pi,\pi) = 2-2 t_1,~~s=0,~ C_0=1,  \nonumber  \\
  \mathrm{S}&:&~~ \varepsilon(\pm x_{\s IC},\pm y_{\s IC}) =1/2t_1,~~s=1,~ C_1=4,  \nonumber  \\
  \mathrm{Max}&:&~~ \varepsilon(0,\pi) =\varepsilon(\pi,0)= 2 t_1,~~s=2,~ C_2=2,
\label{CPtt1}
\end{eqnarray}
where $x_{\s IC}=y_{\s IC}=\arccos(-1/2t_1)$. Evolution of the FS within the BZ is shown in Fig.~\ref{tt1Evol} a,d,e,f. The manifold $M^\varepsilon$ undergoes quite
different transformations along the path $a \rightarrow d \rightarrow e \rightarrow f$ in Fig.~\ref{tt1Evol}, until the BZ $[-\pi,\pi]^2 \leftrightarrow \mathds{T}^2$ gets fully filled. At the second minimum (Lifshitz point) $M^\varepsilon$  becomes disconnected (it gets the second cup), and then at the van Hove critical point (four IC saddle points at $\varepsilon=1/2t_1$, see Fig.~\ref{tt1Evol} e) four handles are attached to the two cups. $M^\varepsilon$  becomes connected again: it can be viewed as a ``distorted'' umbilical torus with two holes. Its cross-section with two holes at the energy level $1/2t_1<\varepsilon<2t_1$ (shown in magenta) and two maxima (black dots) at $\varepsilon= 2t_1$, is schematically drawn in Fig.~\ref{tt1Evol} f.  Two membranes are added when the Morse function reaches the maxima, closing $M^\varepsilon \mapsto \mathds{T}^2$. The Euler characteristic changes at the critical points as follows:
\begin{equation}
\label{ChiEtt1}
 t_1>1/2:~~\chi_E(M^\varepsilon) =
  \left\{
  \begin{array}{lr}
  0,~ &\varepsilon <-2 -2t_1\\ [0.2cm]
  1,~ &-2-2t_1 < \varepsilon <2-2t_1  \\ [0.2cm]
  2,~ &2-2t_1 < \varepsilon <1/2t_1  \\ [0.2cm]
  -2, ~ &1/2t_1 <\varepsilon <2 t_1 \\ [0.2cm]
  0, ~ & \varepsilon > 2 t_1. \\
  \end{array}
  \right.
\end{equation}

%%%%%%%%%%%%%%%%%%%%%%%%%%%%%%%%%%%%%%%%%%%%%%%%%%%%%%%%%%%%%%%%%%%%%%%%%%%%%%%%%
%%%%%%%%%%%%%%%%%%%%%%%%%%%%%%%%%%%%%%%%%%%%%%%%%%%%%%%%%%%%%%%%%%%%%%%%%%%%%%%%%
%%%%%%%%%%%%%%%%%%%%%%%%%%%%%%%%%%%%%%%%%%%%%%%%%%%%%%%%%%%%%%%%%%%%%%%%%%%%%%%%%%
%%%%%%%%%%%%%%%%%%%%%%%%%%%%%%%%%%%%%%%%%%%%%%%%%%%%%%%%%%%%%%%%%%%%%%%%%%%%%%%%%%

From the above analysis of the perturbations \eqref{epsNNN}, \eqref{epst1} of the simplest tight-binding spectrum \eqref{Eps0} we can draw now some important general conclusions. Microscopic details of the Hamiltonian (e.g., lattice type, couplings, hopping, interactions) do change the number, the order of appearance, and the multiplicity of the critical points of the Morse function (it is the energy function of the conduction band from the physical point of view). In $d=1$ only maxima or minima are possible, while in higher dimensions saddle points occur in addition. Among the non-degenerate critical points the maxima and minima correspond to the continuous band, Mott, or Lifshitz transitions, while the saddle points correspond to the van Hove transitions. The indices $s$ of the critical points lie in the range:
\begin{equation}
\label{ds}
  \mathrm{For ~a~ given~} d:~~0 \leq s \leq d~.
\end{equation}
The types \eqref{CPsD1}, \eqref{CPsD2}, and \eqref{CPsD3} exhaust the possibilities, thus expansions near such critical points (band, Mott, Lifshitz, van Hove) will always yield $\beta$, $\gamma$, and $z$ within the FST universality class \eqref{IndAll}. Along with the standard attributes of the theory of continuous phase transitions, like the order parameter ($\mathcal{P}$), compressibility ($\chi$), etc, characterized by exponents  \eqref{IndAll}, at each critical point the topology of $M^\varepsilon$ undergoes a discrete transformation: $n$-cell (or cells) are added, which is signalled by a change of the Euler characteristic. The sequence of those transformations (and $\chi_E(M^\varepsilon)$) depends on the model's details (cf., e.g., Figs.~\ref{Donuts},\ref{1Dmodif}, and  \ref{tt1Evol}), but the totally filled band corresponds to $M^\varepsilon \mapsto \mathds{T}^2$, so $\chi_E=0$ for the band and Mott insulators.

The above analysis is not applicable for the degenerate critical points where the Hessian of $\varepsilon(\mathbf{k})$ vanishes. For example, the critical points of functions \eqref{epsNNN}, \eqref{epst1} are degenerate at $t^\prime=1/4$  and $t_1=1/2$, respectively. Such high-order critical points occur only at some particular (fine-tuned) values of the microscopic parameters, and we assume the situation with non-degenerate critical points of the Morse function as generic.
The critical exponents at the high-order points are different from \eqref{IndAll} and need to be studied on the case-by-case basis. See, e.g., \cite{Yuan:2020,Betouras:2020,Betouras:2025} and more references there.

%xxxxxxxxxxxxxxxxxxxxxxxxxxxxxxxxxxxxxxxxxxxxxxxxxxxxxxxxxxxxxxxxxxxxxxxxxxxxxx

%%%%%%%%%%%%%%%%%%%%%%%%%%%%%%%%%%%%%%%%%%%%%%%%%%%%%%%%%%%%%%%%%%%%%%%%%%%%%%
\section{Special points in the M2-phase (d=2)}\label{M2CPs}
\begin{figure}[h]
\centering{\includegraphics[width=10.0cm]{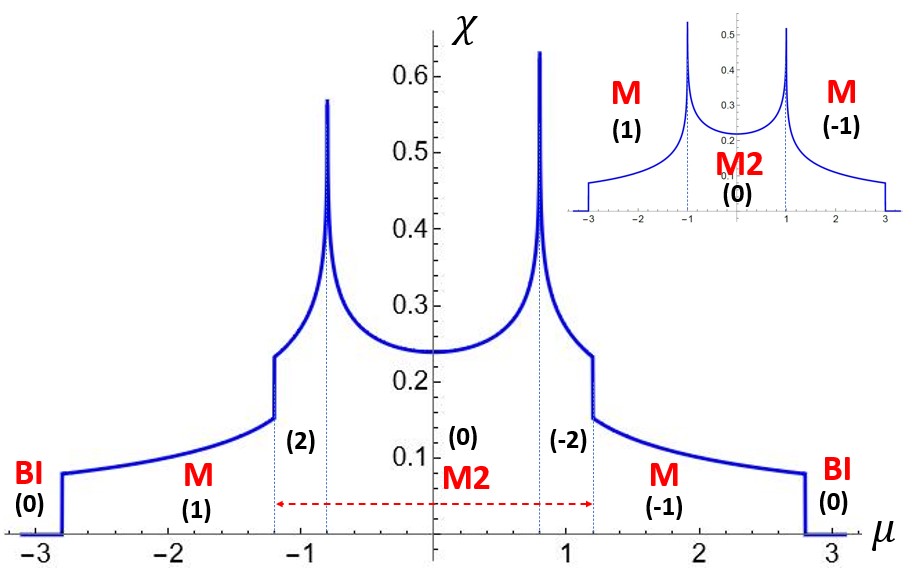}}
\caption{$d=2$, $U<2$ ($U_c=4$). The compressibility $\chi(\mu)$ for $U=1.6$.
The critical values of chemical potential: $\pm \mu_c=\pm 2.8$, $\pm \mu_L=\pm 1.2$.
The van Hove critical points correspond $\pm \mu_{vH}=\pm 0.8$ ($\pm \mu_{\circ,vH}=\pm 0.5898$ ) and
$\bar n(-\mu_{\circ,vH})=0.283547$,  $\bar n(\mu_{\circ,vH})=0.716438$. The special case $U=2$ when the Lifshitz and van Hove critical points merge
is shown in the inset. The Euler characteristics $\chi_E$ for each phase are shown in parentheses.}
\label{M2Chi}
\end{figure}

%%%%%%%%%%%%%%%%%%%%%%%%%%%%%%%%%%%%%%%%%%%%%%%%%%%%%%%%%%%%%%%%%%%%%%%%%%%%%%
%
%
In case $U<2$ the order of the Lifshitz and the van Hove transitions change with the respect to the case shown in Fig.~\ref{2DFilChi} (c).
The van Hove transition occurs inside the M2-phase with two FSs. See Fig.~\ref{M2Chi}.
In case of two FSs (M2-metal), we get from the LT \eqref{LuttG}, \eqref{LuttDiffD} the exact result:
\begin{equation}
\label{DerM2}
  \frac{d \mu}{d \mu_\circ}= \frac{2 \mathbf{K}(1- \mu_\circ^2/4)}{\mathbf{K}(1-\mu_+^2/4)+\mathbf{K}(1-\mu_-^2/4)}~,
\end{equation}
which becomes
\begin{equation}
\label{DerM1}
  \frac{d \mu}{d \mu_\circ}= \frac{2 \mathbf{K}(1- \mu_\circ^2/4)}{\mathbf{K}(1-\mu_\pm^2/4)}
\end{equation}
in the M-phase with only one FS. From comparison of Eqs.~\eqref{DerM2} and \eqref{DerM1} we infer immediately occurrence of a discontinuity in the slope of $\mu(\mu_\circ)$ at the Lifshitz points separating the M2- and M-phases.

%%%%%%%%%%%%%%%%%%%%%%%%%%%%%%%%%%%%%%%%%%%%%%%%%%%%%%%%%%%%%%%%%%%%%%%%%%%%%%%%%
%%%%%%%%%%%%%%%%%%%%%%%%%%%%%%%%%%%%%%%%%%%%%%%%%%%%%%%%%%%%%%%%%%%%%%%%%%%%%%%%%
%%%%%%%%%%%%%%%%%%%%%%%%%%%%%%%%%%%%%%%%%%%%%%%%%%%%%%%%%%%%%%%%%%%%%%%%%%%%%%%%%%
%%%%%%%%%%%%%%%%%%%%%%%%%%%%%%%%%%%%%%%%%%%%%%%%%%%%%%%%%%%%%%%%%%%%%%%%%%%%%%%%%%

At the van Hove critical points $\mu= \pm U/2$ with arbitrary $U$ the LT \eqref{LuttG} yields
\begin{equation}
\label{LuttVH}
   2 g(\pm \mu_\circ)= \pm g_R(U)~,
\end{equation}
and the filling is
\begin{equation}
\label{fillVH}
   \bar n = \frac12 \pm \frac{1}{8 \pi^2} g_R(U)~.
\end{equation}
As folows from Eqs.~\eqref{LuttVH} and \eqref{fillVH}, for $U>2$, when  $g_R(U)=2\pi^2$, the transition occurs at commensurate fillings  $\bar n=1/4,3/4$, corresponding to the values of noninteracting chemical potential $\mp \mu_{\circ,s}= \mp 0.7198$, cf. Eq.~\eqref{134}:
\begin{equation}
\label{muOvHBigU}
 g(\mu_{\circ,s})= \pi^2~\mapsto~n(\mp \mu_{\circ,s})=1/4,3/4~.
\end{equation}
This case is analyzed in the main text, Sec.~\ref{2D}. For $U<2$ the van Hove transition occurs at a generic IC filling
\begin{equation}
\label{fillVHIC}
   \bar n(\pm \mu_{\circ,vH}) = \frac12 \pm \frac{1}{8 \pi^2} g(U)~,
\end{equation}
where the critical values of the noninteracting chemical potential $\mu_\circ =\pm \mu_{\circ,vH}$ are found from the following equation:
\begin{equation}
\label{mu0vH}
   g(\mu_{\circ,vH})= \frac12 g(U)~.
\end{equation}
This is easy to understand qualitatively from comparison of Figs.~\ref{2DFilChi} (c) and \ref{M2Chi}: in case of larger separation of the upper and lower bands
($U>2$), the lower band becomes 1/2-filled at $\mu_{vH,1}=-\mu_{vH}$ when the upper band is still empty, while at $\mu_{vH,2}=\mu_{vH}$ the lower band is full and the upper band gets 1/2-filled. The separation of the bands is smaller at $U<2$, and 1/2-filling of the lower band at $\mu_{vH,1}$ occurs simultaneously
with some partial filling of the upper band (lower $\leftrightarrow$ upper at $\mu_{vH,2}$).

Using in Eq.~\eqref{LuttG} the expansions \eqref{gTaylor} and \eqref{gHalf} near the critical points, we get to lowest order the same equation for $\mu(\mu_{\circ},U)$ as Eq.~\eqref{MuTrans}, with the only difference that on its l.h.s. $\mu_{\circ,s} \mapsto \mu_{\circ,vH}$ and $g_1 \coloneqq g^\prime(\mu_{\circ,s}) \mapsto g^\prime(\mu_{\circ,vH})$. Note that at the van Hove points only one of the terms $\mathbf{K}(1- \mu_\pm^2 /4)$ in the denominator of \eqref{DerM2} is logarithmically divergent, while the rest is regular. So, the results for $\mu(\mu_\circ)$ near the van Hove transition in the M-phase, stated  around Eqs.~\eqref{MuTrans} and \eqref{muVH2D}, hold for the M2-phase, up to inessential differences in the numbers $g^\prime$  ($\propto$ DOS) and $\mu_{\circ,vH}$.
Note also that conclusions in  the main text about the change of the FS topological properties and of the Euler number at the van Hove points, are not affected
wether the latter occur in the M-phase ($U>2$) at special commensurate fillings or in the M2-phase ($U<2$) at rational fillings. In both cases the van Hove transition corresponds to the saddle point at the center of one the two tori, see Appendix \ref{Morse}. For completeness we indicated the Euler characteristic for each phase in Fig.~\ref{M2Chi}.

Analysis of the special point $U=2$, where the Lifshitz and van Hove critical points merge (see inset in Fig.~\ref{M2Chi}), shows that the leading order
solution \eqref{muVH2D} for $\mu(\mu_\circ)$ is applicable in this case with $\mu_{\circ,vH} \to \mu_{\circ,\s L} \to \mu_{\circ,\s s}$, cf. Eqs.~\eqref{mu0vH}, \eqref{mu0L}, \eqref{muOvHBigU}, and \eqref{134}. The occurrence of the van Hove/Lifshitz transitions at either special commensurate, or generic incommensurate fillings does not contradict the plateau theorem, since they take place in the gapless phases.

%
%xxxxxxxxxxxxxxxxxxxxxxxxxxxxxxxxxxxxxxxxxxxxxxxxxxxxxxxxxxxxxxxxxxxxxxxxxxxxxx
\end{appendix}
%xxxxxxxxxxxxxxxxxxxxxxxxxxxxxxxxxxxxxxxxxxxxxxxxxxxxxxxxxxxxxxxxxxxxxxxxxxxxxx
%xxxxxxxxxxxxxxxxxxxxxxxxxxxxxxxxxxxxxxxxxxxxxxxxxxxxxxxxxxxxxxxxxxxxxxxxxxxxxx

%xxxxxxxxxxxxxxxxxxxxxxxxxxxxxxxxxxxxxxxxxxxxxxxxxxxxxxxxxxxxxxxxxxxxxxxxxxxxxx
\section*{Data Availability}
The additional data that support the findings of this article are available
from the author upon reasonable request.
%xxxxxxxxxxxxxxxxxxxxxxxxxxxxxxxxxxxxxxxxxxxxxxxxxxxxxxxxxxxxxxxxxxxxxxxxxxxxxx

%%%%%%%%%%%%%%%%%%%%%%%%%%%%%%%%%%%%%%%%%%%%%%%%%%%%%%%%%%%%%%%%%%%%%%%%%%%%%%
%%%%%%%%%%%%%%%%%%%%%%%%%%%%%%%%%%%%%%%%%%%%%%%%%%%%%%%%%%%%%%%%%%%%%%%%%%%%%%
\bibliography{C:/Users/gchitov/Documents/Papers/BibRef/CondMattRefs}
%\bibliography{C:/Papers/BibRef/CondMattRefs}
%\bibliography{CondMattRefs}
%\bibliographystyle{apsrev4-1} \bibliography{ref2}
%%%%%%%%%%%%%%%%%%%%%%%%%%%%%%%%%%%%%%%%%%%%%%%%%%%%%%%%%%%%%%%%%%%%%%%%%%%%%%
%
%

\newpage

%%%%%%%%%%%%%%%% START OF SUPPLEMENT %%%%%%%%%%%%%%%

% Figures, tables, equations and pages in the supplement are numbered S1, S2 etc.
\renewcommand{\thefigure}{S\arabic{figure}}
\renewcommand{\thetable}{S\arabic{table}}
\renewcommand{\theequation}{S\arabic{equation}}
\renewcommand{\thepage}{S\arabic{page}}
\setcounter{figure}{0}
\setcounter{table}{0}
\setcounter{equation}{0}
\setcounter{page}{1} % not 0 as \newpage already started a supplementary page
% References continue the numbering from the main text.

\section*{Supplementary Materials}\label{SM}

\begin{center}

\centerline

\textbf{\large Topology of the Fermi surface and universality of the metal-metal and metal-insulator transitions:
$d$-dimensional Hatsugai-Kohmoto model as an example}\\[1.0cm]

\large  Gennady Y. Chitov\\[0.1cm]

\large \textit{Bogoliubov Laboratory of Theoretical Physics,
Joint Institute for Nuclear Research, Dubna 141980, Russia}\\[1.5cm]
\end{center}

\begin{figure}[htbp]
\begin{flushleft}
% Main caption for the entire group
  \caption{$1d$ HK model in the weak coupling regime $U<U_c=2$ ($U=1$) at different fillings. (a)-(e): $\mu=-1.2, -0.7, 0.2, 0.7, 1.2$. At Lifshitz/BI points $\mu_L =\pm 0.5$ and $\mu_c =\pm 1.5$, resp. Each item (a)-(e) of the figure contains: top -- spectra $\varepsilon_\pm(k)$ (red/blue), free band $\varepsilon(k)$ (dotted gray) within BZ $k \in [-\pi,\pi]$, and $\mu$ (dashed magenta); middle -- distribution function per spin $n_{k,\sigma}=0,1/2,1$; bottom -- Green's function $G_\sigma(k,0)$.  Red dots indicate locations of the LY zeros (top) $\leftrightarrow$ Fermi momenta $k_F$ (middle) $\leftrightarrow$ poles of $G_\sigma(k)$ (bottom). Zeros of $G_\sigma(k)$ are shown by bold blue crosses.}
\label{PZgls}
    \vspace{5pt} % Optional: adds a tiny bit of space below the caption
%\begin{flushleft}
 %   \centering
    % First subfigure
    \begin{subfigure}[b]{\textwidth}
 %     \centering
        \includegraphics[width=\textwidth]{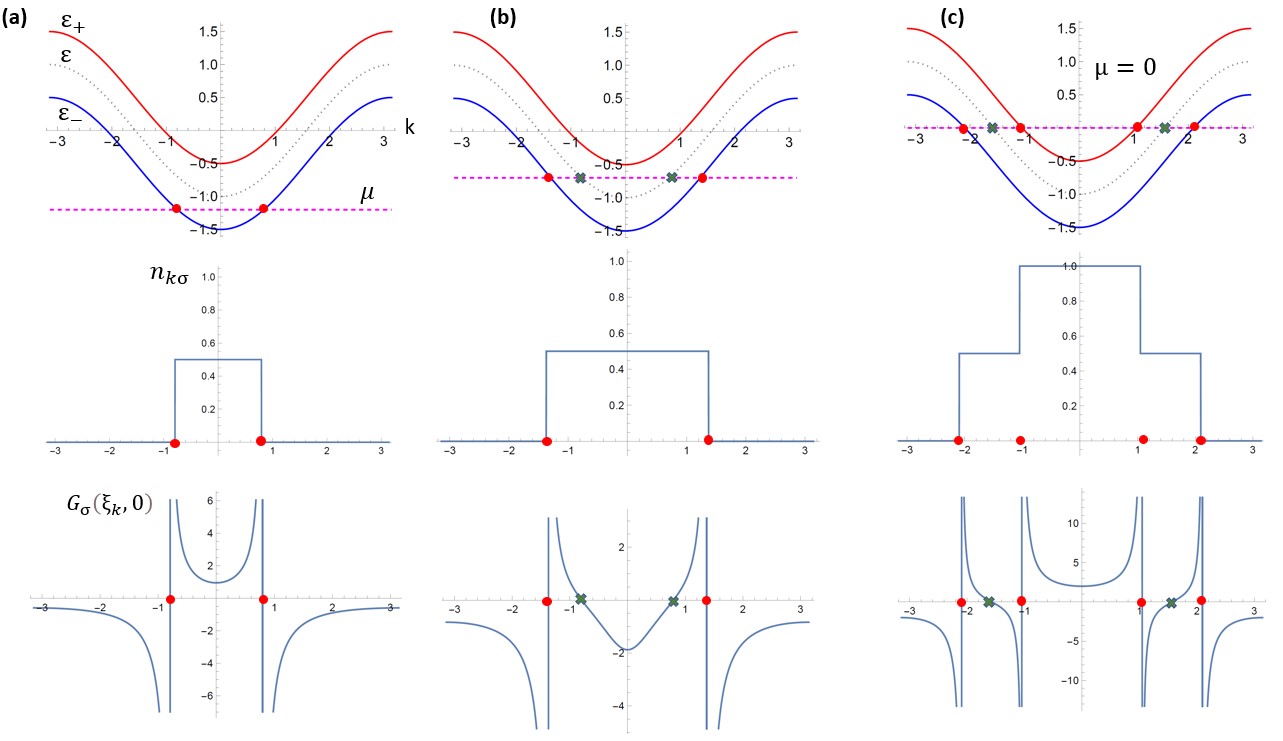}
  %      \caption{First image description.}
  %      \label{fig:sub1}
    \end{subfigure}
   \hfill % Adds horizontal space between figures
    \vspace{5em}
    % Second subfigure
    \begin{subfigure}[b]{0.70\textwidth}
  %      \centering
        \includegraphics[width=\textwidth]{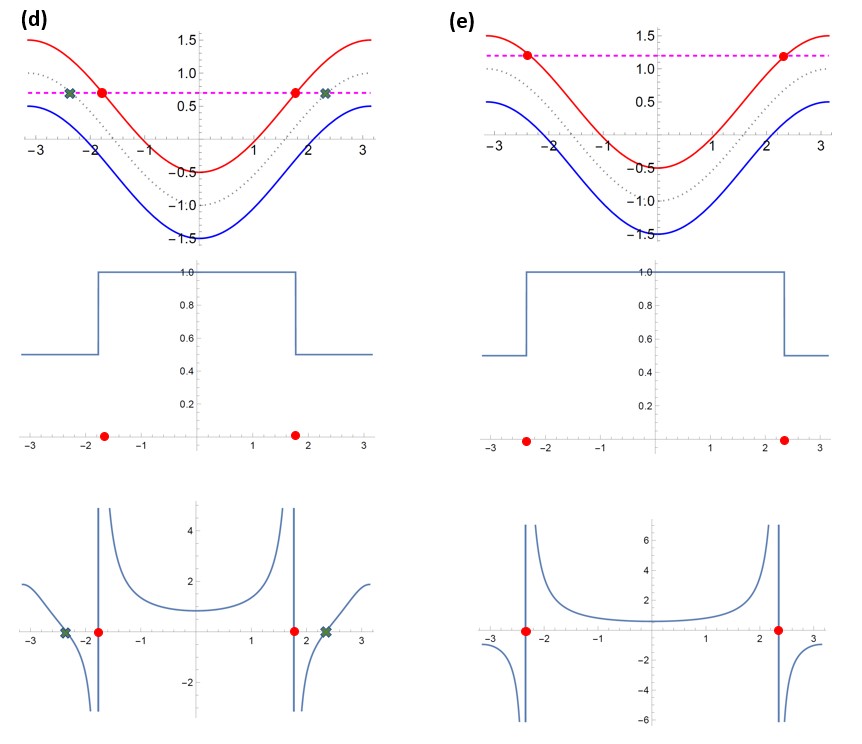}
  %      \caption{Second image description.}
  %      \label{fig:sub2}
    \end{subfigure}
\end{flushleft}
    % Main caption for the entire group
    %\caption{Poles and zeros in the gapless regime $U<U_c$.}
    %\label{PZgls}
\end{figure}

\large
To make a simple visual presentation of the discussion of the Green's function properties in Appendix~\ref{GFApp}, we plot the interacting $\varepsilon_\pm(k)$ and non-interacting $\varepsilon(k)$ spectra along with the chemical potential $\mu$ within the BZ; the distribution functions per spin $n_{k,\sigma}$; and the Green's function $G_\sigma(k,0)$ in the regimes of weak ($U<U_c$, Fig.~\ref{PZgls}) and strong ($U>U_c$, Fig.~\ref{PZMott}) couplings for the simplest case $d=1$.
The collected characteristic plots following growth of $\mu$, show distributions of real zeros and poles of $G_\sigma$ (the latter are also the LY zeros) in the BZ for different phases.

The half-filled case $\mu=0$, $U=1$ in Fig.~\ref{PZgls}~(c), when relation \eqref{FLS-MI} between the Luttinger and two (interacting/non-interacting) Fermi surfaces holds,
corresponds to
\begin{eqnarray}
\label{kFfree}
 k_L&=&k_F(U=0)=\arccos(0)=\frac{\pi}{2}~,  \\
\label{kFmu0}
 k_F^\pm &=& \arccos(\pm U/2)= \frac{2\pi}{3}; \frac{\pi}{3}~.
\end{eqnarray}

The half-filled case $\mu_{M,1} \leq \mu \leq \mu_{M,1}$, $U=3$ in Fig.~\ref{PZgls}~(c)-(f) corresponds to the MI phase. With the redefinition $\mu \equiv 0$ inside the whole MI phase, see Fig.~\ref{PZgls}~(d), Eq.~\eqref{FLS-MI} holds with $k_L$ and $k_F(U=0)$ from Eq.~\eqref{kFfree}, while $\mathcal{S}_d(k_F \to \pi)=2\pi$ comes from the fully filled lower band.

%%%%%%%%%%%%%%%%%%%%%%%%%%%%%%%%%%%%%%%%%%%%%%%%%%%%%%%%%%%%%%%%%%%%%%%%%%%%%%%%%
%%%%%%%%%%%%%%%%%%%%%%%%%%%%%%%%%%%%%%%%%%%%%%%%%%%%%%%%%%%%%%%%%%%%%%%%%%%%%%%%%
%%%%%%%%%%%%%%%%%%%%%%%%%%%%%%%%%%%%%%%%%%%%%%%%%%%%%%%%%%%%%%%%%%%%%%%%%%%%%%%%%
%%%%%%%%%%%%%%%%%%%%%%%%%%%%%%%%%%%%%%%%%%%%%%%%%%%%%%%%%%%%%%%%%%%%%%%%%%%%%%%%%

\newpage

%%%%%%%%%%%%%%%%%%%%%%%%%%%%%%%%%%%%%%%%%%%%%%%%%%%%%%%%%%%%%%%%%%%%%%%%%%%%%%%%%
%%%%%%%%%%%%%%%%%%%%%%%%%%%%%%%%%%%%%%%%%%%%%%%%%%%%%%%%%%%%%%%%%%%%%%%%%%%%%%%%%
% --- FIRST PAGE: Images (a)  (f) ---
\begin{figure}[p]
\begin{flushleft}
% Main caption for the entire group
  \caption{$1d$ HK model in the  in the strong coupling regime $U>U_c=2$ ($U=3$) at different fillings. (a)-(h): $\mu=-1.5, -0.7, 0, 0.4, 0.5, 1.5, 2.5$. At Mott/BI points $\mu_M =\pm 0.5$ and $\mu_c =\pm 2.5$, resp. Each item (a)-(h) of the figure contains: top -- spectra $\varepsilon_\pm(k)$ (red/blue), free band $\varepsilon(k)$ (dotted gray) within BZ $k \in [-\pi,\pi]$, and $\mu$ (dashed magenta); middle -- distribution function per spin $n_{k,\sigma}=0,1/2,1$; bottom -- Green's function $G_\sigma(k,0)$.  Red dots indicate locations of the LY zeros (top) $\leftrightarrow$ Fermi momenta $k_F$ (middle) $\leftrightarrow$ poles of $G_\sigma(k)$ (bottom). Zeros of  $G_\sigma(k)$ are shown by bold blue crosses.}
\label{PZMott}
    \vspace{5pt} % Optional: adds a tiny bit of space below the caption
 %   \centering
    % First subfigure
    \begin{subfigure}[b]{\textwidth}
 %     \centering
        \includegraphics[width=\textwidth]{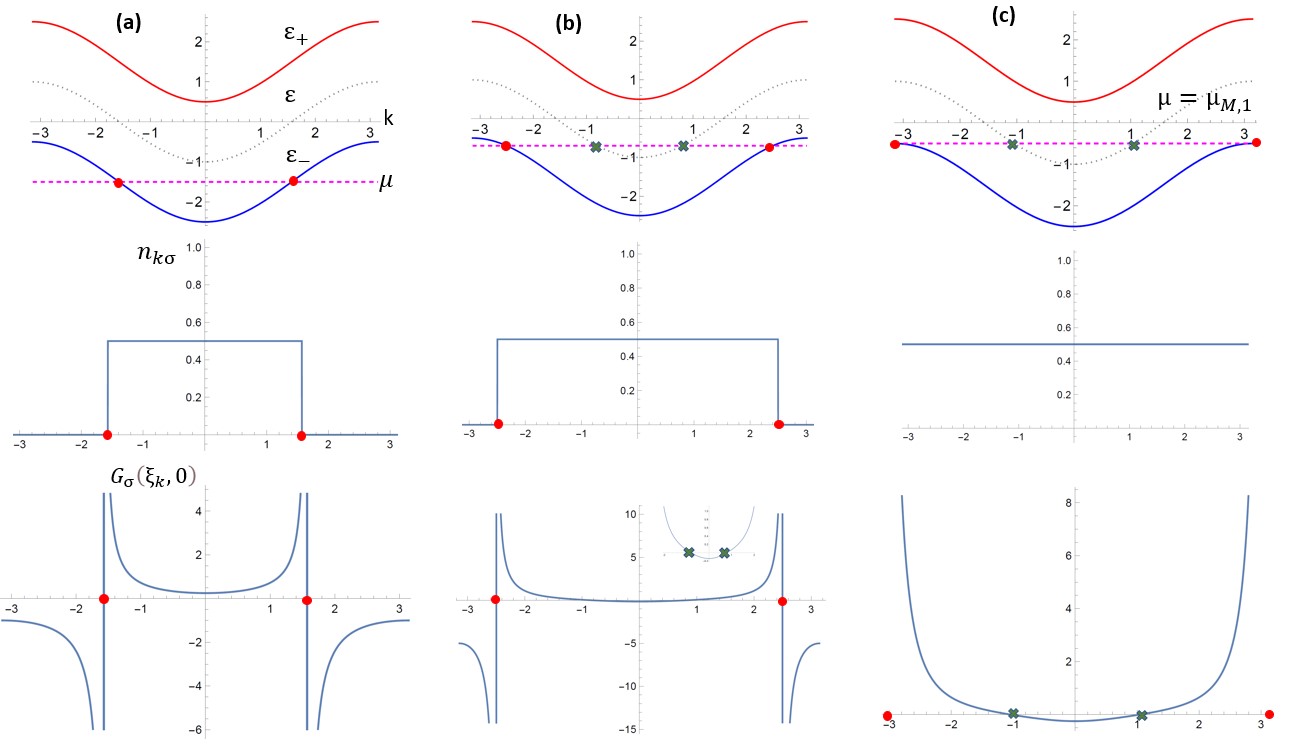}
  %      \caption{First image description.}
  %      \label{fig:sub1}
    \end{subfigure}
   \hfill % Adds horizontal space between figures
    \vspace{1em}
    % Second subfigure
    \begin{subfigure}[b]{\textwidth}
  %      \centering
        \includegraphics[width=\textwidth]{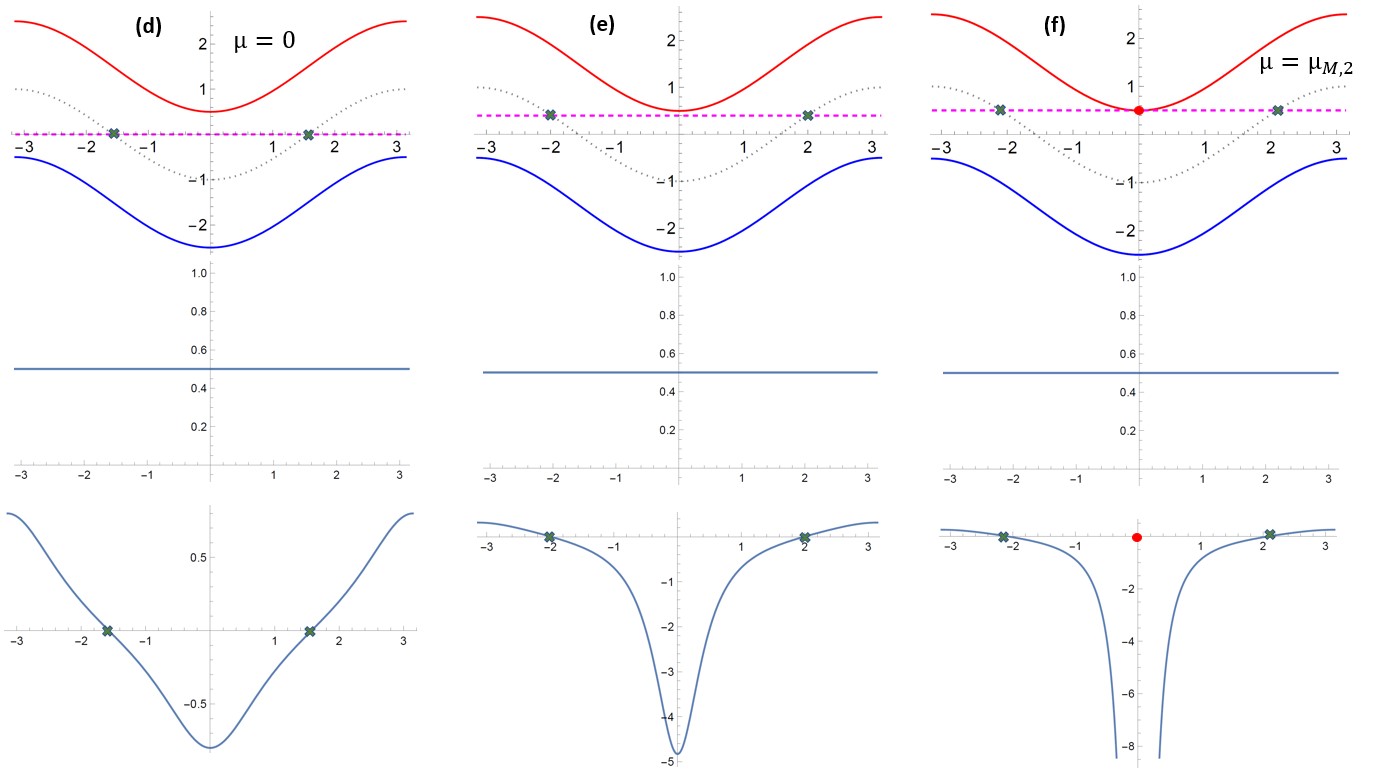}
  %      \caption{Second image description.}
  %      \label{fig:sub2}
    \end{subfigure}
  \end{flushleft}
\end{figure}
\clearpage % Forces LaTeX to go to the next page
%
% --- SECOND PAGE: Images (g) and (h) ---
  \begin{figure}[p]
  \ContinuedFloat % <--- Crucial command! Tells LaTeX this is a continuation.
  \begin{flushleft}
    % Third subfigure
    \begin{subfigure}[b]{0.7\textwidth}
  %      \centering
        \includegraphics[width=\textwidth]{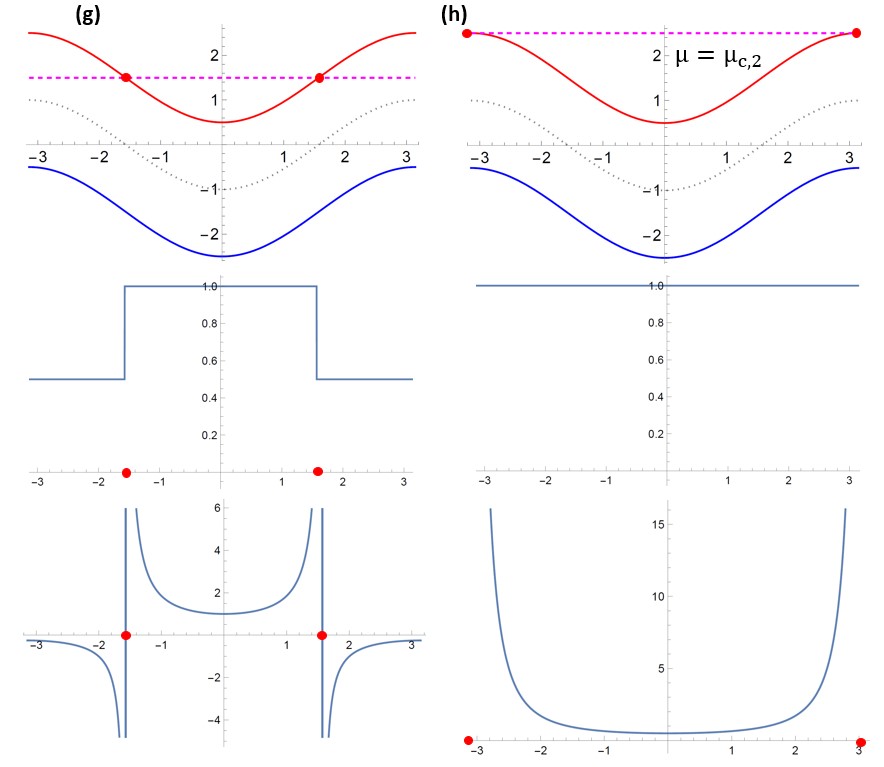}
  %      \caption{Second image description.}
  %      \label{fig:sub2}
    \end{subfigure}
\end{flushleft}
    % Main caption for the entire group
    %\caption{Poles and zeros in the gapless regime $U<U_c$.}
    %\label{PZgls}
\end{figure}
%%%%%%%%%%%%%%%%%%%%%%%%%%%%%%%%%%%%%%%%%%%%%%%%%%%%%%%%%%%%%%%%%%%%%%%%%%%%%%%%%
%%%%%%%%%%%%%%%%%%%%%%%%%%%%%%%%%%%%%%%%%%%%%%%%%%%%%%%%%%%%%%%%%%%%%%%%%%%%%%%%%

\end{document}